\documentclass{aa}
\usepackage[utf8]{inputenc}
\usepackage[varg]{txfonts}
\usepackage{fixltx2e}
\usepackage{color}
\usepackage{amssymb}
\usepackage{amsmath}
\usepackage[colorlinks=true,citecolor=blue,linkcolor=blue,urlcolor=blue]{hyperref}

\bibpunct{(}{)}{;}{a}{}{,} % following A&A style

\begin{document}

\defcitealias{Hitomi16}{H16}
\defcitealias{HitomiDynamics18}{H18}

\title{Measuring bulk flows of the intracluster medium in the Perseus and Coma galaxy clusters using \emph{XMM-Newton}}
\titlerunning{Bulk flows in the Perseus and Coma clusters}
\authorrunning{J.~S.~Sanders et.~al}

\author{J.~S. Sanders \inst{1}
  \and K.~Dennerl \inst{1}
  \and H.~R.~Russell \inst{2,3}
  \and D.~Eckert \inst{4,1}
  \and C.~Pinto \inst{5,3}
  \and A.~C.~Fabian \inst{3}
  \and S.~A.~Walker \inst{6,7}
  \and T.~Tamura \inst{8}
  \and J.~ZuHone \inst{9}
  \and F.~Hofmann \inst{1}
  }
\institute{
  Max-Planck-Institut f\"ur extraterrestrische Physik,
  Giessenbachstraße 1, D-85748 Garching, Germany
  \and School of Physics and Astronomy, The University of Nottingham, University Park, Nottingham, NG7 2RD, UK
  \and Institute of Astronomy, Madingley Road, Cambridge, CB3 0HA, UK
  \and Observatoire de Genève, Chemin des Maillettes, 51, 1290 Versoix, Switzerland
  \and ESTEC, European Space Agency, Keplerlaan 1, 2201 AZ, Noordwijk, Netherlands
  \and Astrophysics Science Division, X-ray Astrophysics Laboratory, Code 662, NASA Goddard Space Flight Center, Greenbelt, MD 20771, USA
  \and Department of Physics and Astronomy, University of Alabama in Huntsville, Huntsville, AL 35899, USA
  \and Institute of Space and Astronautical Science (ISAS), Japan Aerospace Exploration Agency (JAXA), Kanagawa 252-5210, Japan
  \and Harvard-Smithsonian Center for Astrophysics, 60 Garden Street, Cambridge, MA 02138, USA
  }

\date{Received ---, Accepted ---}

\abstract{
  We demonstrate a novel technique for calibrating the energy scale of the EPIC-pn detector on \emph{XMM-Newton}, which allows us to measure bulk flows in the intracluster medium (ICM) of the Perseus and Coma galaxy clusters.
  The procedure uses the fluorescent instrumental background lines present in all observations, in particular, Cu-K$\alpha$.
  By studying their spatial and temporal variations, in addition to incorporating calibration observations, we refined the absolute energy scale of the detector to better than $150\:\textrm{km s}^{-1}$ at the Fe-K line, a large improvement over the nominal calibration accuracy of $550\:\textrm{km s}^{-1}$.
  With our calibration, we mapped the bulk motions over much of the central 1200 and 800~kpc of Perseus and Coma, respectively, in spatial regions down to 65 and 140~kpc size.
  We cross-checked our procedure by comparing our measurements with those found in Perseus by \emph{Hitomi} for an overlapping 65~kpc square region, finding consistent results.
  For Perseus, there is a relative line-of-sight velocity increase of $480 \pm 210\:\textrm{km s}^{-1}$ $(1\sigma)$ at a radius of 250~kpc east of the nucleus.
  This region is associated with a cold front, providing direct evidence of the ICM sloshing in the cluster potential well.
  Assuming the intrinsic distribution of bulk motions is Gaussian, its width is $214 \pm 85 \:\textrm{km s}^{-1}$, excluding systematic uncertainties.
  Removing the sloshing region, this is reduced to $20$--$150\:\textrm{km s}^{-1}$, which is similar in magnitude to the \emph{Hitomi} line width measurements in undisturbed regions.
  In Coma, the line-of-sight velocity of the ICM varies between the velocities of the two central galaxies.
  Maps of the gas velocity and metallicity provide clues about the merger history of the Coma, with material to the north and east of the cluster core having a velocity similar to \object{NGC 4874}, while that to the south and west has velocities close to \object{NGC 4889}.
  Our results highlight the difference between a merging system, such as Coma, where we observe a $\sim 1000$~km~s$^{-1}$ range in velocity, and a relatively relaxed system, such as Perseus, with much weaker bulk motions.
}


\maketitle

\section{Introduction}
The velocity structure of the intracluster medium (ICM), the dominant baryonic component of galaxy clusters, remains poorly observationally constrained except in the core of the \object{Perseus cluster} \citep[][hereafter H16]{Hitomi16}.
Simulations predict that the ICM should contain turbulent or random motions, and bulk flows caused by the merger of other clusters and subcomponents \citep[e.g.][]{Lau09, Vazza11}.
In addition, there can be relative bulk motions of a few hundred km~s$^{-1}$ caused by sloshing of the ICM in the potential well, generated by merging substructures \citep[e.g.][]{Ascasibar06,ZuHone18}.
The central active galactic nucleus (AGN), via the action of its jets and inflation of relativistic bubbles, also likely generates motions of a few hundred km~s$^{-1}$ \citep[e.g.][]{Bruggen05,Heinz10}.

Measuring the velocities of the ICM is important for several reasons.
Turbulent motions provide additional pressure support, particularly at large radii, which affects calculations of hydrostatic equilibrium and cluster mass estimates \citep[e.g.][]{Lau09}.
AGN feedback injects a great deal of energy into the ICM which replaces that which is lost radiatively by X-ray emission \citep[reviewed in][]{Fabian12}.
Feedback has to provide distributed heating but how this works depends on the microphysics of the ICM and the balance between turbulence and sound waves or shocks.
Measuring the associated velocities would help constrain AGN feedback models.
Velocity is an excellent probe of the microphysics of the ICM, such as viscosity, which acts to smooth velocity structure \citep[e.g.][]{Reynolds05,ZuHone18}.
Simulations predict a close connection between turbulent thermodynamic and velocity power spectra \citep{Gaspari14} which should be tested.
In addition, measuring velocities will help explain the processes taking place in cluster mergers and the sloshing of gas in cold fronts, which is believed to last for several Gyr \citep{Ascasibar06,Roediger12}.

\emph{Hitomi} \citep{HitomiTel16} with its high spectral resolution microcalorimeter Soft X-ray Spectrometer detector \citep{HitomiSXS16} directly determined bulk and random motions in the \object{Perseus cluster} by measuring the position and width of the Fe-K emission lines \citepalias{Hitomi16}.
It found a line-of-sight velocity dispersion of $164 \pm 10 \:\textrm{km s}^{-1}$ between radii of 30 and 60 kpc and a gradient of $150\:\textrm{km s}^{-1}$ bulk flow across 60 kpc of the cluster core.
Away from the central nucleus and the northwestern cavity, the dispersion drops to $\sim 100\:\textrm{km s}^{-1}$ \citep[][hereafter H18]{HitomiDynamics18}.
Therefore, the Perseus core is not strongly turbulent, despite the obvious impact of the AGN and its jets on the surrounding ICM \citep[e.g.][]{FabianPer00}.
Unfortunately, \emph{Hitomi} was lost and will not be available to observe more clusters, in particular merging galaxy clusters which would have probed different physical processes.
The next missions equipped with a high-resolution instrument capable of measuring velocities are likely to be \emph{XRISM} (the replacement for \emph{Hitomi}, formerly known as \emph{XARM}), to be launched in 2022, and \emph{Athena} in the early 2030s.

Prior to \emph{Hitomi}, \emph{Suzaku} obtained velocities in several systems using the Fe-K line thanks to its accurate calibration, despite only having a charge-coupled device (CCD) detector.
\cite{Tamura14} examined multiple pointings, placing limits on relative velocities of $300 \:\textrm{km s}^{-1}$ over scales of 400 kpc in Perseus while on larger scales below $600\textrm{ km s}^{-1}$, with hints of detection of velocity structure at $150-300\:\textrm{km s}^{-1}$ in the cold front $45$ to $90$~kpc west of the cluster.
\cite{Ota16} examined several clusters with \emph{Suzaku}, including the central region of the Perseus cluster.
Systematic errors from the \emph{Suzaku} calibration were around $300 \:\textrm{km s}^{-1}$ and it had a large point-spread function (PSF) of around 1.8~arcmin (half power diameter).
\emph{Chandra} has also been used to measure gas velocities, finding evidence of bulk motions in merging massive clusters \citep{Liu16}.

Another direct method for measuring velocities is to use the \emph{XMM} Reflection Grating Spectrometer (RGS) to limit or measure line widths \citep[e.g.][]{Sanders10_A1835, Pinto15}.
Because of the slitless nature of the RGS, this is only effective in the central parts of peaked clusters.
In several systems, limits below $300\:\textrm{km s}^{-1}$ were obtained using this technique \citep{SandersVel13}, which agrees with the low turbulent velocities later found by \emph{Hitomi} in Perseus.

There are other indirect methods for measuring motions, including examining resonant scattering \citep[e.g.][]{Werner09,Ahoranta16,Ogorzalek17}.
Measurements using resonant scattering are currently only possible for low-temperature X-ray emitting gas in elliptical galaxies or the cores of cool-core groups or clusters, except in the Perseus cluster \citep{HitomiRes18}.
The power spectrum of density fluctuations has been used to obtain upper limits on the amount of turbulence in some systems  \citep{ZhuravlevaFluct14a}.  Repeating this analysis on simulated Perseus-like data \citep{WalkerSlosh18}, however, shows that the signal from sloshing gas could contribute strongly outside 60 kpc radius.
Turbulence can also be constrained based on arguments about its radial propagation \citep{Bambic18}.

In this paper, we present a novel technique to calibrate the pn detector of the European Photon Imaging Camera (EPIC) onboard \emph{XMM-Newton}, allowing us to measure bulk motions in clusters of galaxies.
Section \ref{sect:datacal} describes our calibration methods.
Using the calibration and the procedures described in Sect.~\ref{sect:datanalysis}, we analyse observations of the Perseus (Sect.~\ref{sect:perseus}) and Coma (Sect.~\ref{sect:coma}) galaxy clusters to study bulk motions.
Further systematic errors and future improvements are discussed in Sect.~\ref{sect:discuss}.
Uncertainties quoted are at the 1$\sigma$ level.
Any astrophysical velocities quoted have a positive sign for material flowing away from the observer.
All sky images are shown with north to the top and east to the left.

\section{Data calibration}
\label{sect:datacal}
To measure velocities with EPIC-pn we need an excellent characterisation of the energy scale of the instrument.
Because the cluster observations we wish to analyse are spread over the lifetime of the mission, we want to study this entire period.
The currently-quoted accuracy for the detector energy-scale calibration for imaging modes is better than 12.5~eV (XMM-SOC-CAL-TN-0018).
At the energy of Fe-K emission, this is equivalent to roughly $550\:\textrm{km s}^{-1}$.
As we later show, although this quoted accuracy may be an average over the detector at the energy of the Mn-K$\alpha$ calibration line, it appears to underestimate the energy shifts in some locations of the detector at higher energies.

In our analysis we used the \emph{XMM} Science Analysis Software (SAS) version 16.1.0.
When transforming from raw (PHA) to corrected (PI) event energies, SAS applies several different corrections to the energies of the observed events using its calibration database.
This work is done by EPEVENTS, used as part of EPCHAIN, one of the EPIC-pn data-reduction chains.
The effects corrected for include charge transfer inefficiency (CTI), operating along the columns of the detector and gain changes, including those obtained from long-term calibration source monitoring, and the effects of detector temperature variation.
The standard calibration uses the measured position of the 5.89~keV Mn-K$\alpha$ and 1.48~keV Al-K$\alpha$ calibration lines with time.
EPEVENTS applies a linear correction to events between these energies.
Above the energy of Mn-K$\alpha$ it does not introduce non-linearity of the energy scale.
Owing to the nature of the development of the calibration, SAS applies several corrections in sequence when converting from PHA to PI values.
We continue this trend by applying further corrections to the standard PI values.
The current correction procedure is only applicable if the astrophysical line to be examined has an energy above the instrumental Mn-K$\alpha$, because of the energy scale interpolation used below this energy by SAS.
Therefore if using Fe-K to measure redshifts, objects must lie below $z \sim 0.12$.

Our first step is a correction for the average gain of the detector during the observation (Sect.~\ref{sect:1storder}), the second is a correction for the spatial gain variation across the detector with time (Sect.~\ref{sect:2ndorder}) and the final correction is to linearise the energy scale as a function of detector position and time (Sect.~\ref{sect:3rdorder}).

\subsection{The EPIC-pn detector}

\begin{figure}
  \centering
  \includegraphics[width=0.75\columnwidth]{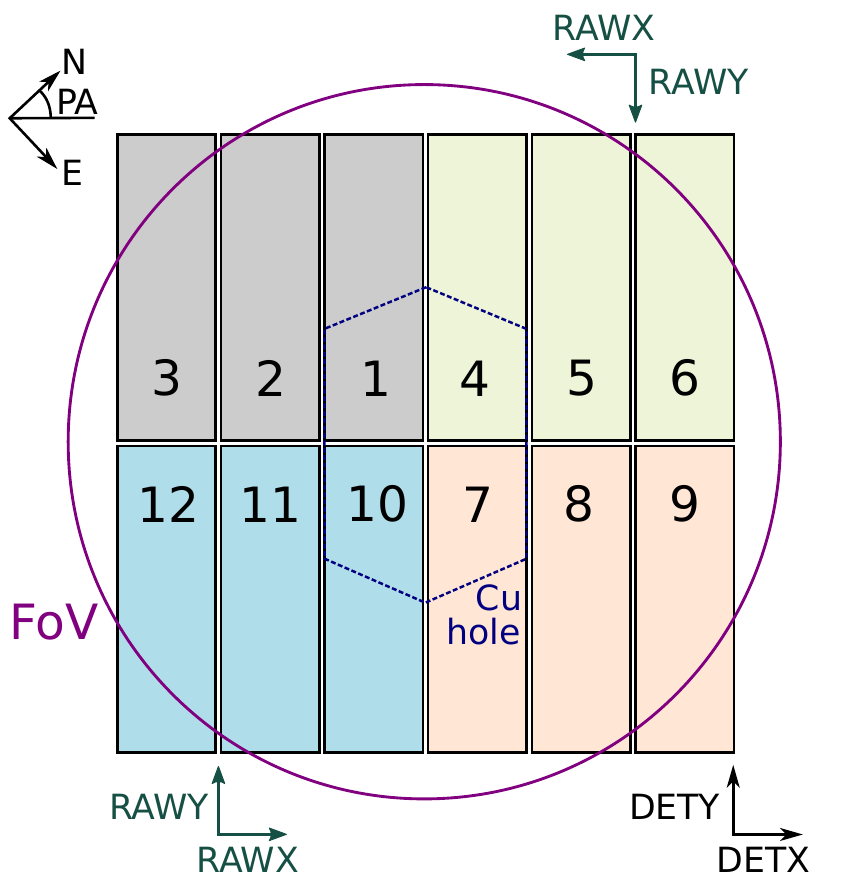}
  \caption{Geometry of the EPIC-pn detector.
    Shown are the numbers of the CCDs, the field of view exposed to the sky (FoV), the raw coordinates (RAWX and RAWY), the detector coordinates (DETX and DETY) and the approximate location of the `Cu hole'.
    The position angle (PA) is the angle between the detector and sky coordinates, an example of which is shown.
    The four quadrants are shaded in different colours.
  }
  \label{fig:epicpn}
\end{figure}

The EPIC-pn camera \citep{Struder01} contains a monolithic silicon CCD array, made up of $4\times3$ CCDs (Fig.~\ref{fig:epicpn}).
Each of the individual CCDs has $64$ columns and $200$ rows.
The detector is read out along the columns as quadrants, each containing three CCDs.
There are a variety of operating modes of the detector, but we are primarily interested in Full Frame (FF) and Extended Full Frame (EFF) modes, which read out the whole detector.
In the EFF mode the fraction of out-of-time events (Sect.~\ref{sect:oot}) is reduced by using longer frame times, at the expense of being more susceptible to pile-up (more than one photon arriving within a frame time).
At an energy of 5.89~keV the detector achieved in its first nine months an average of $\sim 150$~eV full width at half maximum (FWHM) spectral resolution.

\subsection{Instrumental background emission lines}
\begin{figure}
  \includegraphics[width=\columnwidth]{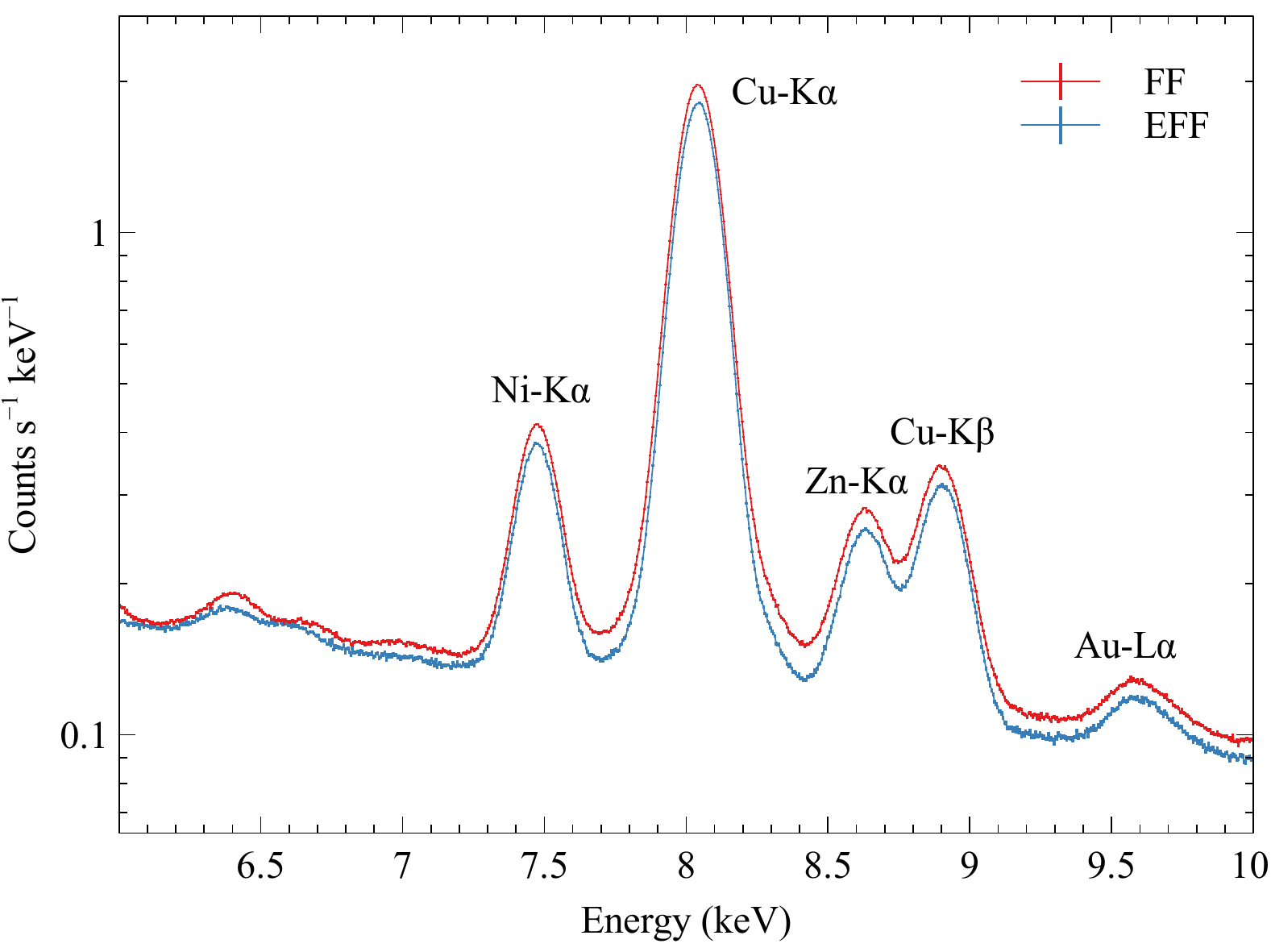}
  \caption{Total stacked spectra of the fluorescent background emission lines for 32.6~Ms of FF (red) and 19.7~Ms EFF (blue) data.
    Before stacking the individual spectra, the three orders of correction described in Sect.~\ref{sect:datacal} were applied.
    Only events with PATTERN==0 were included.
    Au-L$\alpha$ was not used in our analysis because of its weakness.
  }
  \label{fig:fluspec}
\end{figure}

\begin{figure*}
  \includegraphics[width=\textwidth]{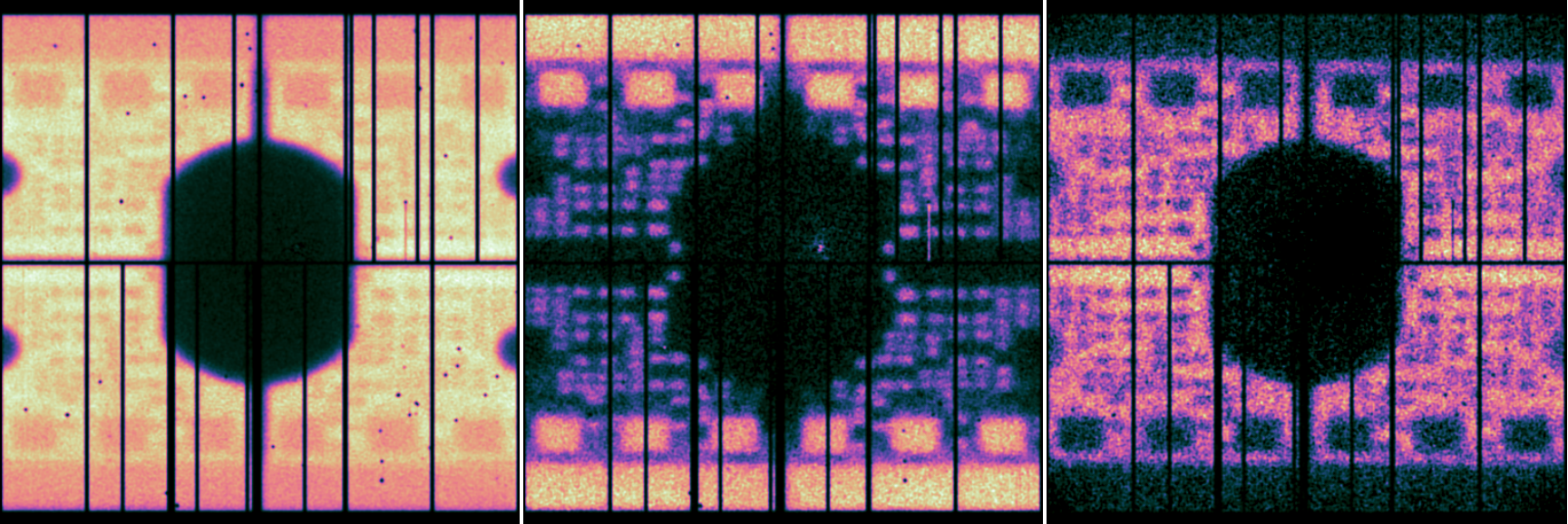}
  \caption{Images of the detector in different fluorescent lines, created from stacked astrophysical observations. Shown are images in Cu-K$\alpha$ (left; 7.805 to 8.285 keV), Ni-K$\alpha$ (centre; 7.280 to 7.680 keV) and Cu-K$\beta$ and Zn-K$\alpha$ (right; 8.455 to 9.075 keV). Approximate background has been subtracted using scaled continuum images on either side of the lines. Some residual astrophysical emission can be seen in the centre of the image. The total exposure is around 64~Ms after cleaning for flares, combining both FF and EFF data and using PATTERN$\le$4. The image was binned by 64 detector coordinate units and smoothed by a Gaussian of $\sigma=1$ pixel width.}
  \label{fig:lineimages}
\end{figure*}

To calibrate the energy scale of the detector, we measured the position of fluorescent emission lines produced by elements present in the electronics boards behind the detector, and visible in all observations \citep{Freyberg01}.
The dominant emission line is the Cu-K$\alpha$ line at 8.04~keV (Fig.~\ref{fig:fluspec}).
Weaker surrounding lines include Ni-K$\alpha$ (7.47~keV), Zn-K$\alpha$ (8.63~keV) and Cu-K$\beta$ (8.90~keV).
The spatial variation of the lines depends on the variation in the composition of the electronics boards.
The distribution of Cu is relatively flat, except for a lack of emission in the centre of the detector (Fig.~\ref{fig:lineimages}), which we call the `Cu hole'.
The presence of this hole leads us to exclude the central part of the detector in our astrophysical analysis, as it cannot be calibrated through our procedure, although there is some residual signal from out-of-time events.
Ni-K$\alpha$ and Cu-K$\beta$/Zn-K$\alpha$ emission shows spatial variation which appears to be anti-correlated.

\subsection{Datasets used for calibration}
\begin{figure}
  \includegraphics[width=\columnwidth]{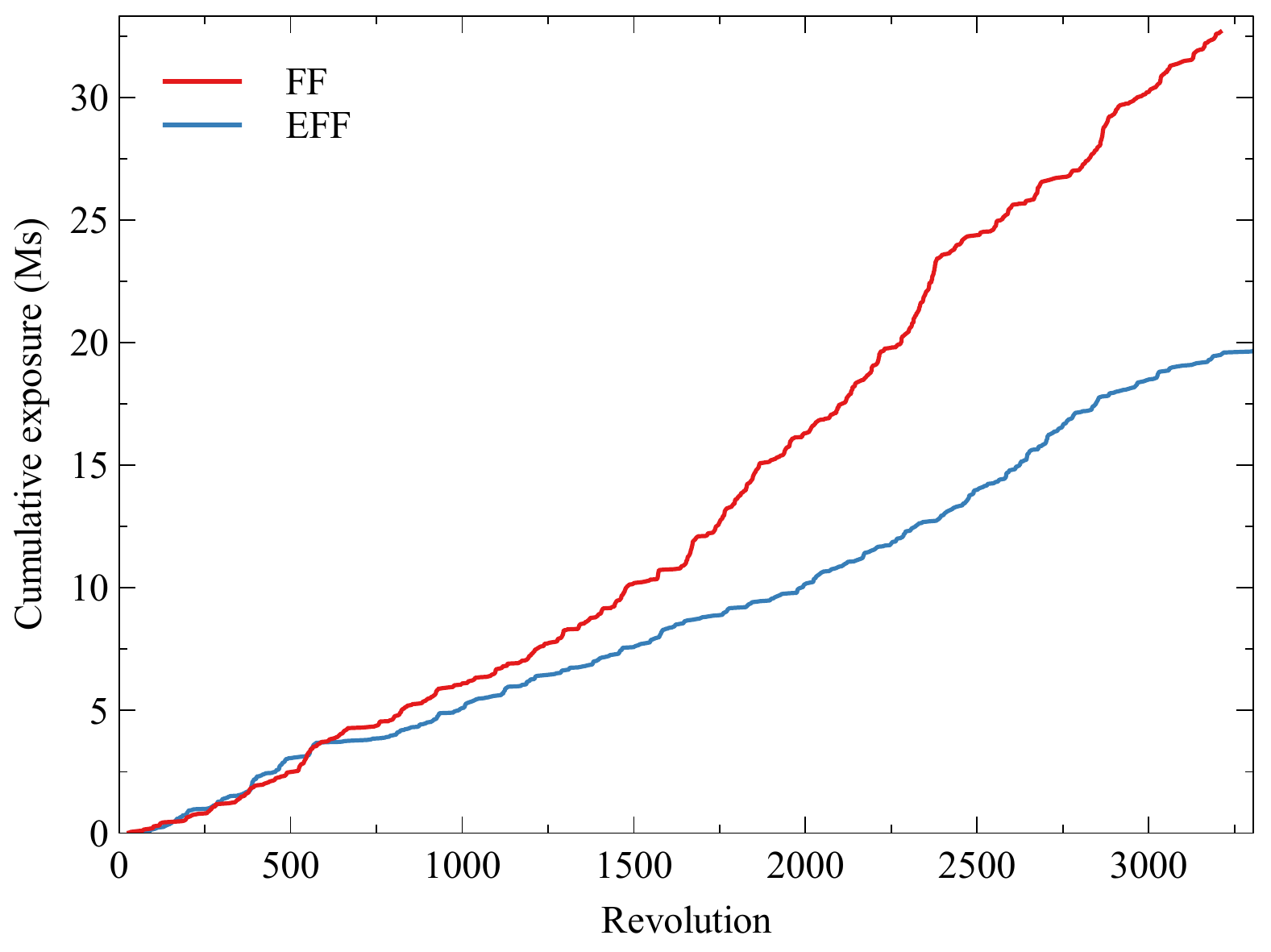}
  \caption{
    Cumulative FF (red) and EFF (blue) cleaned exposure of the astrophysical datasets used in the calibration as a function of time (in \emph{XMM} revolutions).
  }
  \label{fig:expos}
\end{figure}

Our calibration analysis made use of two different sets of data.
The first was a set of observations of astrophysical objects taken from the \emph{XMM} archive (Table \ref{tab:datasets}).
We obtained as many long public observations using EPIC-pn full-frame modes as possible, manually filtering them by examining the preprocessed images to avoid those with bright X-ray point sources and optical sources.
The observations were chosen to cover the full mission lifetime (from 2000 to 2019), although proprietary periods prevented us from analysing some recent data.
These astrophysical datasets contain the background Cu-K$\alpha$, Ni-K$\alpha$, Zn-K$\alpha$ and Cu-K$\beta$ fluorescent lines.
The observations were split into FF and EFF sets for separate analysis. Those with a cleaned exposure of less than 10~ks were ignored.
The cumulative exposures as a function of revolution for the two modes are shown in Fig.~\ref{fig:expos}, where the \emph{XMM} revolution is around 47.9 hours, or two days.

For the energy scale correction (Sect.~\ref{sect:3rdorder}) we also made use of detector calibration observations (Table \ref{tab:calib_datasets}).
We ignored those datasets where the cleaned exposure was less than 10 ks (the filtering is described in Sect.~\ref{sect:cal_selevt}).
EPIC-pn includes a $^{55}$Fe radioactive calibration source which can be viewed by the detector, illuminating much of the field of view with a spectrum of Al-K$\alpha$, Mn-K$\alpha$ and Mn-K$\beta$ emission lines, with energies of $1.48$, $5.89$ and $6.49$~keV, respectively.
The half life of the source is 2.7 years, so it has become much fainter over the mission lifetime.
The illumination of the detector by the source is also not flat, with some CCDs having poor coverage.

\subsection{Spectral models}
When fitting the spectral lines, we used the line shape measurements of \cite{Holzer97} for Mn-K$\alpha$, Ni-K$\alpha$, Cu-K$\alpha$, Mn-K$\beta$ and Cu-K$\beta$.
We implemented an XSPEC model for each line, summing the provided Lorentzian fit sub-components.
The shift in the energy of a line is parametrised by a redshift, measuring the fractional energy shift (not a real physical redshift).
For the weaker Zn-K$\alpha$ line we made a model containing two Lorentzians with the energies taken from \cite{Bearden67}, assuming a flux ratio of 0.52 for the ratios of the K$\alpha_2$ and K$\alpha_1$ lines, similar to the other K$\alpha$ lines in \cite{Holzer97}, and choosing line widths of 3.3 and 2.4~eV, respectively.

\subsection{Selected events}
\label{sect:cal_selevt}
In our spectral analysis, we only considered single-pixel events (PATTERN==0), which have the best energy resolution.
When examining real cluster emission, we only examined events with FLAG==0, which discards bad pixels and neighbouring regions, and the outer part of the detector which is not illuminated by the sky.
When examining the fluorescent lines and calibration source, we instead filtered with (FLAG \& 0xfa097d)==0, which also includes events outside the sky field of view, enabling us to improve our understanding of the detector near its corners.

To reduce the effect of the flaring component of the \emph{XMM} background, we applied a uniform maximum count rate threshold of $1.0$~ct~s$^{-1}$ in 100s bins, in the 10 to 15 keV band with FLAG==0 and PATTERN==0.

When extracting spectra, we binned them by the standard factor of five PI channels (i.e. 5~eV bins).
Slightly improved uncertainties on measured line energies could have been obtained using a factor of one, although this would have meant we could not have used the standard response matrices.

\subsection{Out-of-time events}
\label{sect:oot}
If a photon hits the detector while it is being read out, it will be recorded with a wrong RAWY position, producing an out-of-time (OoT) event.
For bright sources these events can be seen as a stripe along the RAWY direction.
In FF mode, the fraction of events which are OoT is 6.3 per cent, while it is 2.3 per cent in EFF mode.

In our analysis we accounted for OoT events with an OoT event file, created by running EPCHAIN a second time with the parameter `withoutoftime=yes'.
This option randomly shifts events along the RAWY axis, before the standard gain and CTI corrections are applied.
We extracted OoT spectra using the same extraction regions as the sources.
These spectra were used as background spectra when spectral fitting, scaling the exposure times up by the appropriate fractions shown above.
If we added the source spectra together, these background spectra were similarly added.
The effect of correcting for OoT events in our analysis was small; the galaxy cluster astrophysical results differed by less than their statistical uncertainties.

\subsection{Response matrix}
\label{sect:response}
When fitting spectra we used a response matrix with fine energy bins (0.3 eV) created with RMFGEN.
This energy bin size is sufficient to resolve velocities down to $\sim15\:\textrm{km s}^{-1}$.
With larger binning, line energies tend to be discretised onto the energy grid and error bars are often asymmetric.
For the stacked and calibration observations, the response matrix was generated using an early observation, 0114120101, and for small RAWY coordinates, giving the highest achievable energy resolution.
A variable Gaussian smoothing component was used when fitting, to account for the change in energy resolution over the detector position and over time.
EPIC-pn response matrices only differ between spatial positions and observations in their spectral resolution.
Fitting for the energy resolution significantly reduces the storage and computational time which would be required to make individual responses.

The ancillary response file used for fitting the calibration observations and fluorescent line data was generated from the same observation, disabling the effect of the mirrors and the detector filter (using parameters modelfiltertrans=false and modeleffarea=false), which do not affect the background lines.

\subsection{First order gain correction (mean Cu-K$\alpha$)}
\label{sect:1storder}
The first order energy correction we made was to measure the average `redshift', $z$, of the Cu~K$\alpha$ line in each observation and correct it back to zero.
The total Cu-K$\alpha$ spectrum was extracted from each observation, including events outside the sky region, but excluding the central Cu hole. The hole was defined as a polygon with DETX and DETY coordinate pairs of $7601$, $4390$, $-6209$, $5255$, $-4531$, $5978$, $-2361$, $6383$, $-480$, $6181$, $995$, $5573$, $2094$, $4994$, $3133$, $3905$, $3136$, $-6116$, $1747$, $-7447$, $300$, $-8286$, $-1232$, $-8459$, $-3431$, $-8459$, $-4646$, $-8228$, $-5688$, $-7794$, $-6874$, $-6926$, $-7616$, $-6052$.

We fitted the spectrum between $7.00$ and $9.25$ keV by a model with Cu-K$\alpha$, Cu-K$\beta$, Ni-K$\alpha$, Zn-K$\alpha$ and powerlaw components.
The redshift of each line component was allowed to vary independently.
We applied smoothing by a common variable-width Gaussian component, to allow for changes in energy resolution.
The spectra were fitted by minimising the C-statistic in XSPEC.

\begin{figure}
  \includegraphics[width=\columnwidth]{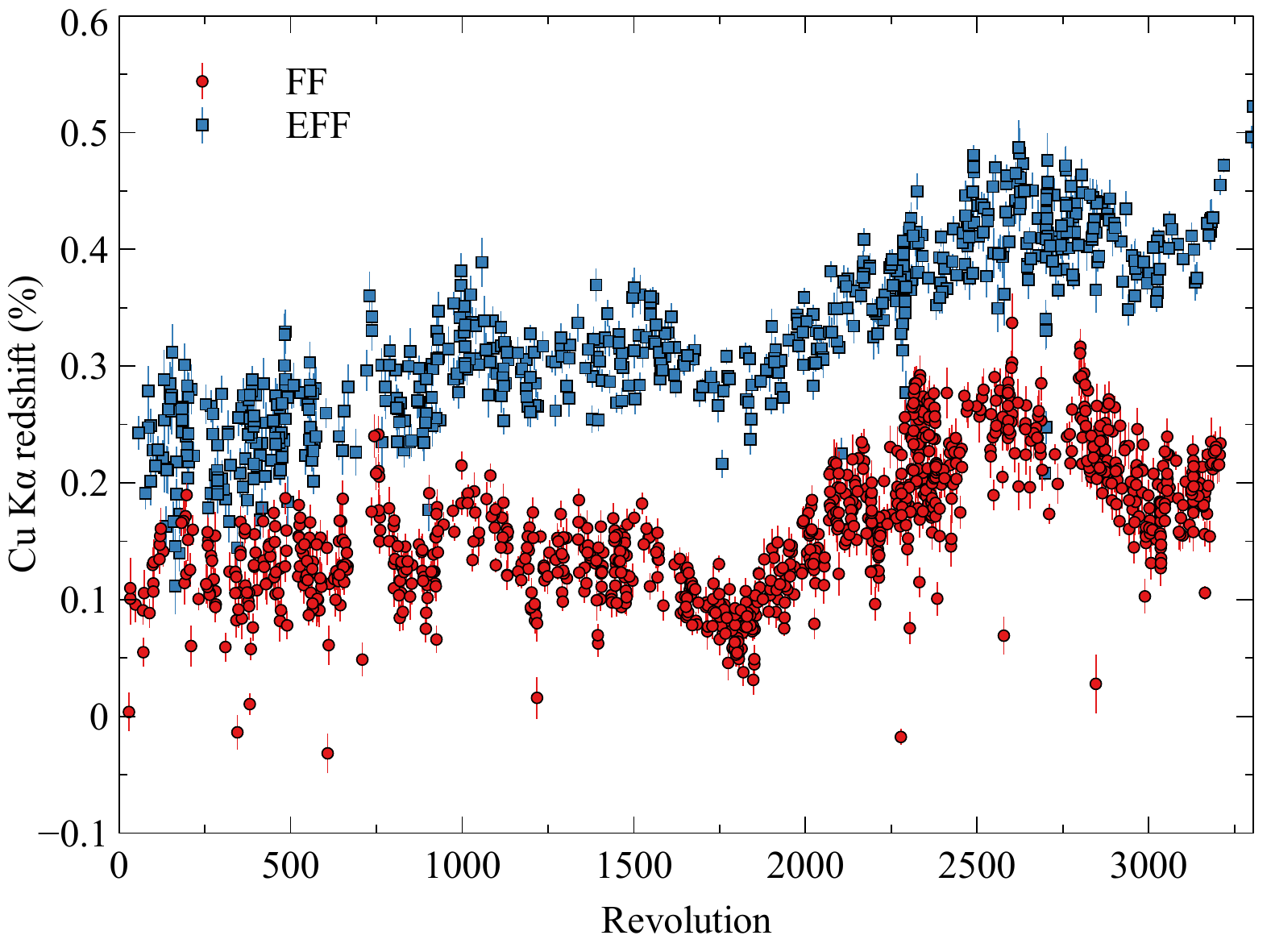}
  \caption{Best-fitting Cu-K$\alpha$ redshifts for individual astrophysical observations as a function of time (in \emph{XMM} revolutions).
    A 0.1 per cent correction is 8 eV at the energy of Cu-K$\alpha$.
  }
  \label{fig:cukaz_time}
\end{figure}

Figure \ref{fig:cukaz_time} shows the Cu-K$\alpha$ redshifts of the astrophysical observations as a function of time.
It can be seen that there is a systematic offset between FF and EFF observations, which widens with time.
At the 8~keV energy of Cu-K$\alpha$, the maximum shift is around 20~eV for FF and 35~eV for EFF modes.

We corrected observations for their average Cu-K$\alpha$ gain shift by scaling the PI values of the observation by $1+z$.
In detail, we take the integer PI value in an event file, add a uniformly-distributed random number between zero and one, scale by $1+z$, then round downwards back to an integer value.

\subsection{Second order gain correction (stacked Cu-K$\alpha$)}
\label{sect:2ndorder}

\begin{figure*}
  \centering
  \begin{tabular}{cc}
    \includegraphics[width=0.46\textwidth]{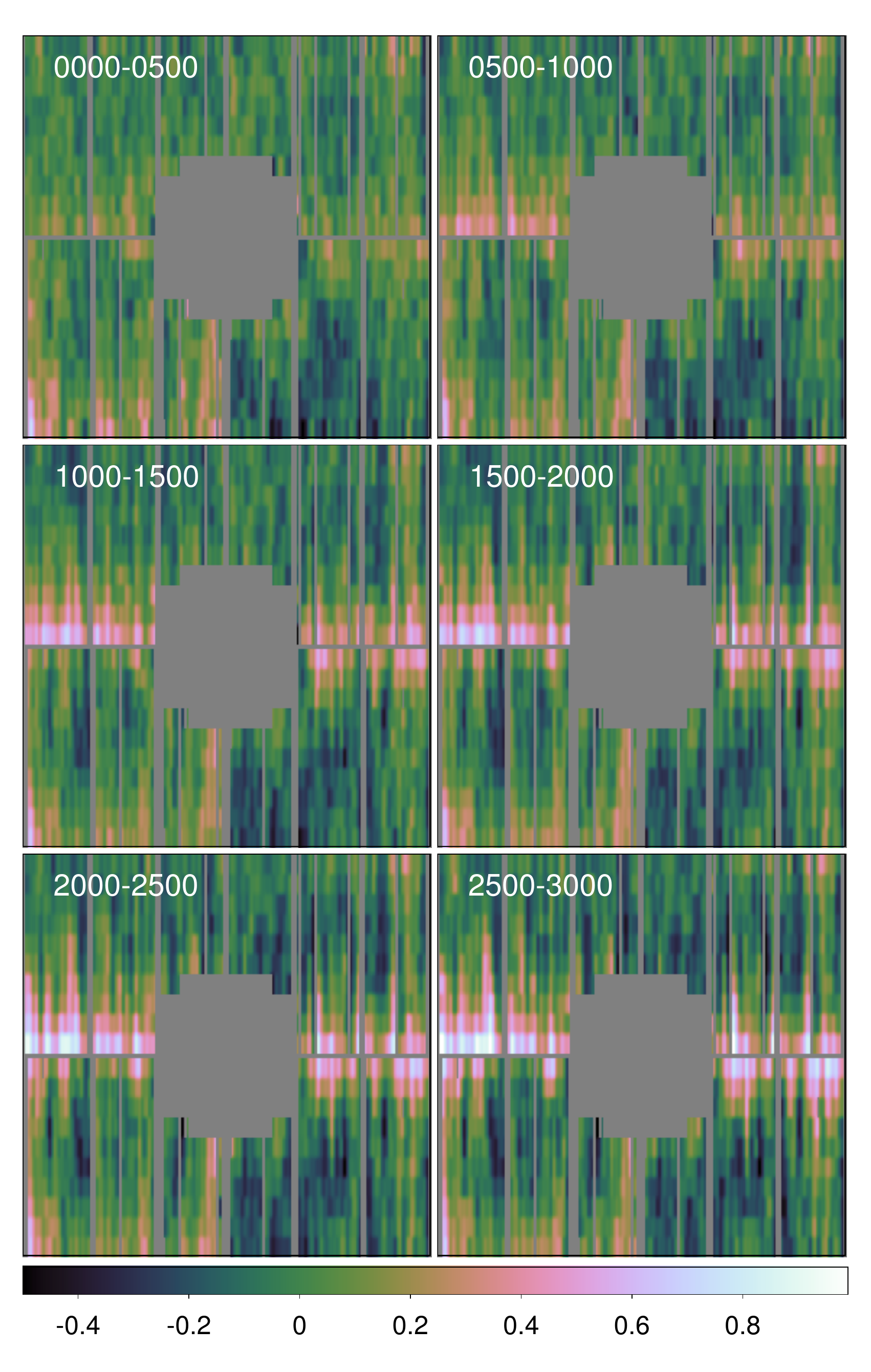} &
    \includegraphics[width=0.46\textwidth]{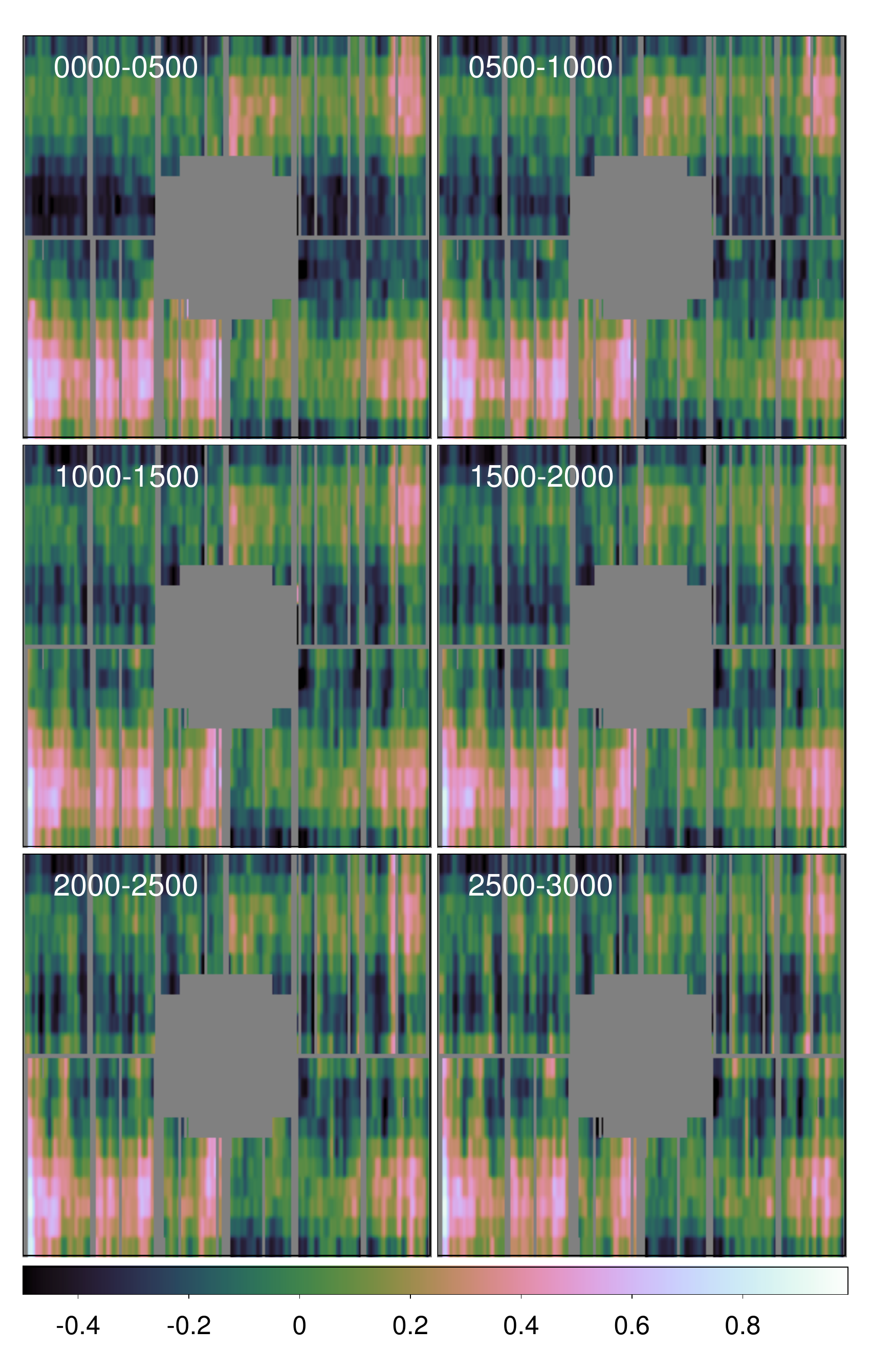} \\
  \end{tabular}
\caption{
Redshift of the Cu-K$\alpha$ line (in per cent) as a function of detector position over the mission lifetime in FF (left) and EFF (right) modes, after correcting for the average value.
The range shown in the top-left of each panel is the \emph{XMM} revolution.
The shift has been measured in $1\times20$ pixel regions and a $\sigma=2$ pixel Gaussian smoothing applied for display.
Detector regions with a large uncertainty in a combined error map have been masked out.
}
\label{fig:cuka_pos_time}
\end{figure*}

After correcting for the average Cu-K$\alpha$ redshift in each observation, we measured the Cu-K$\alpha$ redshift as a function of detector position and time.
We stacked the spectra from the astrophysical observations in time bins of 500 revolutions (overlapping, as the bins start in 250 revolution intervals), examining the FF and EFF modes separately.
As the detector is read out in columns, we stacked the spectra for each CCD in bins of $1\times20$ raw detector pixels.
We fitted each spectrum between 7.7 and 8.4 keV, by the Cu-K$\alpha$ model with variable redshift ($z$) and smoothing, plus a powerlaw.

Figure \ref{fig:cuka_pos_time} shows the redshift in six non-overlapping revolution intervals as a function of position, for the FF and EFF modes.
Excluding the central Cu hole region, in FF mode the position of the Cu-K$\alpha$ line is shifted between around $-0.2$ and $+0.8$ per cent ($-15$ to $+65$ eV), with shifts most positive furthest away from the readout, and most negative in the centre of the CCDs.
In EFF mode there is a different distribution in shift over the detector, with a range of around $-0.5$ to $+0.6$ per cent ($-40$ to $+50$~eV), with the most positive shifts in the centre of each CCD, and the most negative shifts furthest from the readout.
The difference between the FF and EFF maps is fairly constant in time.

The ranges of the shift appear to increase with time, but change fairly gradually, allowing us to correct for this effect.
To correct the gain for an observation, we took the average Cu-K$\alpha$ line shift, when measured from the uncorrected observation (as in Sect.~\ref{sect:1storder}) and the correction map with the central revolution closest in value with the observation revolution.
Each PI value was corrected by the combined shifts, looking up the correction on the map according to CCD number and pixel coordinate.

We tested that the PI scaling procedure works to correct the stacked Cu-K$\alpha$ line energy by correcting the input event files and repeating the procedure in this section.
We found that the produced maps are centred around zero shift, with a standard deviation close to the average statistical uncertainty on each line measurement.

\begin{figure}
  \includegraphics[width=\columnwidth]{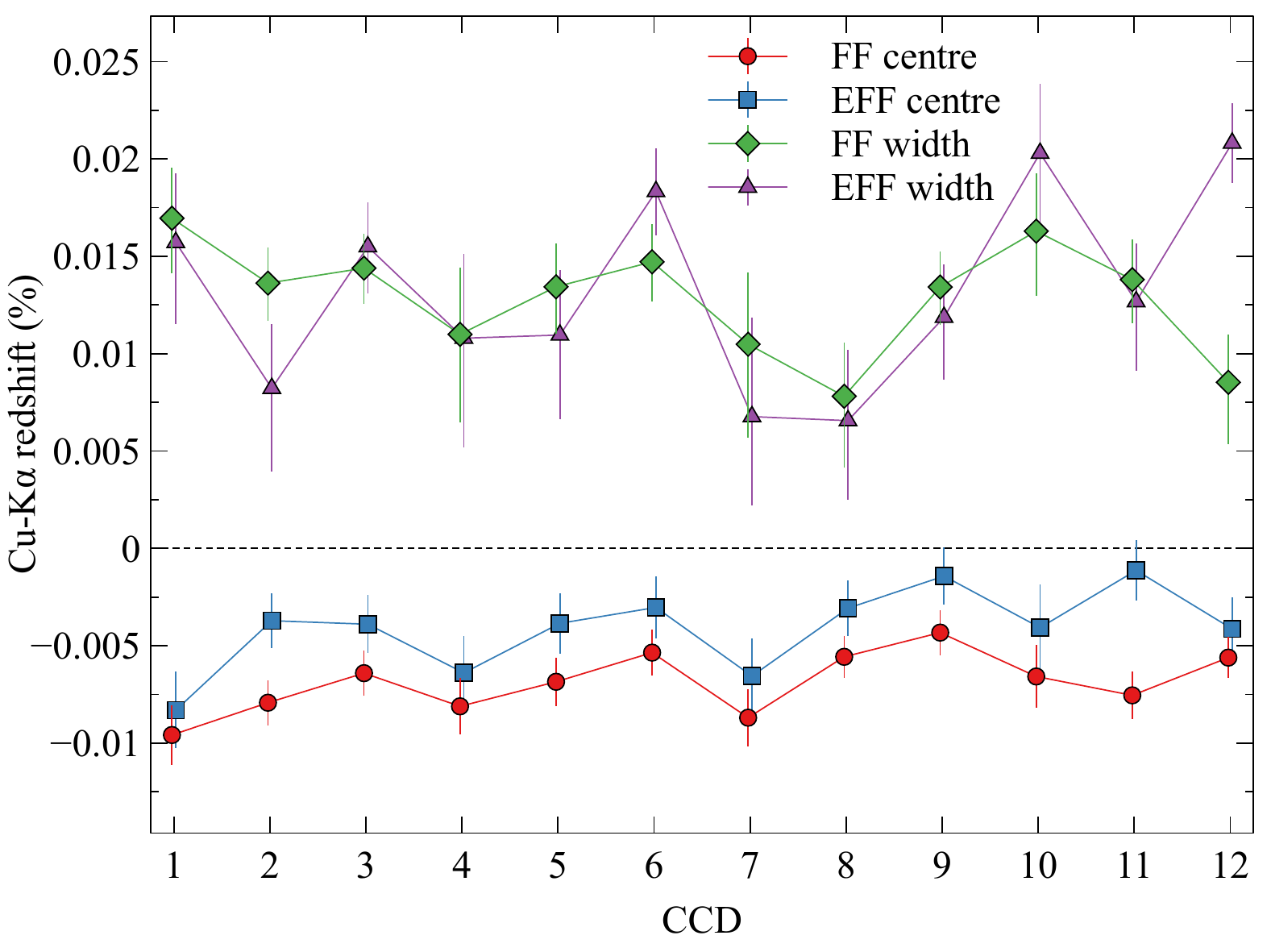}
  \caption{Intrinsic Gaussian centre and width ($\sigma$) of the distribution of Cu-K$\alpha$ redshifts in each CCD for the observations which went into the stacking analysis, after correcting each observation for the average Cu-K$\alpha$ value and the time-dependent maps (Fig.~\ref{fig:cuka_pos_time}).
    This analysis excludes observations with statistical uncertainties on the Cu-K$\alpha$ redshift of greater than 0.1 per cent.
    Results for FF and EFF modes are shown separately.
  }
  \label{fig:ccds-average-width}
\end{figure}

Another test is to look for position-dependent energy changes which are not corrected by the average Cu-K$\alpha$ for an observation and the time-dependent average detector maps we have produced here.
This may occur, for example, if there are random fluctuations in the gain of each CCD within single observations.
To test for this, we extracted and fitted the spectra from individual CCDs for each observation which went into the stacking analysis, after applying our correction for the average Cu-K$\alpha$ and the time-dependent detector maps in this section, excluding the central Cu-K$\alpha$ hole.
The distribution of residual Cu-K$\alpha$ redshifts was well modelled by a Gaussian distribution, after excluding points with large error bars.
We modelled the intrinsic distribution of the points before measurement uncertainties for each CCD as a Gaussian.
Figure \ref{fig:ccds-average-width} shows the obtained Gaussian centre of the distribution and the intrinsic distribution width ($\sigma$) for each CCD in the two different detector modes.

This analysis shows no evidence of strong variation within observations of the Cu-K$\alpha$ line energy.
We see a systematically low Cu-K$\alpha$ energy after correction, but this is only of the order of $0.005$ per cent.
This may be because when we corrected for the average Cu-K$\alpha$ energy for an observation, we fitted the whole fluorescent line complex energy range, but when we examined spatial bins, we only fitted for the Cu-K$\alpha$ line.
The shape of the underlying continuum may affect the Cu-K$\alpha$ line energy slightly.
The corrected Cu-K$\alpha$ values have a distribution width of around $0.01$ per cent, which is very good for our purposes ($\sim 30\:\textrm{km s}^{-1}$).

We also tested a modified version of the 1st and 2nd order corrections, where an initial 1st order correction was done to each CCD, rather than to the detector as a whole.
The changes to the galaxy cluster astrophysical results, using these corrections instead of those we present, were much smaller than the statistical uncertainties.

\subsection{Energy scale correction}
\label{sect:3rdorder}

\begin{figure}
  \includegraphics[width=\columnwidth]{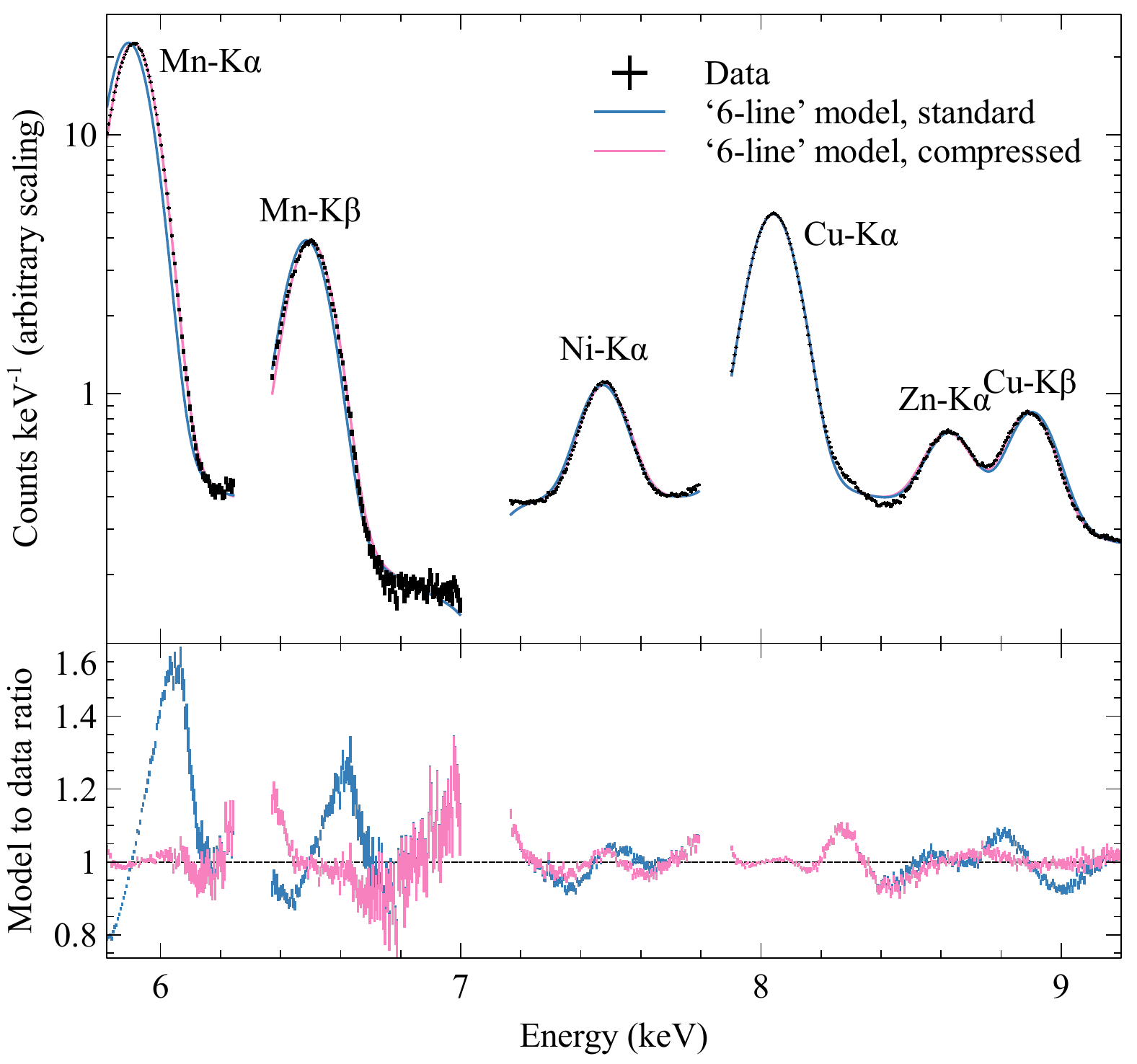}
  \caption{
    Demonstration of the necessity of an energy scale correction.
    Shown in black are the total spectra of calibration observations between revolutions of 1000 and 1500 ($<7$ keV) and the total spectra of calibration observations plus stacked observations in the same time period ($>7$ keV), outside the Cu hole region.
    The count axis for data $>7$~keV has an arbitrary scaling.
    The top panel shows the models and data, while the bottom panel shows their ratio.
    There are two `6-line' models.
    One (blue) assumes that the lines are at their standard energies.
    The other (pink) has a linear compression model for the line energies about Cu-K$\alpha$ where $\delta E/E=0.0013$ keV$^{-1}$.
    The poorly-modelled line wings do not affect the best-fitting compression factors.
  }
  \label{fig:spec-compress}
\end{figure}

After applying the correction for the observation-specific Cu-K$\alpha$ position and our correction maps (the gain corrections), it became apparent that the energies of other lines required further correction.
The data preferred Zn-K$\alpha$ and Cu-K$\beta$ line redshifts greater than zero, and Ni-K$\alpha$ line redshifts less than zero (Fig.~\ref{fig:spec-compress}).
We examined calibration observations, also finding that the Mn-K$\alpha$ and $\beta$ lines were at too high energies.

\begin{figure}
  \includegraphics[width=\columnwidth]{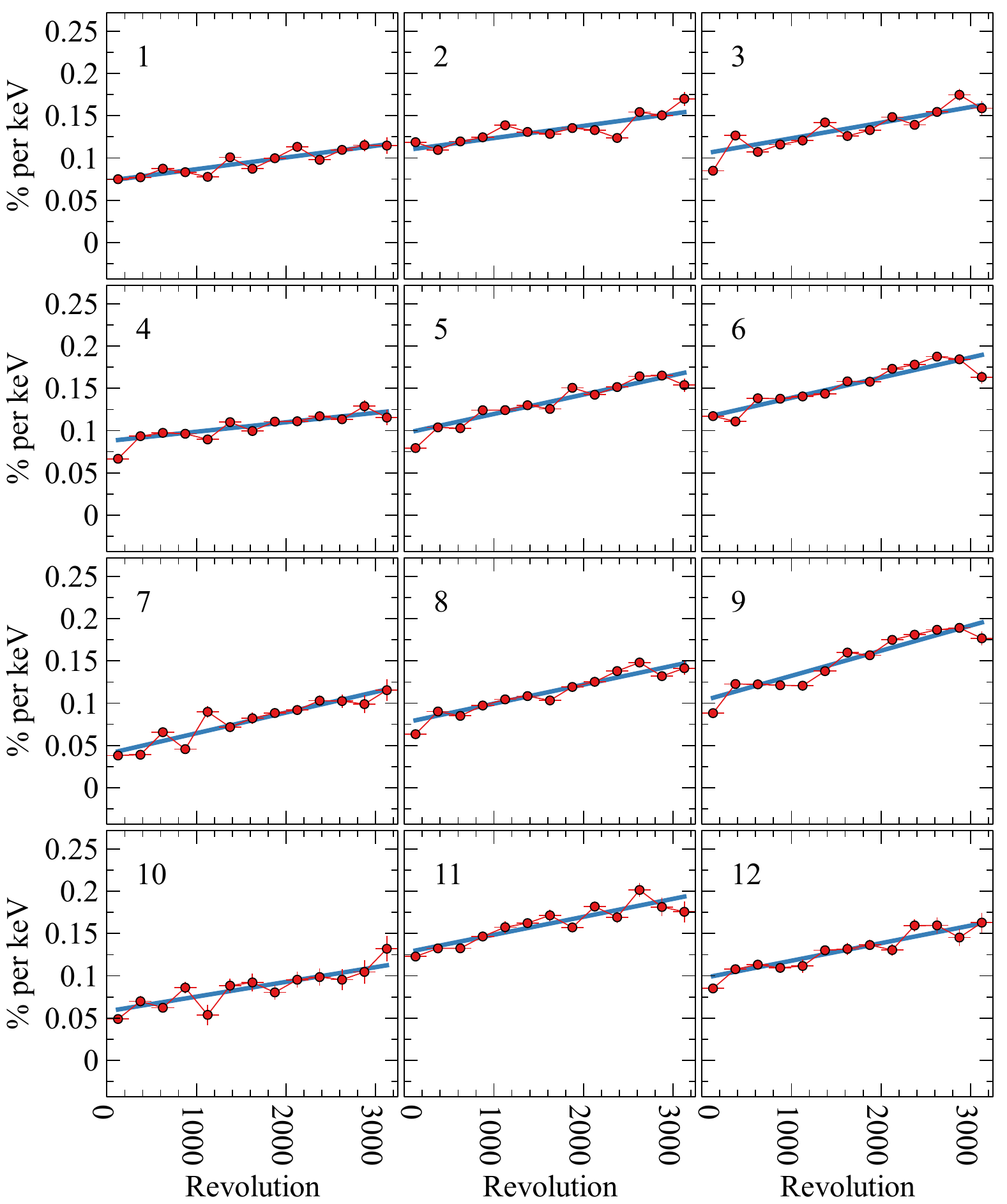}
  \caption{Energy scale correction factor for each CCD as a function of time.
    Positive values mean that events have too low an energy at high energies and too high an energy at low energies.
    This plot shows the results of fitting the `6-line' model to stacked calibration observations and observational datasets after gain correction, excluding the Cu hole.
    The lines show linear plus constant fits to the results excluding the first revolution bin.
  }
\label{fig:compress-time}
\end{figure}

We found that a linear model, where observed line positions are compressed in energy by some factor proportional to their energy difference from the Cu-K$\alpha$ line, modelled these shifts well.
Unfortunately the correction factor has both spatial and time dependence.
Figure \ref{fig:compress-time} shows this energy scale correction factor as a function of time for each CCD.
These results were obtained by fitting a spectral model made of Mn-K$\alpha$, Mn-K$\beta$, Ni-K$\alpha$, Cu-K$\alpha$, Zn-K$\alpha$ and Cu-K$\beta$ lines (the `6-line' model).
The redshifts of the lines were parametrised by an energy correction factor multiplied by the difference in the expected energy of each line to the Cu-K$\alpha$ line energy ($-2.150$, $-1.555$, $-0.570$, $0.590$ and $0.865$~keV, respectively).
The Cu-K$\alpha$ line position was free in these fits.
We also included powerlaw components (each with a free slope) between energies of 5.2 and 7.1 keV, 7.1 and 8.0 keV, and 8.0 and 9.5 keV to account for continuum and broad line components.
The line components were convolved by a Gaussian broadening component parametrised by its energy width.
As the calibration source becomes weaker with time and has spatial variation, we fitted both stacked calibration observations (below 7 keV) and stacked calibration plus real observations (above 7 keV), in bins of revolution.
We fitted the spectrum between 5.825 and 6.245 keV, 6.375 and 7.000 keV, 7.170 and 7.795 keV, and 7.905 and 9.200 keV (as in Fig.~\ref{fig:spec-compress}).
The energy ranges chosen were designed to exclude the asymmetries seen in the Mn-K$\alpha$ and Mn-K$\beta$ line shapes by excluding the broader lower energy part of the line below the peak.
We used the same response and ancillary response as previously (Sect.~\ref{sect:response}) and minimised the C-statistic in the spectral fits.
We did not find any difference when comparing FF and EFF mode data, so we combined both in this analysis.

Figure \ref{fig:compress-time} shows that there is an increasing energy scale compression (PI values are compressed around the Cu-K$\alpha$ line) in each CCD with time.
There appears to be an offset at zero revolutions, and in some CCDs a jump is seen at around revolutions of 250.
The plot includes a constant plus linear with time model, fitted excluding datasets below a revolution of 250 and by minimising the $\chi^2$.
To avoid the fits being strongly biased by the points with small error bars (particularly those early in the mission when the calibration source was strongest), we increased the error bars on the data points by adding $10^{-4} \:\textrm{keV}^{-1}$ in quadrature.

We do not know the physical origin of this non-linearity in the PI values from the detector.
It is affected by the ageing of the detector unit in space, although it was apparent from launch.
By examining spectra in different spatial bins, we found that differences were clearer in the RAWX axis of the CCDs (i.e. groups of columns were more different).
There is also some variation in the RAWY axis, with the steepness of the slope in time being stronger towards large RAWY values.

To remove this effect from our astrophysical data and calibration observations, we fitted stacked spectra in bins of 250 revolutions, and $4\times2$ spatial bins in each CCD (using inclusive pixel ranges in RAWX of 0-16, 17-32, 33-48 and 49-64 and in RAWY 0-110 and 111-200).
As above, the spectra were fitted (taking both stacked calibration and observational data) with the `6-line' model and the energy scale correction factor was obtained.
We fitted the energy correction factors as a function of time for each spatial region with a linear model in time as in Fig.~\ref{fig:compress-time}, excluding the first time bin.

To correct the PI values for an event file, we compute the correction factors as a function of position given its revolution number.
For revolutions below 250, we just use the results for that time bin, but for the others use the linear relation fitted to all the time bins.
We choose a correction factor given by which RAWX bin the event is in, but use linear interpolation of the correction factors in RAWY.

\begin{figure}
  \includegraphics[width=\columnwidth]{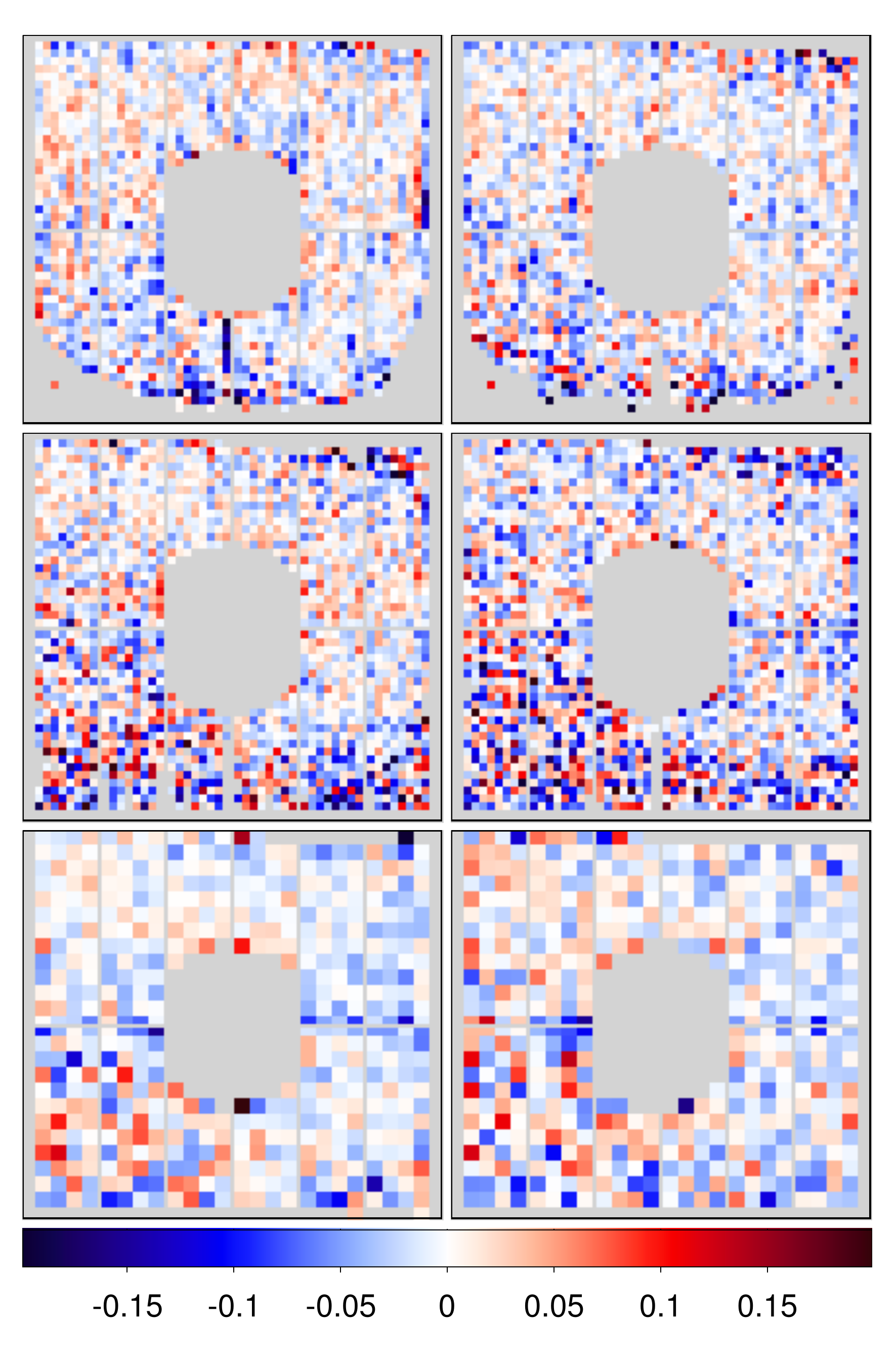}
  \caption{Maps of the residual energy correction factor in stacked observations, after applying our correction to this in the individual observations.
    The units are per cent per keV.
    The panels show revolution bins of 0-500, 500-1000, 1000-1500, 1500-2000, 2000-2500 and 2500-3000 (left to right and top to bottom).
    The uncertainty of these values is not the same across the detector or in time, so we use 8 and 16 pixel binning.
    The statistical uncertainty is much larger in the bottom-left area because of the non-uniformity of the calibration source.
    Pixels where the uncertainty is larger than $10^{-3} \:\textrm{keV}^{-1}$ are not shown.
  }
\label{fig:ecorrmap}
\end{figure}

To test how well the energy scale correction worked, we corrected our input event files and stacked them in revolution bins, including calibration observations (for the Mn-K$\alpha$ and $\beta$ lines) and standard observations (for Ni-K$\alpha$, Zn-K$\alpha$ and Cu-K$\beta$).
Figure \ref{fig:ecorrmap} shows the maps of the residual energy correction factor, after all our corrections.
The corrections are well-scattered around zero.

\begin{figure}
  \includegraphics[width=\columnwidth]{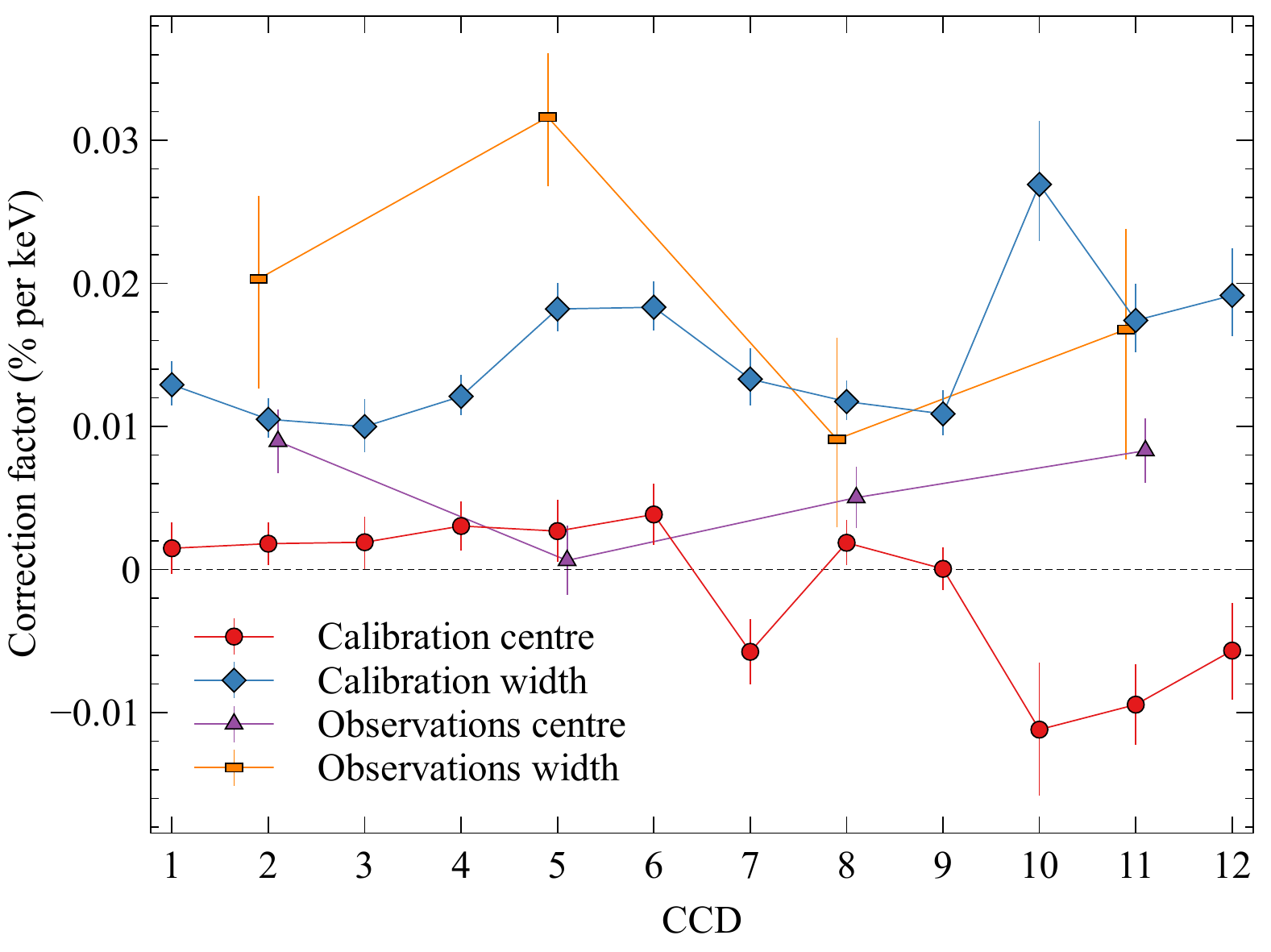}
  \caption{Distribution of residual correction factors (centre and width) obtained for each CCD, after correcting the calibration observations and refitting them.
    These results primarily use the Mn-K$\alpha$ and $\beta$ lines.
    Also shown are the distribution of correction factors obtained in each quadrant (three CCDs) for the individual astrophysical observations after correction, obtained from the Ni-K$\alpha$, Zn-K$\alpha$ and Cu-K$\beta$ lines.
  }
  \label{fig:residual-corrfactor}
\end{figure}

There may be some variation in the correction factor between observations.
We checked for this by fitting the spectra extracted from CCDs in the corrected calibration observations for the residual correction factor.
Assuming the distribution can be represented as a Gaussian, we measured the width beyond what is expected from the statistical uncertainties.
Figure \ref{fig:residual-corrfactor} shows the centre and width of the distribution for each CCD.
We obtain centres close to zero, with offsets less than about $10^{-4}\:\textrm{keV}^{-1}$ and a width around $1.5\times 10^{-4}\:\textrm{keV}^{-1}$.
We excluded observations 0060740801 and 0108860301 in this analysis, as they showed very anomalous values of the correction factor.
Including these observations in this analysis makes little difference to the results, but we suspect there is some problem with these individual observations, a small fraction of the sample.

We also examined the distribution of the residual correction factor for individual astrophysical observations.
This is much harder to measure as the Ni-K$\alpha$, Zn-K$\alpha$ and Cu-K$\beta$ lines are weaker and closer to the Cu-K$\alpha$.
We therefore measure the residual factor in CCD quadrants (Fig.~\ref{fig:epicpn}) rather than in CCDs, and limit the analysis to those observations where it can be measured in each quadrant to better than $2 \times 10^{-3} \:\textrm{keV}^{-1}$.
This signal is difficult to measure, as the individual measurements have an average statistical error of $1.1 \times 10^{-3} \:\textrm{keV}^{-1}$ and the statistical width is an order of magnitude smaller.
Using the same Gaussian distribution assumption, we find consistent values of the width parameter.
The centre parameter appears a bit larger, on average around $7 \times 10^{-5}\:\textrm{keV}^{-1}$, perhaps indicating that the linear correction model becomes inadequate at this high level of precision, or there is slightly inaccurate modelling of the details of the lines.

If we instead examine whole observations rather than separate quadrants, we decrease the measurement errors by a factor of two.
In this case we obtain an average residual correction factor of $(7\pm1)\times 10^{-5} \:\textrm{keV}^{-1}$ and a distribution width of $(9^{+3}_{-4})\times 10^{-5} \:\textrm{keV}^{-1}$.

In summary, these tests show that we have accurately calibrated the energy scale of the detector in different spatial regions.
Summing the residual correction factor offset and distribution width for the calibration observations ($10^{-4}\:\textrm{keV}^{-1}$ and $1.5\times10^{-4}\:\textrm{keV}^{-1}$, respectively), our energy scale appears accurate to better than $2.5\times 10^{-4}\:\textrm{keV}^{-1}$, or $150\:\textrm{km s}^{-1}$ (rounding up) at the energies of cluster Fe-K emission.
When we simultaneously fit the spectra from observations at different epochs and with different detector positions, we may obtain better precision, as some of the calibration differences will cancel out.

The overall accuracy of the correction for a single observation is determined by the precision to which the Cu-K$\alpha$ line energy can be measured.
In archival astrophysical observations (Table \ref{tab:datasets}), for an exposure time of  $T_\mathrm{exp}$, the Cu-K$\alpha$ redshift uncertainties have a functional form of $10^{-4} \: A (T_\mathrm{exp} / 10 \: \mathrm{ks})^{-1/2}$, where $A$ is between $1.6$ and $2.5$.
This individual uncertainty is significantly smaller than the statistical uncertainty on astrophysical redshift for a single observation.
This systematic error will be reduced if multiple observations are combined, as it is random and Gaussian.
Therefore, this systematic error is unimportant, except perhaps for extremely bright regions, such as the core of the Perseus cluster, which we cannot examine here owing to a lack of offset observations.

We applied further checks on the calibration procedure using the Perseus and Coma datasets in Appendix \ref{sect:checks}.

\section{Data analysis}
\label{sect:datanalysis}
Described here are the data analysis procedures applied to the observations of the Perseus and Coma clusters (observation details are listed in Table \ref{tab:obs}).
Firstly, we processed each observation data file with the standard EPCHAIN SAS tool.
To filter bad time intervals from flares, we applied the same $1.0$~ct~s$^{-1}$ rate threshold as described in Sect.~\ref{sect:cal_selevt}.
The event files were energy corrected using the three steps described in Sect.~\ref{sect:datacal}, where we applied an overall gain correction using the observation Cu-K$\alpha$ redshift, the spatial and time-dependent gain map correction and finally the energy scale correction.
When analysing spectra, we used only single-pixel events (PATTERN==0) to be consistent with our calibration analysis.
We also filtered using FLAG==0 to avoid regions close to CCD edges or bad pixels.

When spectral modelling within XSPEC, we used the APEC spectral code \citep{SmithApec01} to model cluster emission, absorbed with a PHABS \citep{BalucinskaChurchPhabs92} component to account for Galactic photoelectric absorption.
We used APEC version 3.0.9, which includes corrections derived from the \emph{Hitomi} observations \citep{HitomiAtom18} and assumed the Solar abundance ratios of \cite{Lodders09} used in the \emph{Hitomi} analysis.
In the fits the temperature, metallicity, redshift and normalisation were allowed to be free.
The spectra were fitted between $4.00$ and $9.25$ keV, avoiding low energies where our energy corrections are most likely incorrect, but allowing us to derive a temperature using the continuum.
We assumed a Galactic column density of $1.4 \times 10^{21} \:\textrm{cm}^{-2}$ for Perseus and $8.54\times 10^{19} \:\textrm{cm}^{-2}$ for Coma \citep{Kalberla05}, although the absorbing component has only a weak effect in the energy range examined.
As background components we included  Cu-K$\alpha$, Ni-K$\alpha$, Zn-K$\alpha$ and Cu-K$\beta$ lines, and a powerlaw component with its photon index fixed at 0.136 (an average taken from our archival observations).
The redshifts of the background lines were tied together and allowed to vary.
Although these background redshifts are not astrophysically interesting as they should be zero, they allowed us to test our gain correction.
The normalisations of the lines and powerlaw were free to vary.
For the background components we used an ancillary response file which did not include the effect of the filter or mirrors.
We minimised the C-statistic when fitting the spectra.

We used the response matrices generated for each observation and spatial region.
Spatial weighting maps with the number of counts between $3.00$ and $7.00$ keV were used when making the responses.
We used the same 0.3 eV energy bin size in the response matrices as for our calibration analysis.
No additional broadening was used in these fits, however.
The ancillary response files were also calculated using the same spatial weighting maps.

There were multiple observations for each spatial region we examined.
We tried two techniques for fitting the spectra.
The first, with which we show the results for the individual spectral regions, was to do a joint simultaneous fit of the spectra.
In this analysis we only included a spectrum for an observation if there were more than 2000 counts between $4.00$ and $9.25\:\textrm{keV}$, to be able to sufficiently constrain the velocity.
We allowed the normalisations of each component to be separate in each observation, but forced metallicities, temperatures and APEC redshifts to be the same.
The background line energies were allowed to vary between observations.
The second technique was to add the spectra from the different observations together and to create average weighted response and ancillary response files, using the number of counts between $4.00$ and $9.25$~keV as a weighting factor.
The total spectrum was fitted, reducing the number of free parameters, improving the speed of the fitting procedure and making it easier for XSPEC to find the best-fitting parameters.
We used this technique when fitting to produce the spectral maps of the clusters.
As we will show, there is generally good agreement between the `joint' and `total' fitted results.

We also repeated the fits for the total spectra with the SPEX spectral code \citep{KaastraSPEX96} instead of APEC, using version 3.04 with version 3.04.00 of its atomic code and tables.
As the SPEX plasma model was designed to be used within SPEX only, we tabulated the lines and continuum produced by the model at 0.02 dex temperature intervals and exported it into XSPEC in APEC table format\footnote{Code available at \url{https://github.com/jeremysanders/spex_to_xspec}}.
This allowed us to use exactly the same modelling procedure for the two codes.
As we show later, there is very good agreement between the APEC and SPEX results.

\section{The Perseus cluster}
\label{sect:perseus}
The \object{Perseus cluster}, \object{Abell 426}, is the X-ray brightest galaxy cluster in the sky.
There is clear evidence of feedback in the centre of the cluster, where the lobes of the central AGN, 3C\,84 are displacing the ICM \citep{BohringerPer93,FabianPer00}.
The AGN is also generating weak shocks in the ICM. Ripples, which may be sound waves generated by the AGN activity, are observed \citep{FabianPer03}.
If sound waves, these ripples could combat a significant fraction of the radiative cooling taking place in a distributed fashion \citep{SandersPer07}.
Alternatively, the surface brightness fluctuations in the cluster, if interpreted as turbulence, could energetically offset the radiative cooling taking place \citep{ZhuravlevaFluct14a}, although sloshing of the gas in the potential well could be a significant contribution to the density variations with increasing radius \citep{WalkerSlosh18}.

There is a well-known east-west asymmetry in the X-ray surface brightness \citep{ChurazovPer03}.
The cluster is also extended in this direction on the largest scales.
These edges in surface brightness and temperature are seen out to 700~kpc radius \citep{Simionescu12,Urban14,WalkerPerFront18}.
An explanation for these structures is that the gas is sloshing in the potential well of the system \citep{Markevitch01,ChurazovPer03,WalkerSlosh17}, caused by a disturbance to the potential because of a minor-merger, as is also seen in simulations of clusters \citep[e.g.][]{Ascasibar06}.
In Perseus there is a chain of galaxies in the western side of the cluster which may be responsible for the sloshing \citep{ChurazovPer03}.

Perseus was the main target for observation by \emph{Hitomi} \citepalias{Hitomi16}, providing an excellent opportunity to compare our technique with high resolution spectroscopy.

\begin{figure*}
  % one fig to left, two to right
  \centering
  \begin{minipage}[c][][c]{.63\textwidth}
    \includegraphics[width=\textwidth]{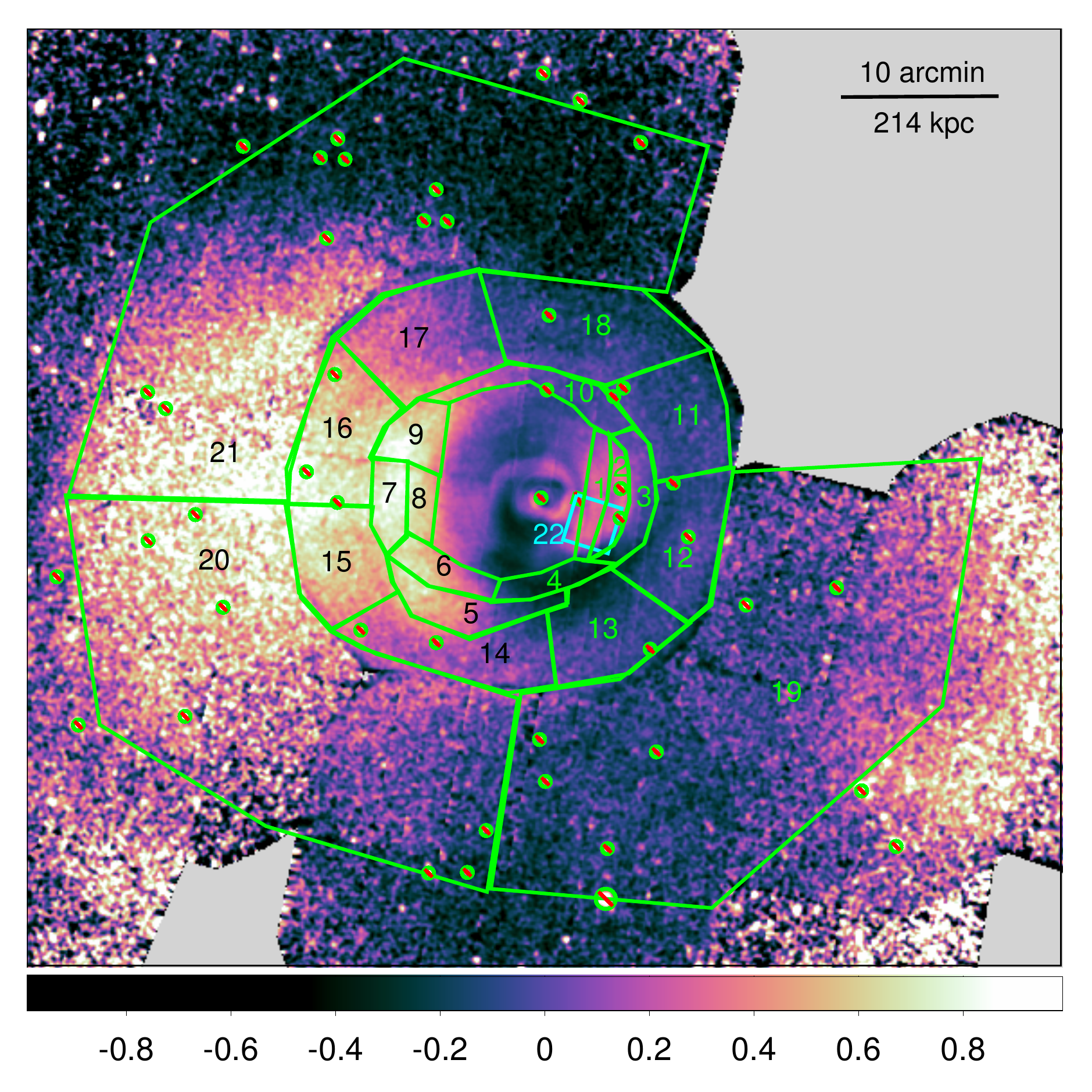}
  \end{minipage}
  \hfill
  \begin{minipage}[c][][c]{.3\textwidth}
    \includegraphics[width=\textwidth]{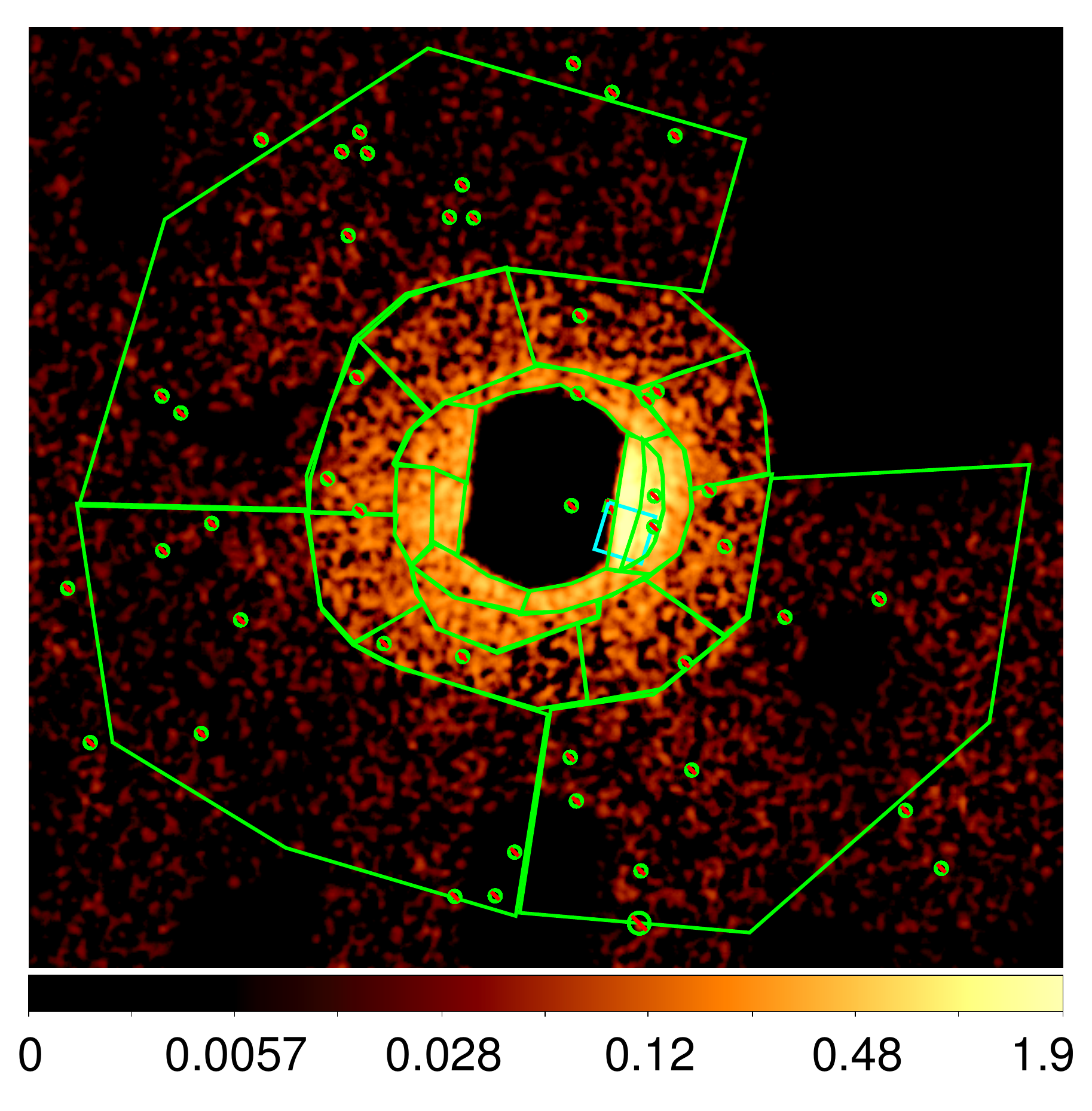}
    \vfill
    \includegraphics[width=\textwidth]{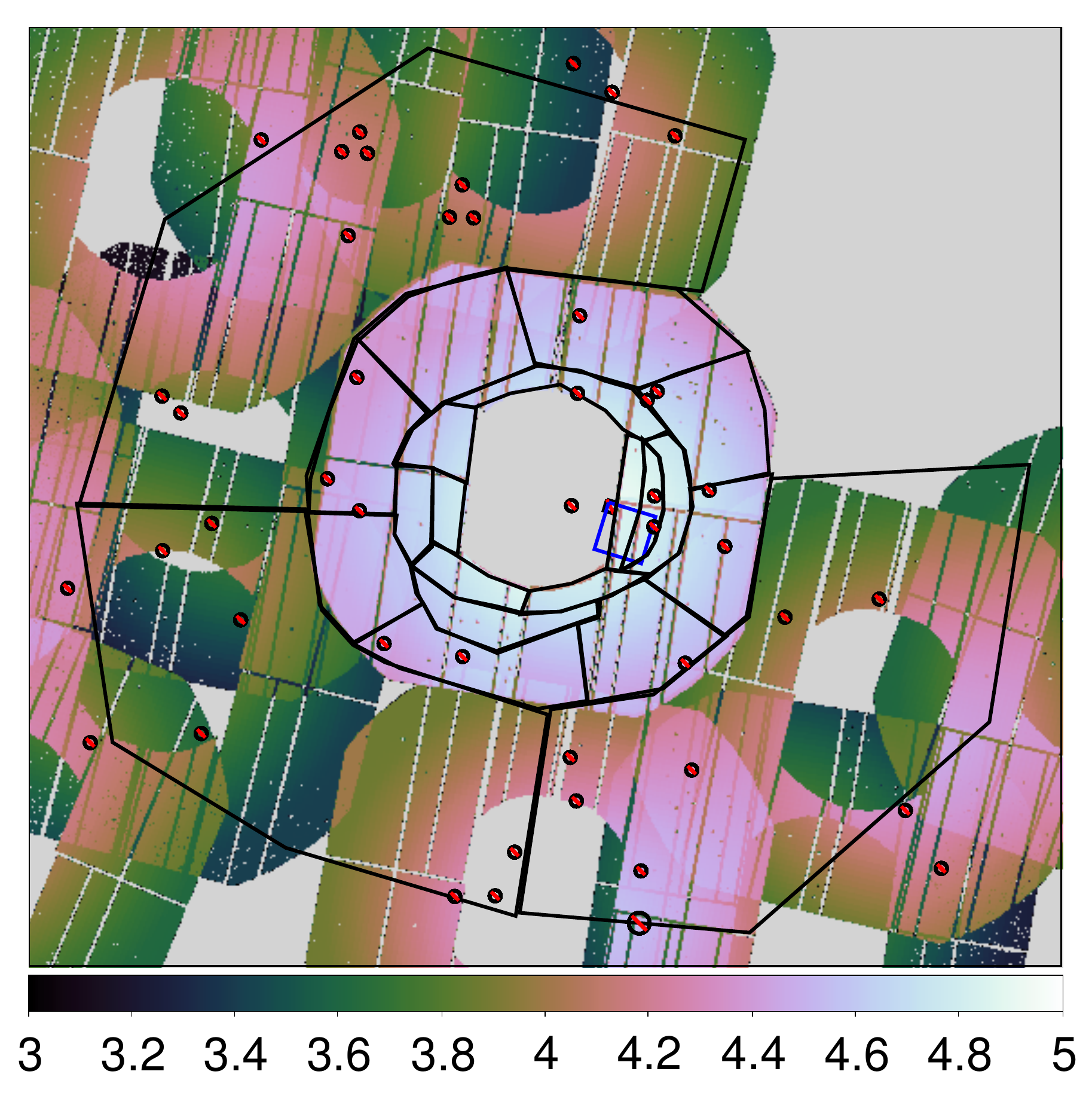}
  \end{minipage}
  \caption{
Regions examined in Perseus, chosen to cover areas with similar numbers of counts in the Fe-K complex.
(Left panel)
Regions plotted on the fractional difference in 0.5 to 2.0 keV surface brightness from the average at each radius.
Region 22 overlaps with other regions as it covers the outer \emph{Hitomi} pointing.
(Top-right panel)
Fe-K count map, showing the number of counts in each 3.2 arcsec pixel in the Fe-K complex (6.50 to 6.90 keV, restframe), after subtracting neighbouring scaled continuum images (6.06 to 6.43 and 6.94 to 7.16 keV, restframe).
Gaussian smoothing of $\sigma=4$ pixels was applied.
(Bottom-right panel)
Exposure map, plotting the total Fe-K exposure (log s) excluding the Cu hole and accounting for vignetting.
}
\label{fig:per_reg}
\end{figure*}

\begin{figure*}
  \includegraphics[width=\textwidth]{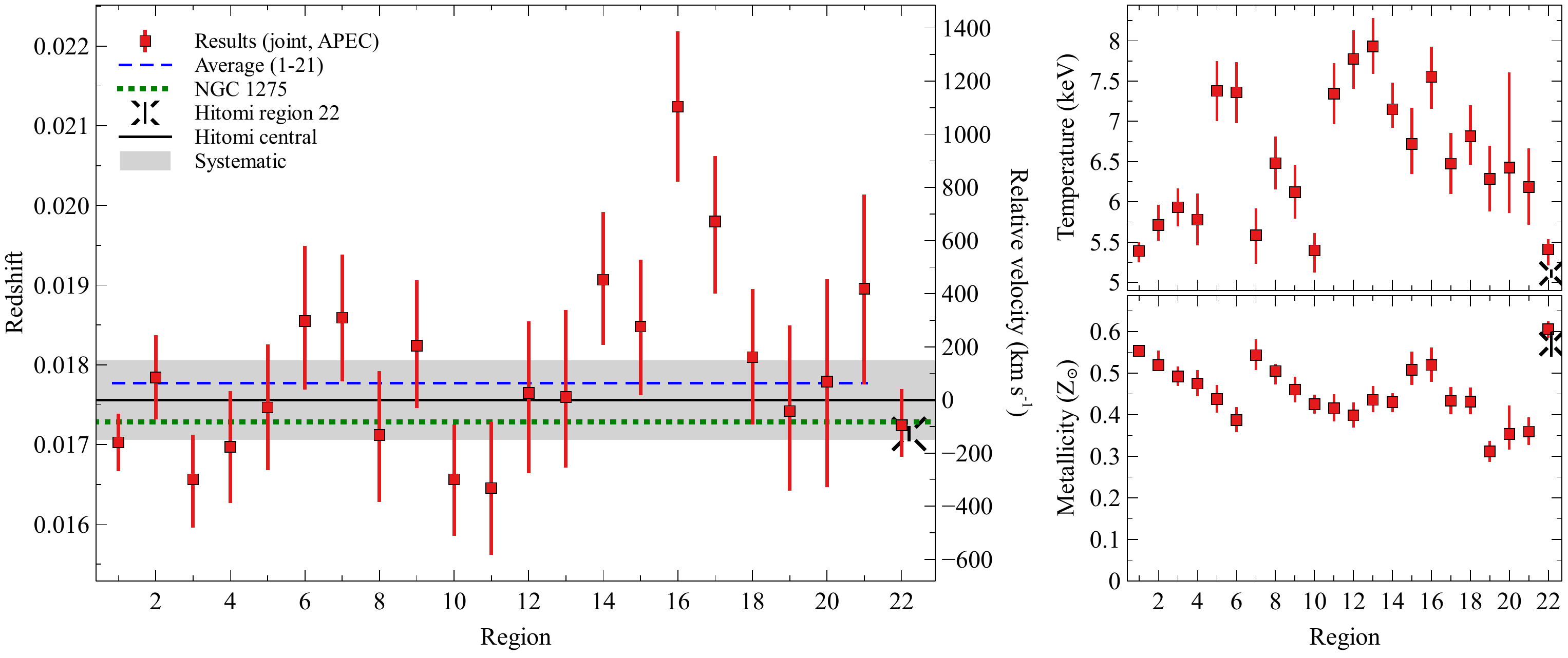}
\caption{
  Results for the Perseus regions.
  Shown are the best-fitting values and statistical uncertainties for the joint fit using the APEC model.
  (Left panel) Redshifts and velocities.
  The average shown is the weighted average for the non-overlapping regions 1 to 21.
  The right axis shows the redshifts as velocities relative to the \citetalias{Hitomi16} result ($z=0.01756$).
  The \object{NGC 1275} velocity of $z=0.017284$ is from \citetalias{HitomiDynamics18}.
  Region 22 was chosen to overlap with the outer \emph{Hitomi} pointing, for which we also show their result.
  (Right panels)
  Temperatures and metallicities. Also shown are the values for the outer \emph{Hitomi} pointing from \protect\cite{HitomiTemp18} and \protect\cite{Simionescu18}.
}
\label{fig:per_vel}
\end{figure*}

\begin{figure*}
  \centering
  \includegraphics[width=0.75\textwidth]{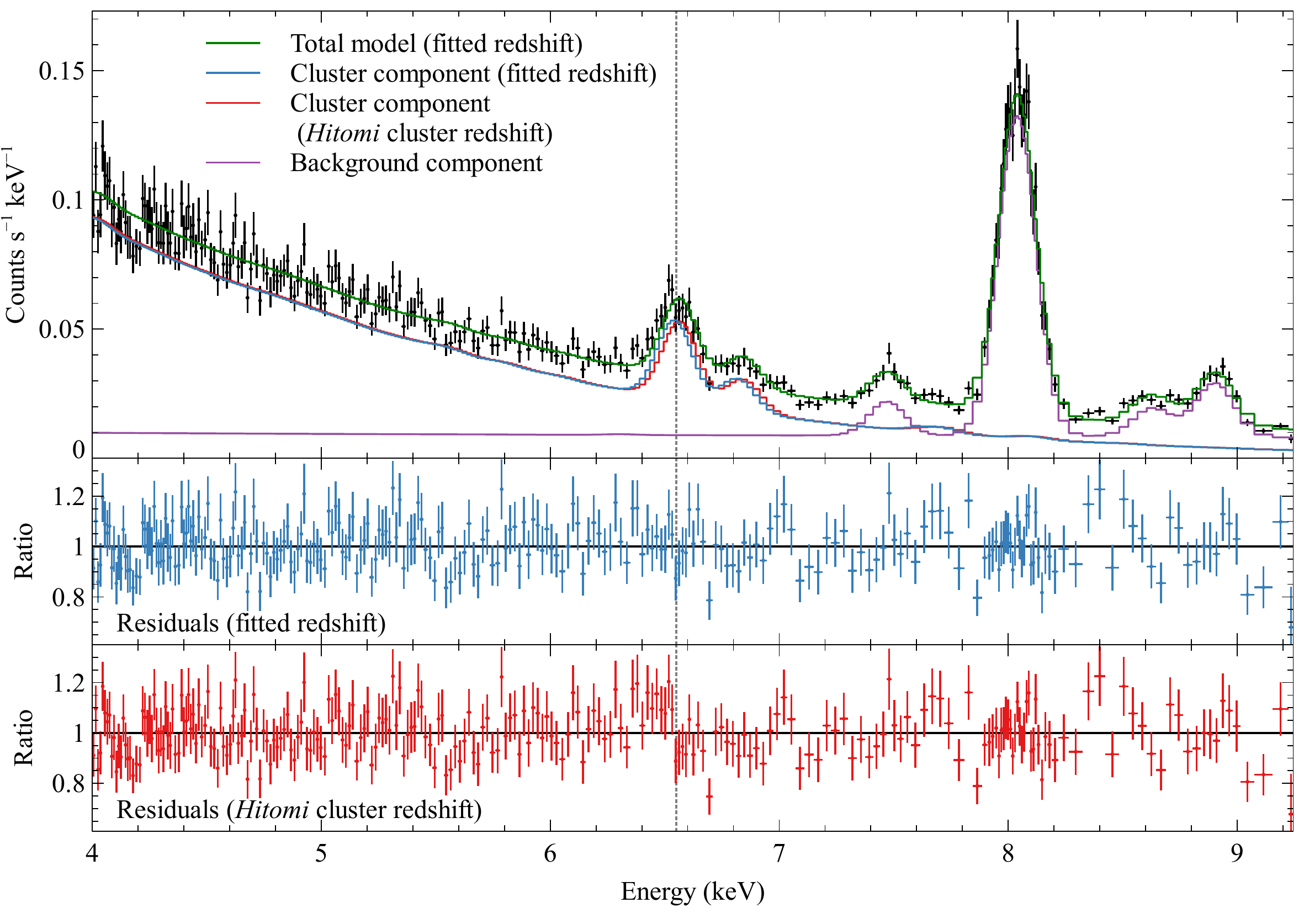}
  \caption{
Example fit of the total spectrum for region 16 in Perseus, rebinned for display to have a signal to noise ratio of 10 in each spectral bin.
The vertical line is at an energy of 6.55~keV.
(Top panel) Fitted data and best-fitting cluster model (fitted redshift), showing its cluster and background components.
Also shown is the cluster component with the redshift changed to the \emph{Hitomi} cluster value.
(Centre panel) Residuals of the total best-fitting model.
(Bottom panel) Residuals of the total model with the APEC redshift changed to the \emph{Hitomi} value.
}
\label{fig:per_reg16}
\end{figure*}

\begin{figure*}
  \centering
  \includegraphics[width=0.7\textwidth]{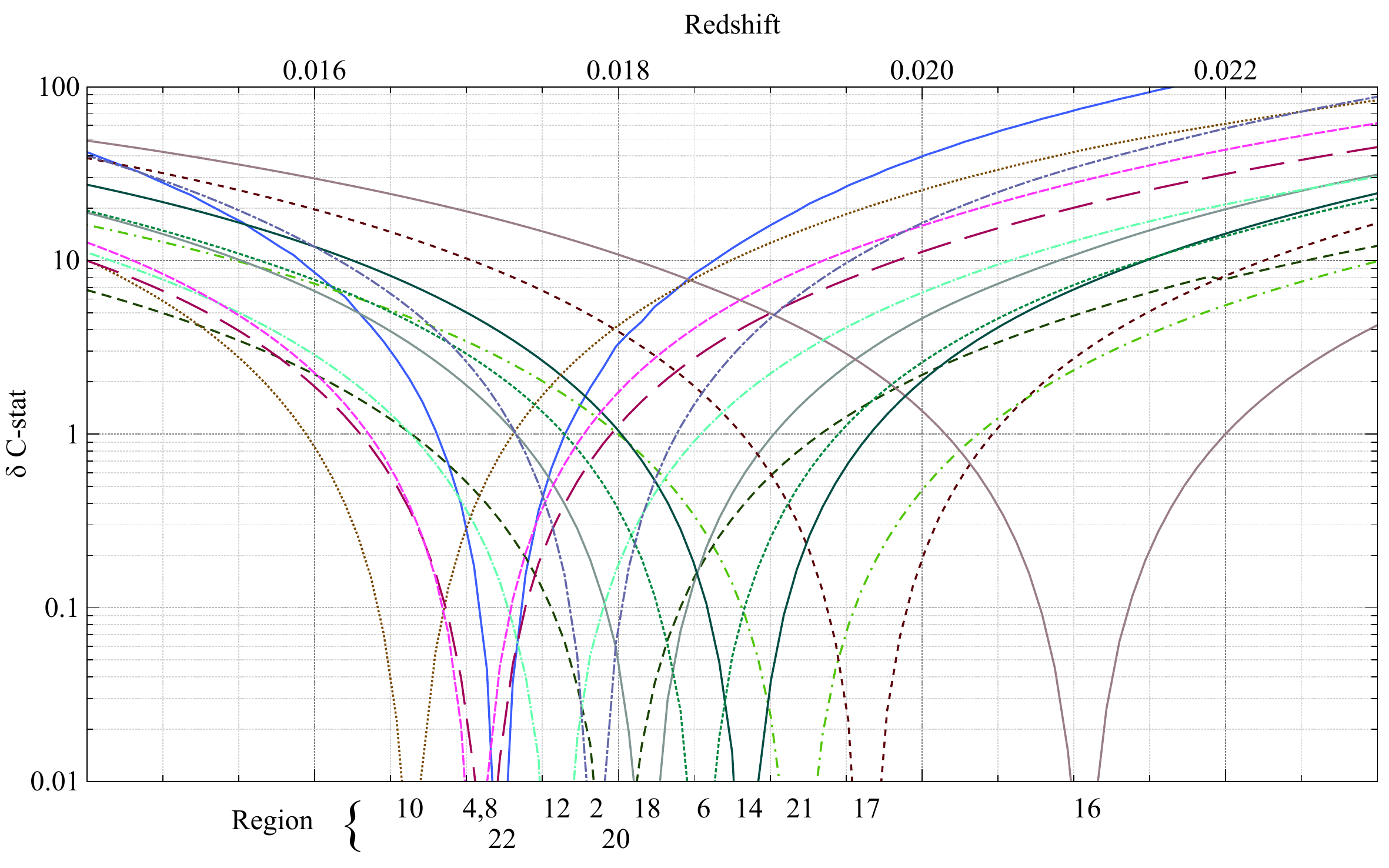}
  \caption{
Change in fit statistic as a function of redshift for selected regions in the Perseus cluster, for the total spectral fits.
}
\label{fig:per_steps}
\end{figure*}

\begin{figure*}
  \centering
  \begin{tabular}{cc}
    \includegraphics[width=0.4\textwidth]{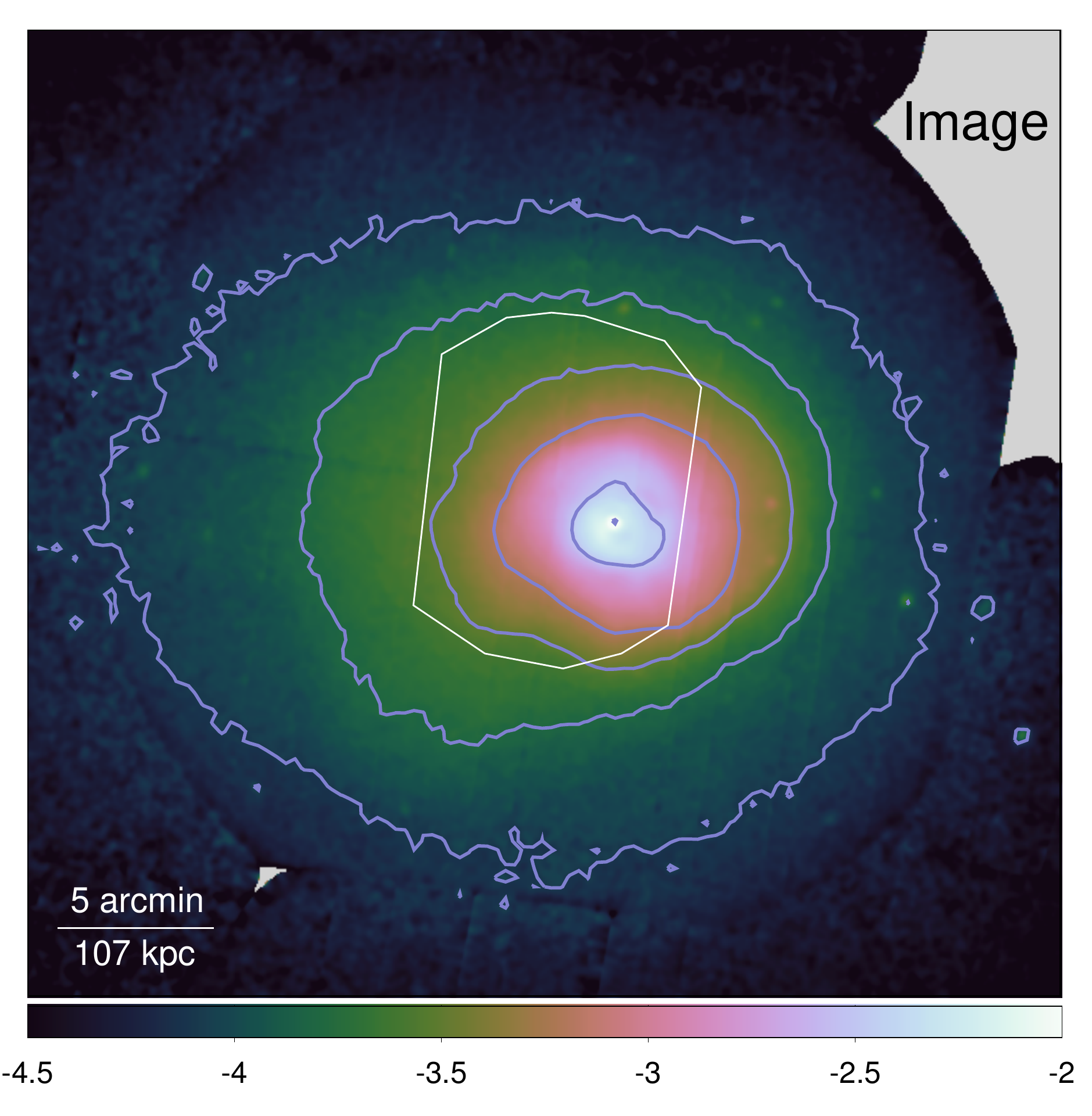} &
    \includegraphics[width=0.4\textwidth]{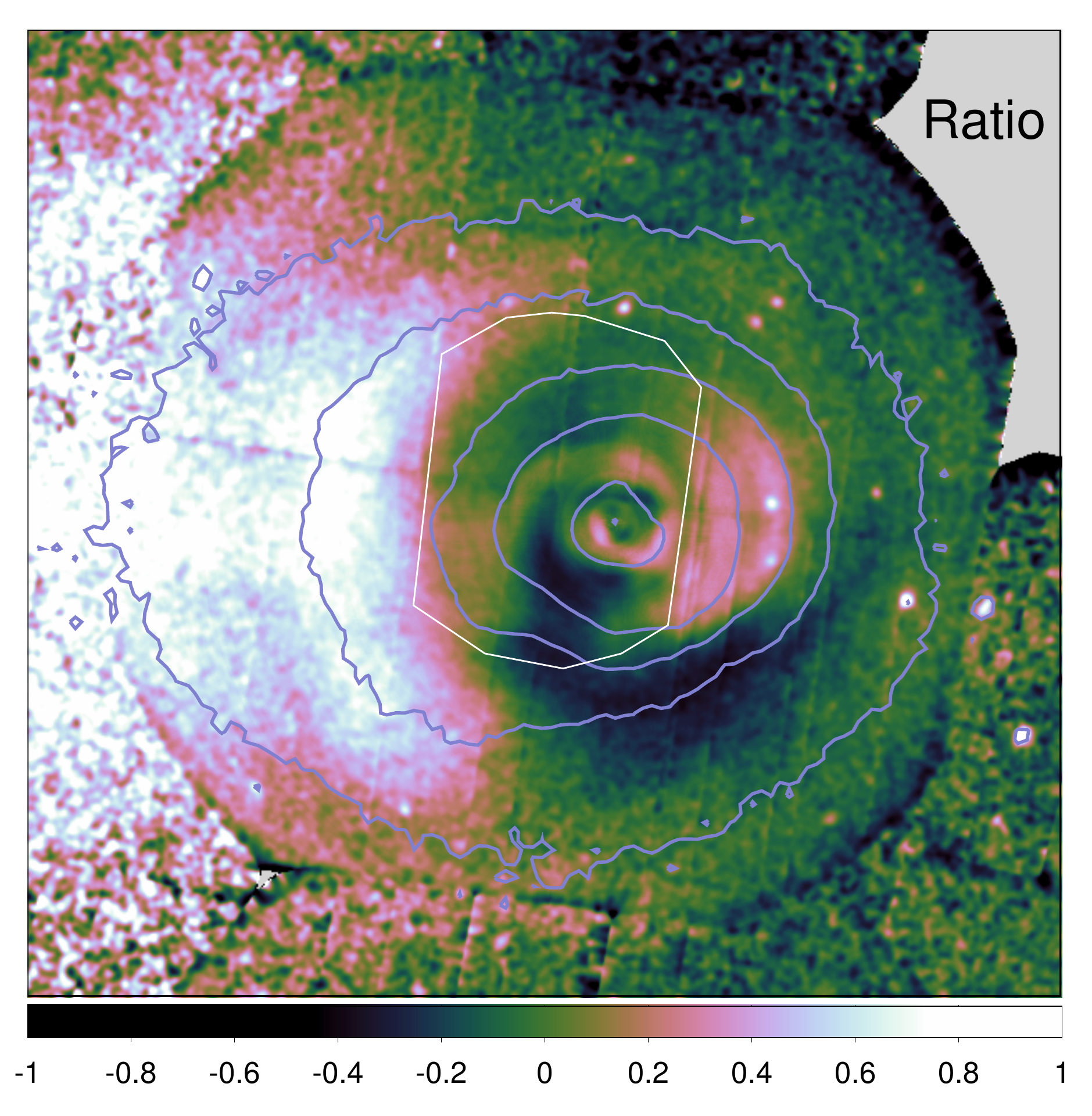} \\
    \includegraphics[width=0.4\textwidth]{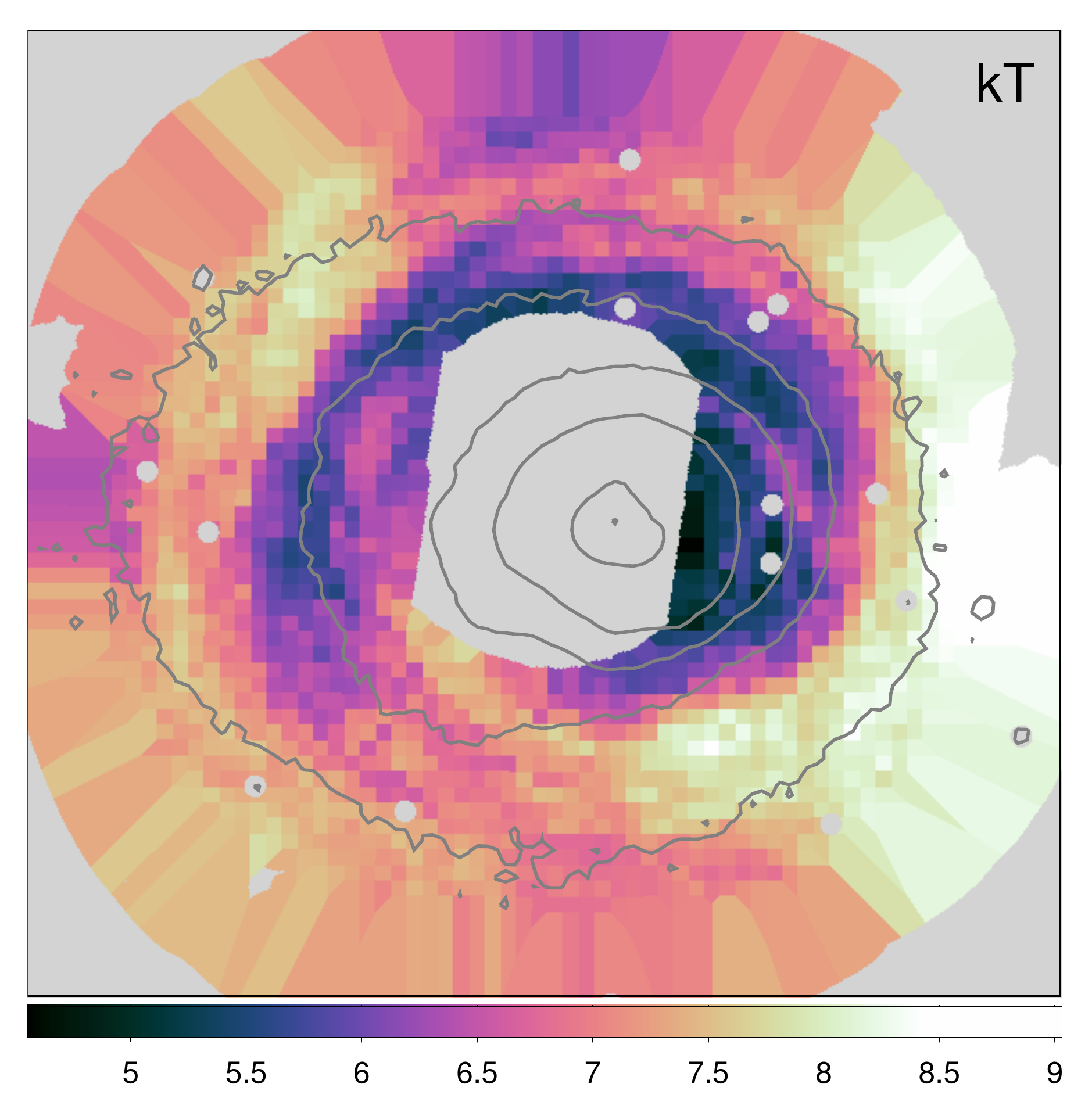} &
    \includegraphics[width=0.4\textwidth]{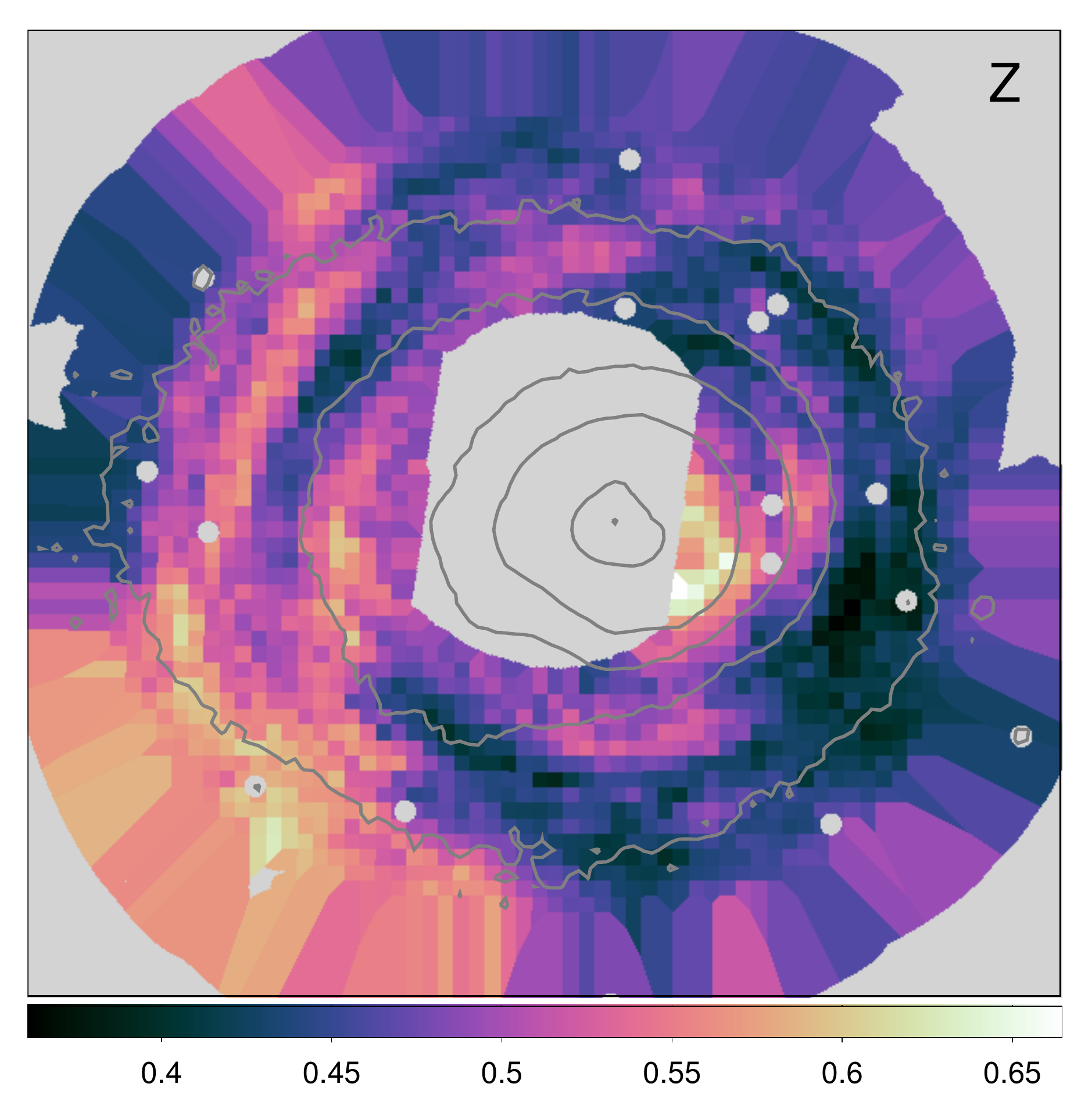} \\
    \includegraphics[width=0.4\textwidth]{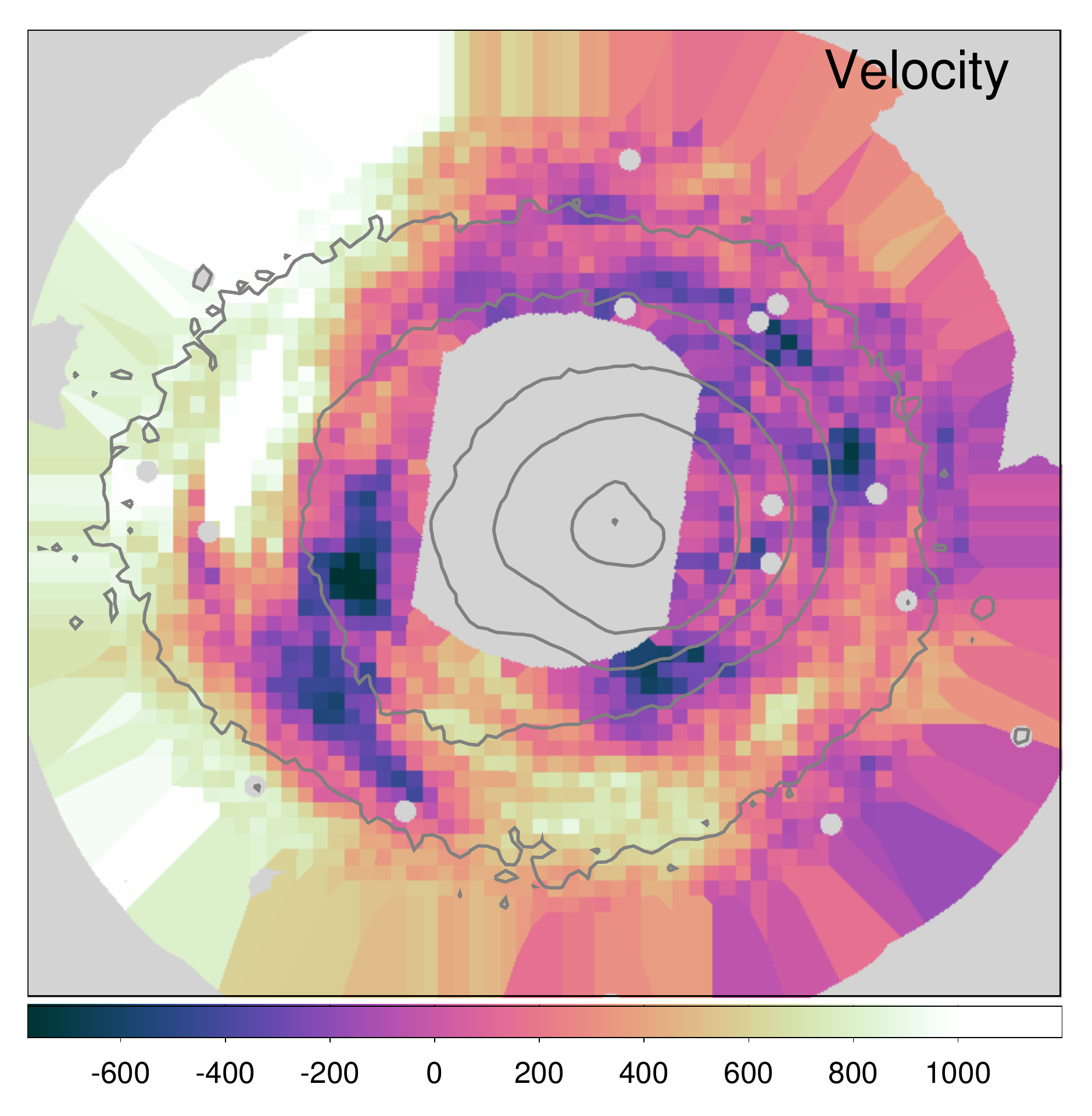} &
    \includegraphics[width=0.4\textwidth]{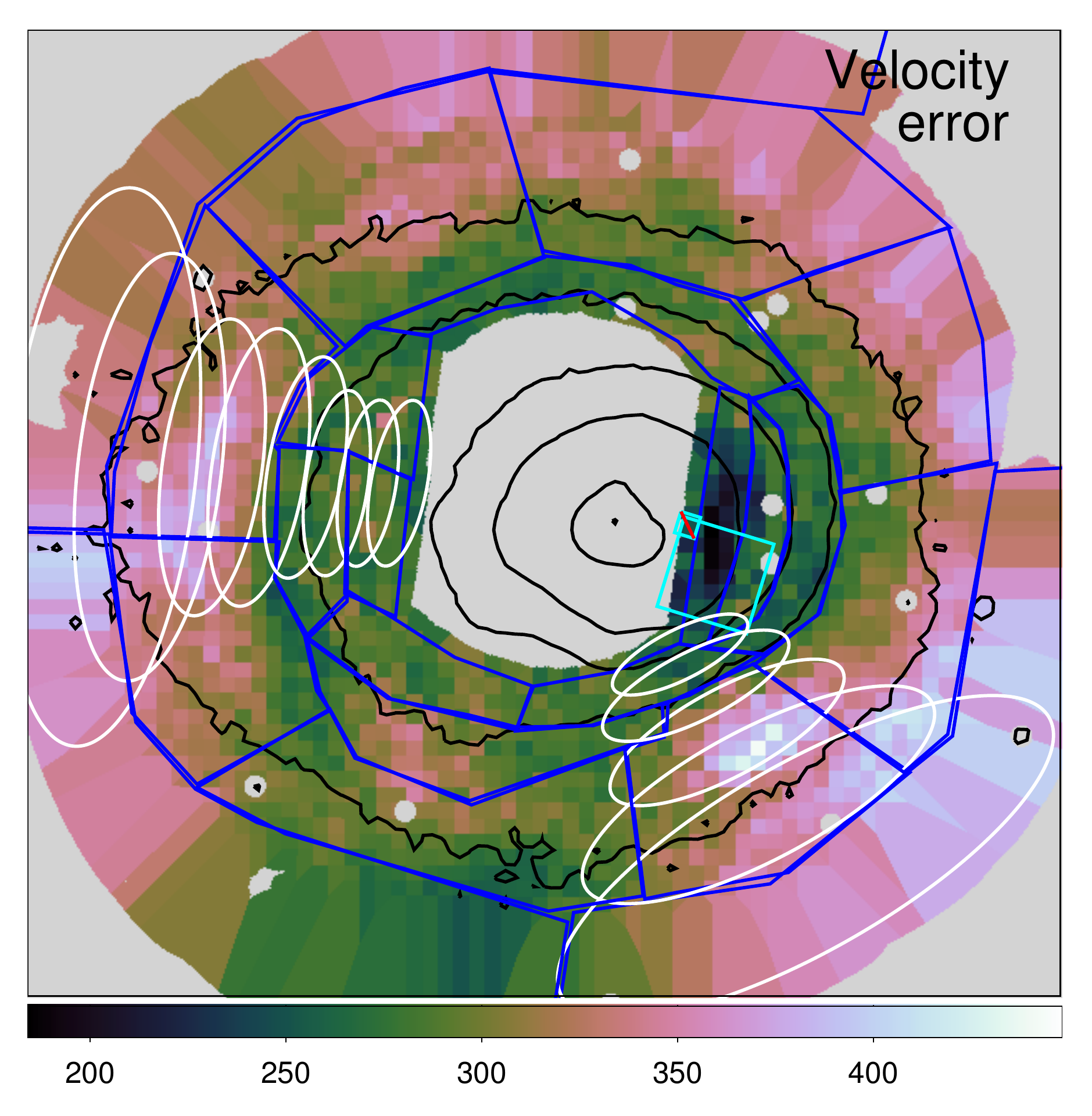} \\
  \end{tabular}
\caption{
Perseus cluster maps. Shown is the  0.5 to 2 keV X-ray surface brightness and contours (log cts~s$^{-1}$ per 1.6 arcsec pixel; top left), fractional difference of surface brightness to radial average (top right), temperature (keV; centre left), metallicity ($\textrm{Z}_{\odot}$; centre right), velocity relative to the cluster value in \citetalias{Hitomi16} (km~s$^{-1}$; bottom left) and $1\sigma$ statistical uncertainty on velocity (km~s$^{-1}$; bottom right).
Maps were created by a moving 3:1 elliptical region (rotated to lie tangentially to the nucleus) containing $\sim 700$ Fe-K counts (see example ellipses in velocity error plot).
The pixels are not statistically independent.
}
\label{fig:per_maps}
\end{figure*}

\begin{figure}
  \includegraphics[width=\columnwidth]{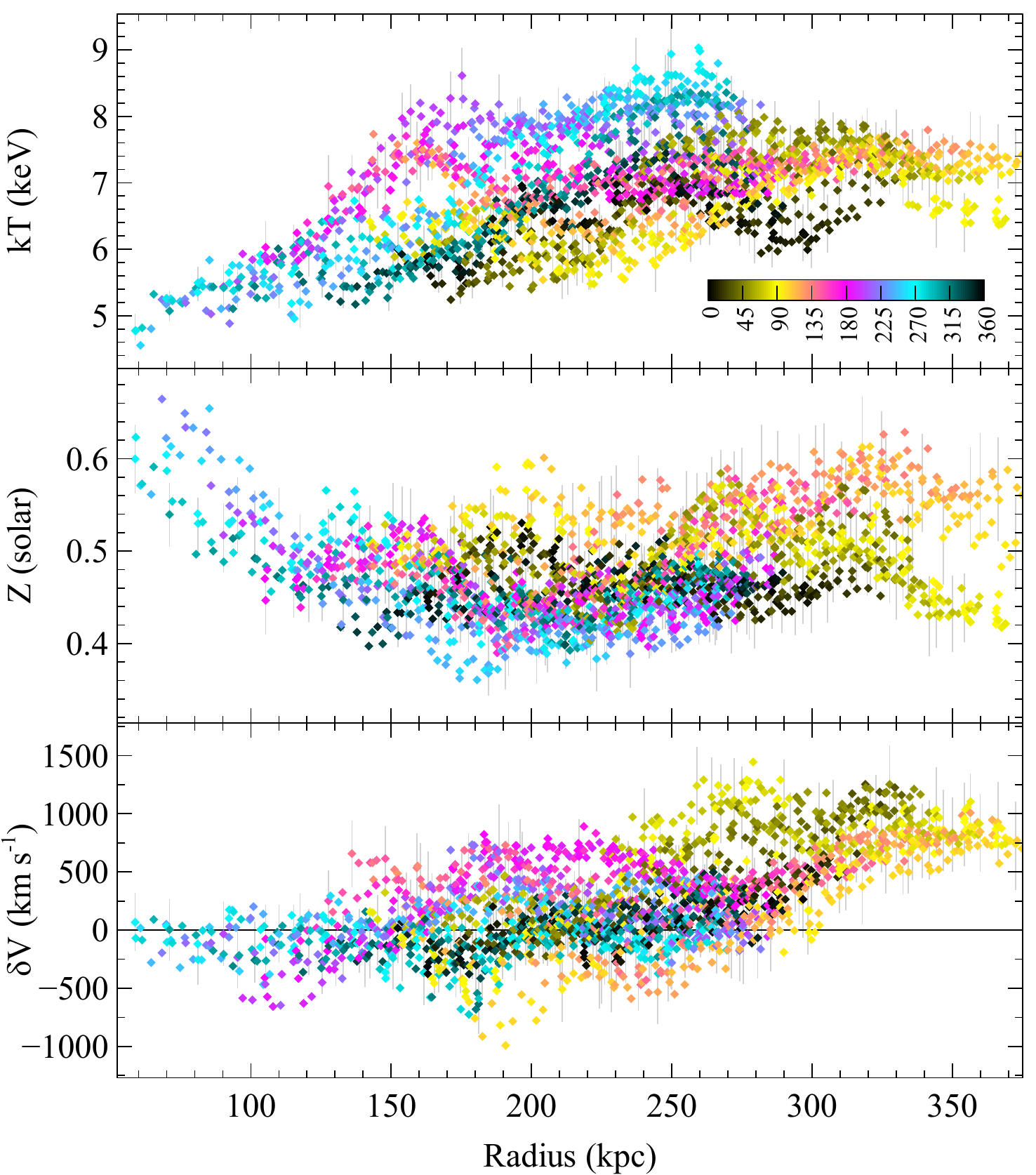}
  \caption{
Radial profiles of temperature, metallicity and velocity pixels from the Perseus maps in Fig.~\ref{fig:per_maps} coloured by position angle.
Error bars are only shown for every 10 points.
Radii are from the position of 3C~84.
Nearby points are not statistically independent.
  }
  \label{fig:per_prof}
\end{figure}

\subsection{Fitting spectra from regions}
In Perseus we made 21 non-overlapping regions by hand (Fig.~\ref{fig:per_reg} left panel) containing similar numbers of counts in the Fe-K complex (around 1000, but increasing to 4500 in the centre; Fig.~\ref{fig:per_reg} top-right panel), and following the surface brightness contours, such as those generated by cold fronts.
We excluded bright point sources by hand,
as automatic source detection was confused by the strong spatial variation of the source.
Unfortunately, with the archival pointings the main AGN feedback region and main \emph{Hitomi} pointing is within the Cu hole, so we cannot examine this area.
However, the offset pointing of \emph{Hitomi} mostly overlaps with our data, so we created an additional region for comparison with this (Region 22), which overlaps with our other regions 1 and 2.
Region 22 is very similar to region 6 in \citetalias{HitomiDynamics18}, although the PSF of \emph{XMM} is much smaller than \emph{Hitomi} so our result will be less effected by flux from neighbouring regions.

Figure \ref{fig:per_vel} (left panel) shows the redshifts and velocities (relative to the \citetalias{Hitomi16} average velocity) we obtained for the regions in Perseus.
Plotted are the results from joint simultaneous fits with the APEC model.
In Fig.~\ref{fig:per_vel} (right panels) we also show the best-fitting temperatures and metallicities in each region.
We list the results of the spectral fits in Table \ref{tab:perfit}.

For our region 22 which overlaps with the outer \emph{Hitomi} pointing, we obtained a redshift of $0.0172 \pm 0.0004$ (all uncertainties quoted only include statistical errors), while a value of $0.0171 \pm 0.0001$ was found using \emph{Hitomi} \citepalias{HitomiDynamics18}, with a model for the PSF.
The difference between these is within the statistical uncertainties, even if systematic effects are not included.
We note, however, that there is a small portion of the \emph{Hitomi} pointing region we cannot examine  and that the PSF of \emph{Hitomi} could blur out some signal over the cluster despite the applied correction.
We can also compare our temperature value for that region ($5.42_{-0.15}^{+0.12}\:\textrm{keV}$) with the \emph{Hitomi} PSF-corrected continuum ($5.11\pm 0.05\:\textrm{keV}$) and line temperatures ($5.00\pm0.16 \:\textrm{keV}$) from \cite{HitomiTemp18}.
Our temperature value is 6 to 8 per cent greater, although again our region does not  overlap exactly with the outer \emph{Hitomi} pointing, which also includes cooler gas at smaller radii.
We obtain a metallicity (dominated by the Fe lines) of $0.61\pm0.02\:\textrm{Z}_{\odot}$, while \cite{Simionescu18} obtained a consistent Fe metallicity of $0.57\pm 0.03\:\textrm{Z}_{\odot}$ with \emph{Hitomi}.

The average value of the redshift in the non-overlapping regions 1 to 21 is $0.01767 \pm 0.00015$, agreeing with the value obtained by \citetalias{Hitomi16} for the central region ($0.01756$).
The fit of a constant value for the regions is poor (reduced $\chi^2=37.3/20=1.86$).
If we assume the intrinsic bulk velocities come from a Gaussian distribution and obtain the model parameters, we find a central redshift of $0.01778 \pm 0.00025$ and a distribution width of $0.00071 \pm 0.00030$ or $214 \pm 85 \:\textrm{km s}^{-1}$.

The offset between the average and Gaussian centre is probably caused by the higher velocities in regions 14-18, which are 200-300 kpc from the nucleus to the east, appearing to have systematically larger velocities than those on the other side (regions 11-13).
The eastern regions are on average $480 \pm 210 \:\textrm{km s}^{-1}$ higher than the \citetalias{Hitomi16} average result, while those to the west are consistent at $20 \pm 190 \:\textrm{km s}^{-1}$ relative to \emph{Hitomi}.
Regions 14 to 18 have systematically higher surface brightnesses at that radius (Fig.~\ref{fig:per_reg} left panel).
The cluster is also cooler towards the eastern side (Fig.~\ref{fig:per_vel} top-right panel).
To see the significance of the signal in this region, we show an example spectrum for region 16 (Fig.~\ref{fig:per_reg16}).
We show the residuals for a model with the best-fitting redshift and those using the reference cluster \emph{Hitomi} value.
If the cluster redshift is used, there are clear positive and negative residuals either side of the He-like Fe \textsc{xxv} line complex, indicating a line shift.
In Fig.~\ref{fig:per_steps} we show the change in fit statistic if the redshift is varied from the best-fitting value and the spectra refitted, for selected spatial regions.
The curves are smooth and only have a single minimum, and only become asymmetric far from the best-fitting redshifts, showing that the best-fitting values and statistical uncertainties are robust.

If we model the distribution of velocities in only the central region (regions 1 to 10) as a Gaussian we find a central redshift of $0.01739 \pm 0.00021$, while the distribution width is small ($28$ to $173\:\textrm{km s}^{-1}$; at the $1\sigma$ level).

There are further possible systematic errors, which we discuss in Sect.~\ref{sect:systematics}, although they appear unimportant compared to the energy calibration and statistical uncertainties.

\begin{figure*}
\centering
\includegraphics[width=0.31\textwidth]{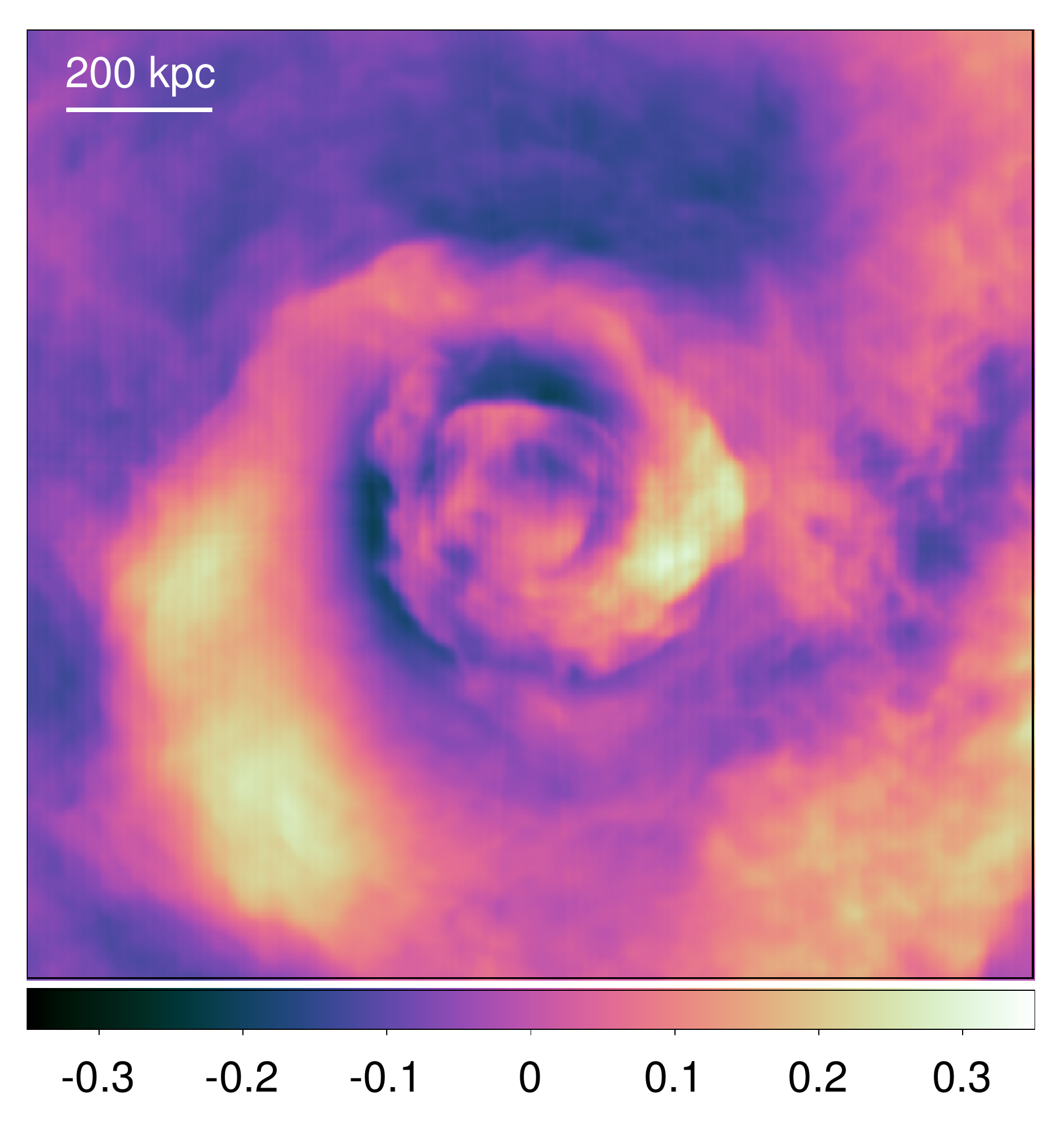}
\includegraphics[width=0.31\textwidth]{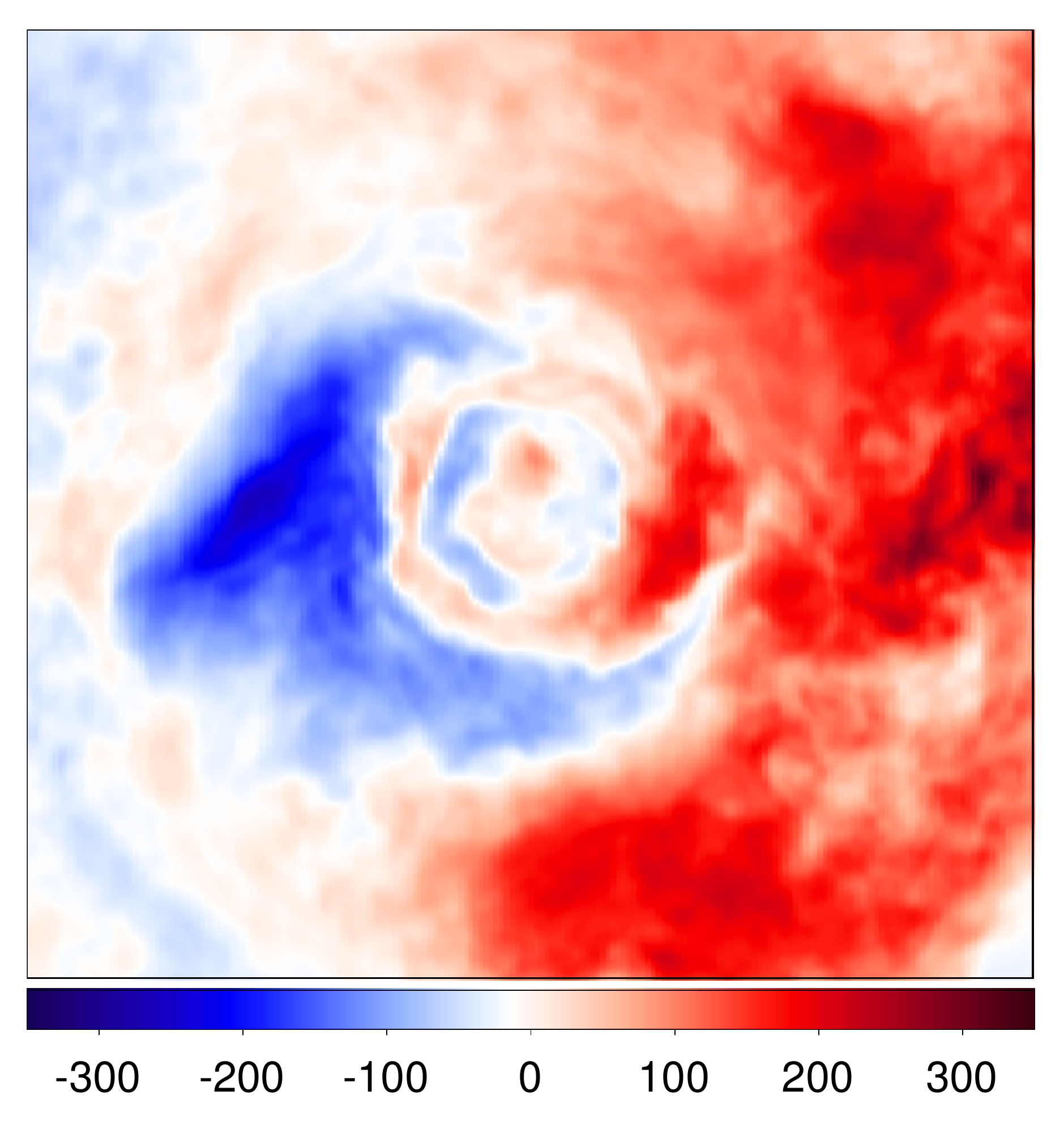}
\includegraphics[width=0.31\textwidth]{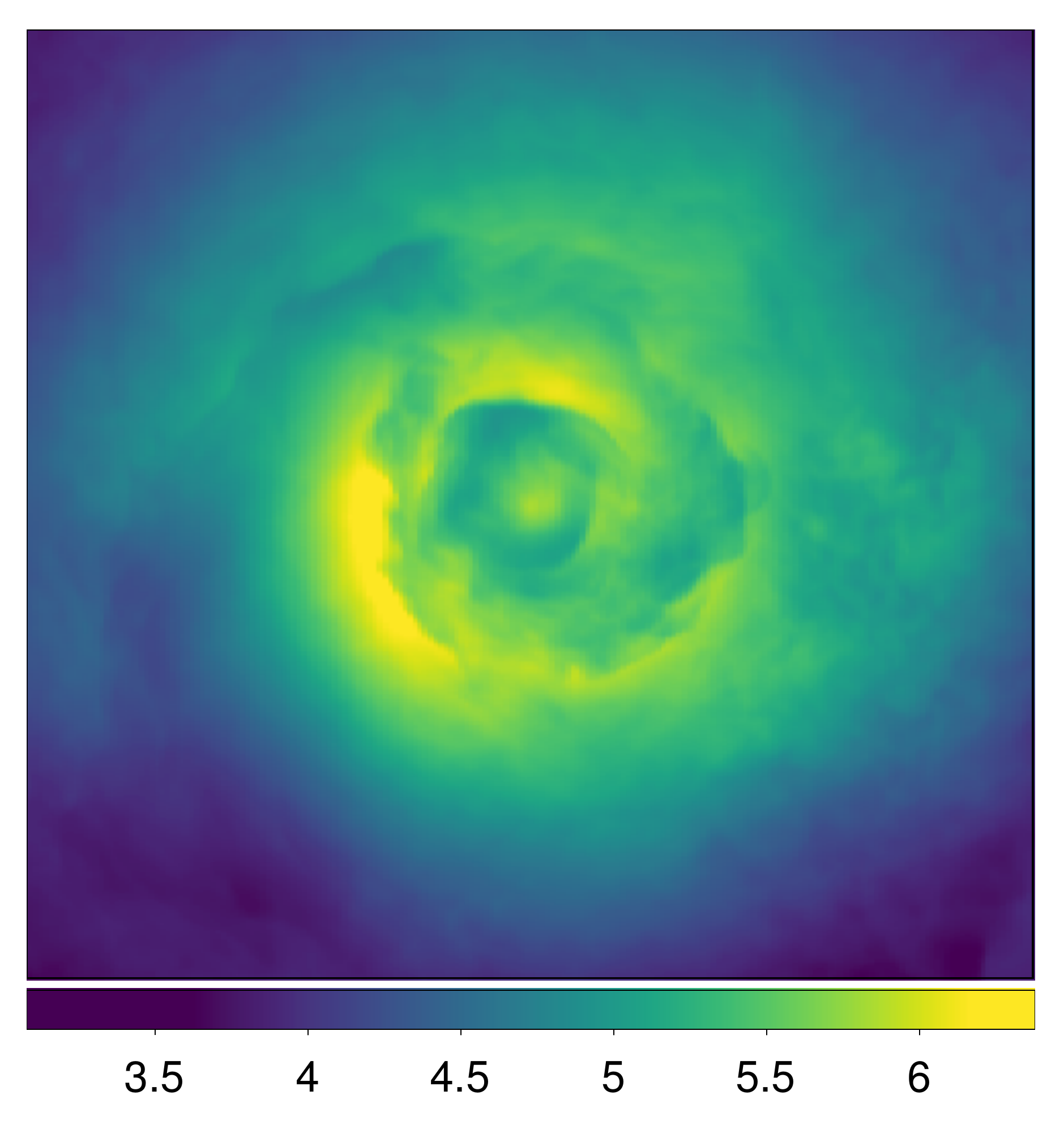}\\
\includegraphics[width=0.31\textwidth]{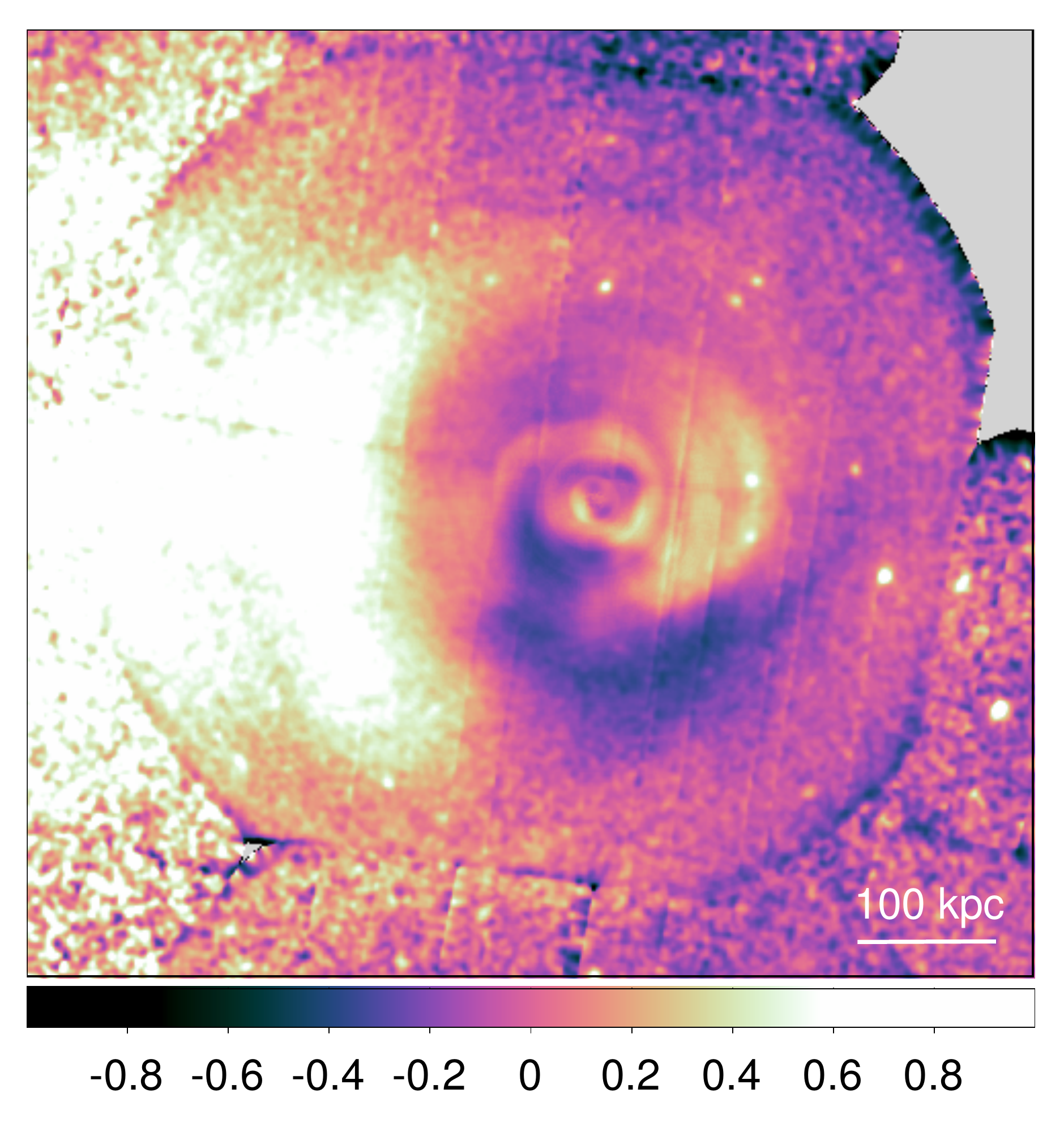}
\includegraphics[width=0.31\textwidth]{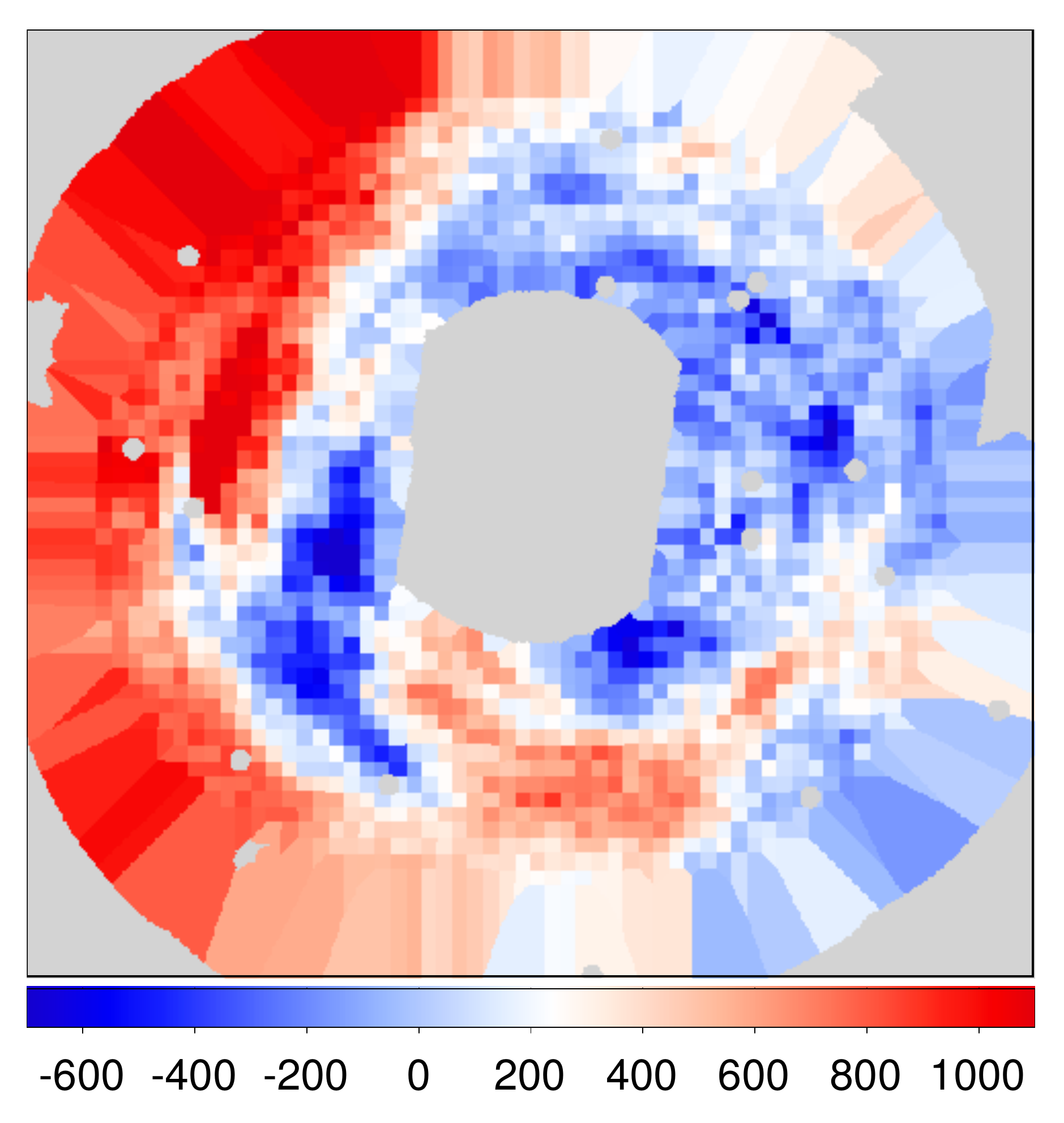}
\includegraphics[width=0.31\textwidth]{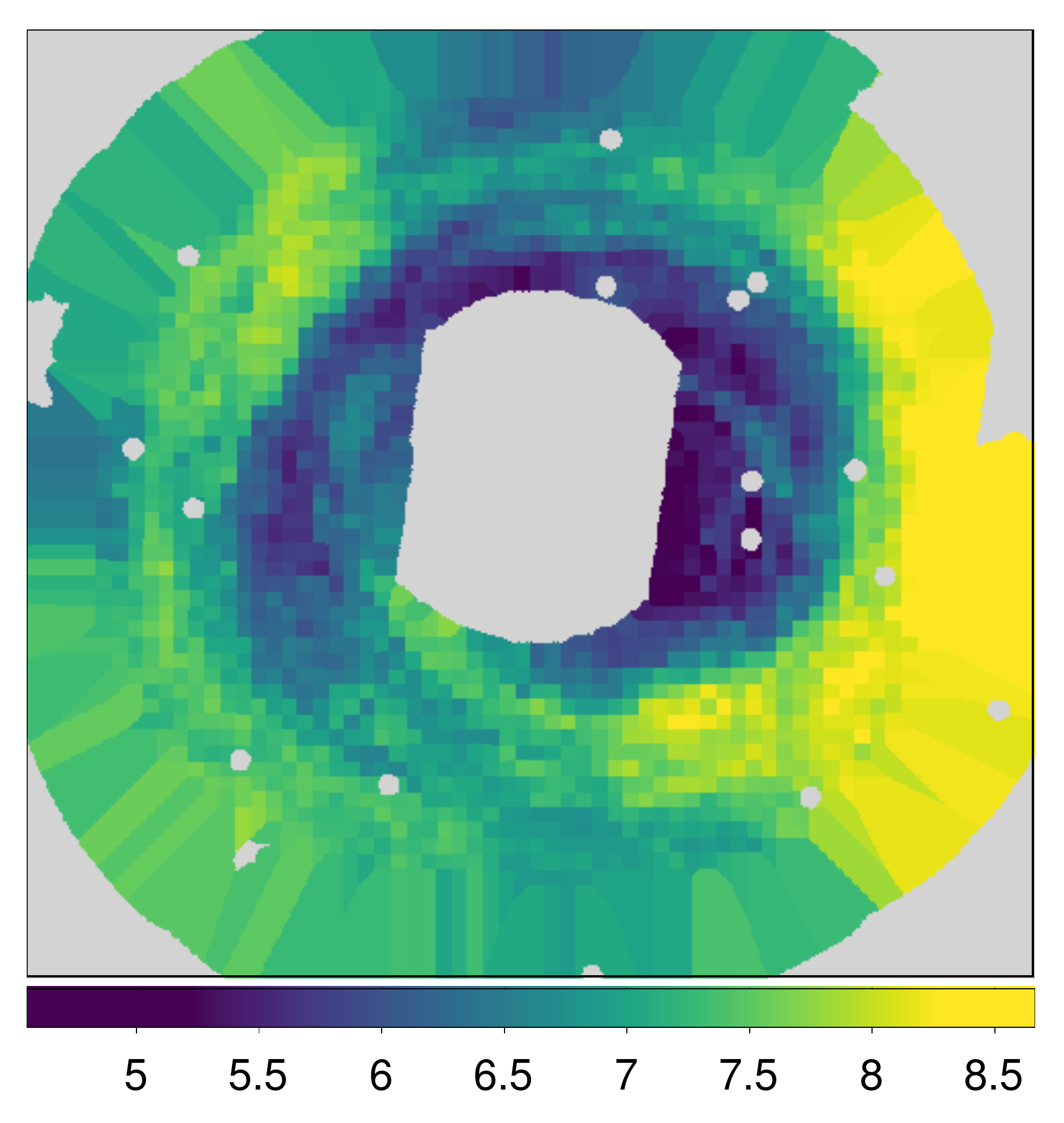}\\
\includegraphics[width=0.31\textwidth]{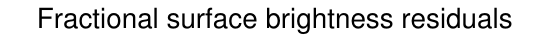}
\includegraphics[width=0.31\textwidth]{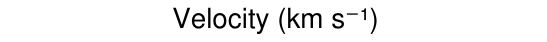}
\includegraphics[width=0.31\textwidth]{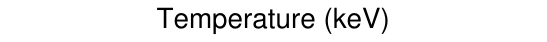}\\
\caption{
Comparison of a sloshing simulation (top row) and data for the Perseus cluster (Fig.~\ref{fig:per_maps}; bottom row).
The columns show surface brightness residuals (left), velocity (centre) and temperature (right).
The simulation is of a Perseus-like cluster which has bulk sloshing motions and turbulence initiated by an off-centre encounter with a subcluster of 1/5 its mass, from \protect\cite{ZuHone10,ZuHone18}.
}
\label{fig:per_sim}
\end{figure*}

\subsection{Spectral maps}
We can see the velocity, temperature and metallicity structure better by making a smooth map of the cluster.
To do this we extracted and fitted spectra from regions over the cluster, moving in grid with a spacing of 0.5 arcmin.
To make the regions better follow the steeply-peaked surface brightness profile of the cluster, we
used elliptical regions with a 3:1 axis ratio, rotating them so that the longest axis lay tangentially to the vector to the central AGN.
We adaptively changed the radii of the ellipses to have $\sim 700$ counts in the Fe-K complex after continuum subtraction,  to have similar uncertainties on the velocities.
Point sources were excluded from the regions.
The map was created by fitting the total spectra rather than by a joint fit.
We used the same response matrix for all the pixels, but created weighted-average ancillary responses for each ellipse, created from the individual ones for each observation.

Figure \ref{fig:per_maps} shows the resulting maps of the velocity, temperature and metallicity.
For comparison we also show an X-ray image of the cluster from \emph{XMM}, and the fractional difference of the X-ray image to the average at each radius from the nucleus.
The temperature plot shows that there is a cool spiral of gas going from the west of the cluster to the east via the north.
Its outer edge varies from $\sim 170$ to $260$~kpc radius.
The outer edge of the cool spiral can be seen in the surface brightness image, although there is another region of higher surface brightness which lies at larger ($\sim 330$~kpc) radius to the east.
There is a spiral metallicity structure associated with the cooler spiral.
To the east of the core, the low temperature spiral gas has low metallicity, with the hotter material at smaller and larger radii having higher metallicities.

Examining the velocity structure, there appears to be a clear correlation between the temperature and velocity maps.
These correlations can also be seen in radial profiles of the maps (Fig.~\ref{fig:per_prof}).
The cool spiral has lower velocities, while the material lying outside it to the east and south has larger velocities.
There is possibly higher velocity material to the south-west, although that location is close to the edge of the field of view and the elliptical regions are rather large, smoothing out small scale structure.
We do not see clear velocity structure within the cool spiral.
These maps suggest that some of the regions in Fig.~\ref{fig:per_reg} could contain velocity structure within them which is being averaged out.
Better regions could be defined from the velocity map itself, but this is likely to bias the resulting velocities towards the extremes.

\subsection{Perseus discussion}
Other than \emph{Hitomi}, previous measurements of the velocity structure in Perseus include those of \cite{DupkeBregmanPer01}, who claimed a $1000 \:\textrm{km s}^{-1}$ velocity difference on 30 arcmin scales on an axis inclined with a position angle of $135^{\circ}$, $45^{\circ}$ to the major east-west axis of the cluster.
This is on larger scales than we probe here, so we cannot make comparisons.

\cite{Tamura14} examined multiple pointing positions in the Perseus cluster and in the core using \emph{Suzaku}.
On larger scales they limited motions to $<600 \:\textrm{km s}^{-1}$.
Our results are consistent with this on limit.
They found a hint of bulk motions around $-150$ to $-300\:\textrm{km s}^{-1}$ around $45$ to $90$~kpc to the west of the cluster centre.
This is a high surface brightness region inside the inner western cold front and includes our regions 1, 2 and 22.
Their result is roughly consistent with our results, with regions 22 and 1 being lower than the \emph{Hitomi} average by $\sim 220$~km~s$^{-1}$, although the lower surface brightness region 2 is greater than \emph{Hitomi} by $\sim 50\:\textrm{km s}^{-1}$.
More detailed comparisons are difficult as their results are measured relative to each other and their reference region is inside the Cu-hole in our data.
\cite{Ota16} placed limits on the velocities in Perseus of $860\:\textrm{km s}^{-1}$ at the 90 per cent level, which is consistent with our results.

We found evidence of a bulk motions of $\sim 480 \pm 210\:\textrm{km s}^{-1}$ away from the observer, at a radius $250$~kpc to the east of the Perseus cluster core.
At this radius the gas velocities are unlikely to be affected by the central AGN.
High spatial resolution maps (Fig.~\ref{fig:per_maps}) suggest that there is a good correlation between the structure of the cold front to the east, as seen in temperature map, and the gas velocity.
The material at smaller radius with lower temperature appears to have a low negative velocity, while the hotter gas outside this radius has a high positive velocity.
This cooler material has a spiral-like twisted structure.
Such structures are seen in simulations of ICM sloshing in cluster potential wells.
Figure \ref{fig:per_sim} compares a simulation of a Perseus-like cluster with bulk-sloshing motions and turbulence initiated by an off-centre encounter with a subcluster, with our Perseus maps.
The surface brightness residuals, temperature and velocity have similar morphologies to what we see in Perseus, and are indicative of sloshing, although the simulations are not an exact match to what is observed.

Unusually, the higher surface brightness, cooler material has a lower metallicity than the hotter material at larger radius and at smaller radius, which is somewhat contrary to simulations of sloshing of gas in potential wells \citep[e.g.][]{Roediger12}, where high metallicities are correlated with high surface brightnesses, and as seen in other clusters (e.g. Centaurus; \citealt{SandersCent16}).
There is also an unusual surface brightness structure at the temperature interface.
Moving across the front, the X-ray surface brightness is higher than the radial average, but drops temporarily at the interface between the hotter and colder gas.
Usually across sloshing structures there are an alternate series of cooler, metal rich relatively-brighter, and hotter, metal poor and relatively-fainter regions.
The lack of a typical alternating structure in Perseus may indicate that the structure is not simple sloshing.
It is possible that there are additional structures induced by a merging subcluster, as is indicated by the east-west chain of galaxies \citep{ChurazovPer03}.

\cite{ZuHone18} simulated the emission-weighted velocity field in Perseus from sloshing motions, projecting it along different lines of sight.
The range of line shifts found in the simulations was between $\sim-250$ and $300 \:\textrm{km s}^{-1}$.
We find velocities which are a bit larger than this ($500 \pm 100\:\textrm{km s}^{-1}$), although the measurement uncertainties and $\sim 100\:\textrm{km s}^{-1}$ systematic uncertainties could bring these into agreement.
Alternatively, if there is an undergoing merger, then this could also be responsible for the increased velocities, although the excess chain of galaxies lies to the western and not the eastern side.

If the gas were sloshing in the potential well, it might be expected to see a corresponding negative velocity in the cold front on the other side of the core at smaller radius \citep{ZuHone18}.
The average velocity for regions 1 to 3 on the western side of the core where the X-ray surface brightness is high, is $-150 \pm 90 \:\textrm{km s}^{-1}$.
This magnitude of this velocity is much lower than our detection to the east.

Assuming a Gaussian distribution, the width of the velocities for all the non-overlapping regions is  $260^{+85}_{-70} \:\textrm{km s}^{-1}$.
If we examine the distribution of velocities in the inner regions (1 to 10), the width is lower ($110^{+100}_{-75}\:\textrm{km s}^{-1}$) and close to the $100\:\textrm{km s}^{-1}$ Doppler width of the Fe-K line as observed by \emph{Hitomi} regions where there is no AGN disturbance \citepalias{HitomiDynamics18}, although the spread of bulk velocities and the line width are not necessarily measuring the same set of motions.

Examining the velocity map structure inside the cold front to the west shows little evidence of structure.
This western cold front appears much more typical, with cooler, more metal rich and higher surface brightness material inside the front and the opposite outside.
There is little evidence of a change in velocity across this edge to the west, but there is a possible high velocity region to the south outside the front, seen in region 14.

\section{The Coma cluster}
\label{sect:coma}
The \object{Coma cluster}, \object{Abell 1656}, contains two dominant galaxies, which are \object{NGC 4874} and \object{NGC 4889}.
The two galaxies have a large velocity difference of $\sim 700\:\textrm{km s}^{-1}$ between them \citep{FitchettWebster87}, likely caused by a major merger in the cluster.
The flat central X-ray surface brightness supports the idea that the cluster is unrelaxed \citep{Briel92,White93}.
The cluster hosts a giant radio halo \citep{Willson70}, indicating it is a merging system.
In addition, an analysis of the velocity distribution of the galaxies in Coma shows that there are subgroups associated with the two central galaxies \citep{Adami05} and up to 15 further subgroups in the system.
A weak lensing analysis \citep{Okabe10} also suggests multiple substructures within the cluster, two of which are associated with the central galaxies.
It is unclear which of the two central galaxies is the newest arrival in the cluster, although the velocity structure of the Coma galaxies appears to better match \object{NGC 4874} \citep{Adami05}.

Previous \emph{XMM} observations have shown that the group \object{NGC 4839} to south of the cluster \citep{Neumann01} appears to be merging, as indicated by a temperature enhancement between the group and the cluster.
There is also a radio relic in the direction of the merging group \citep{JaffeRudnick79,Giovannini85}.

In addition, there is another group to the east seen by its X-ray emission \citep{Vikhlinin97}, associated with \object{NGC 4911} and \object{NGC 4921}.
This system has an estimated mass of $5-8 \times 10^{12}\:\textrm{M}_{\odot}$ \citep{Neumann03}.
The outer edge of its X-ray emission lies at the edge of the radio halo \citep{Simionescu13}.

There are also a number of cooler denser arms of X-ray emitting gas that extend into the core of the cluster, which may be stripped material from merging subclusters \citep{SandersComa13}, some of which appear connected to the NGC\,4911/4921 structure.

\subsection{Results}

\begin{figure*}
  \centering
  \begin{minipage}[c][][c]{.69\textwidth}
    \includegraphics[width=\textwidth]{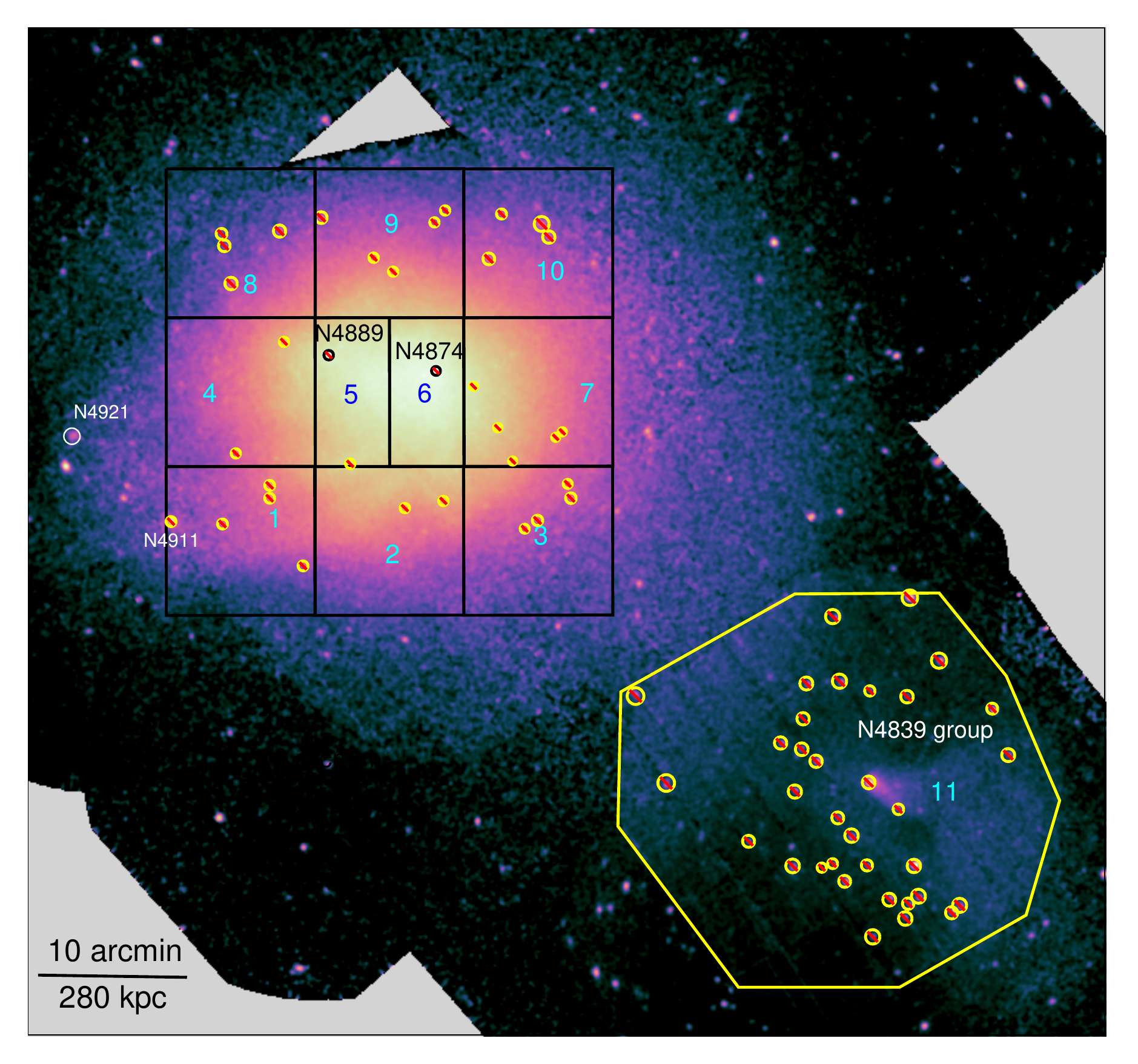}
  \end{minipage}
  \hfill
  \begin{minipage}[c][][c]{.3\textwidth}
    \includegraphics[width=\textwidth]{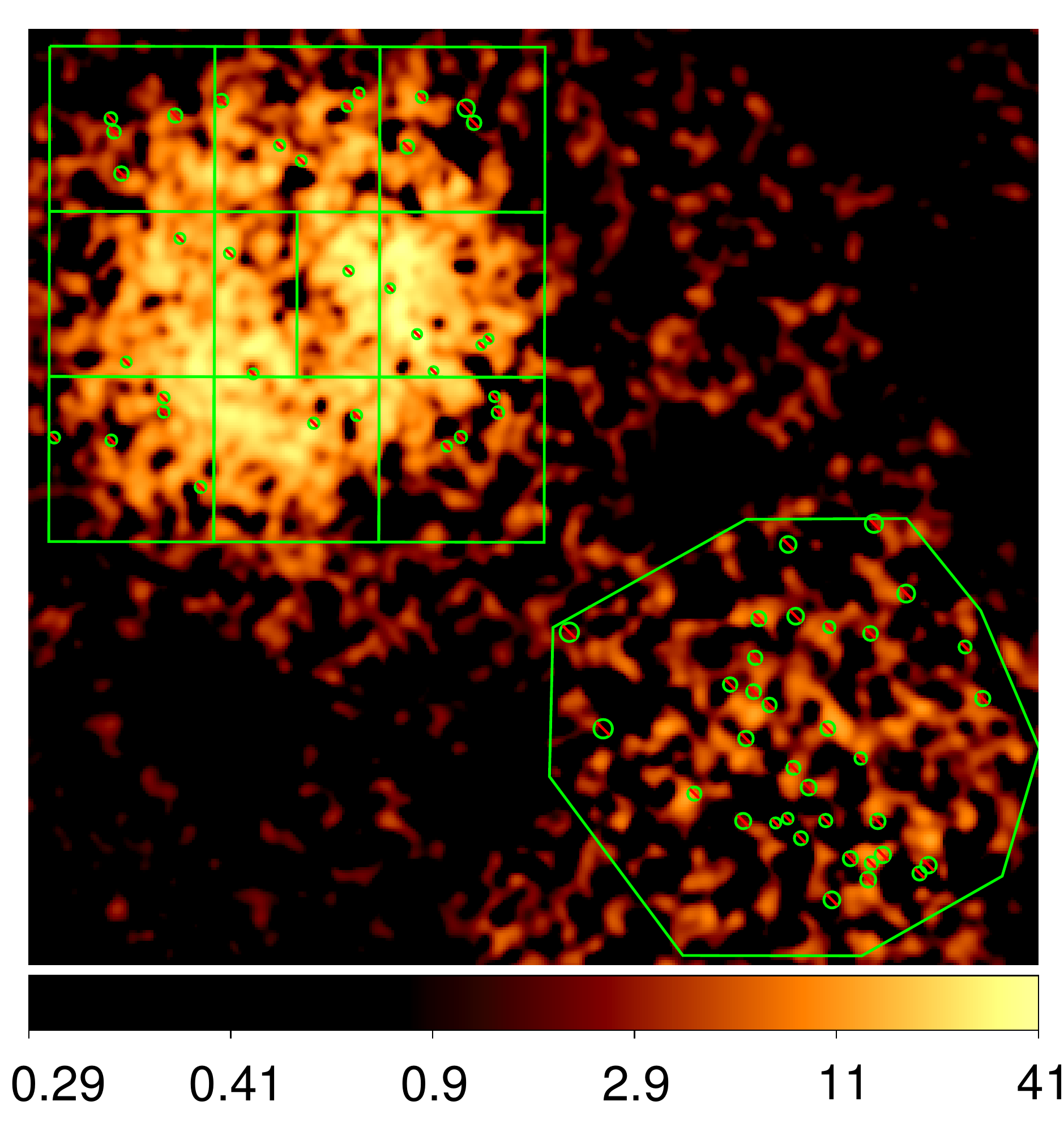}
    \vfill
    \includegraphics[width=\textwidth]{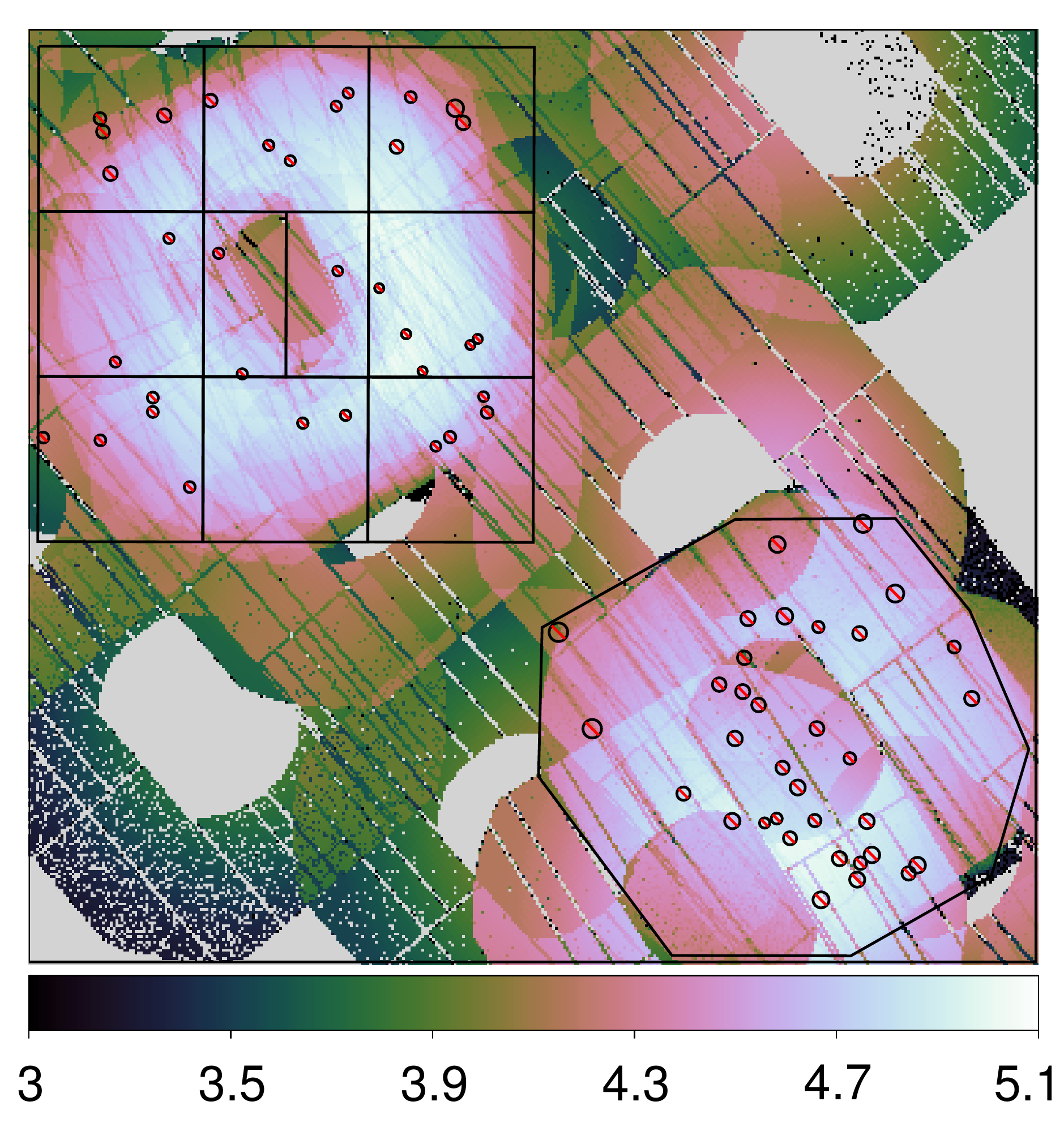}
  \end{minipage}
  \caption{
    Regions examined in Coma.
    (Left panel) The regions plotted on an exposure-corrected surface brightness image in the 0.5 to 2.0 keV band.
    (Top-right panel) Image in the Fe-K band, after subtracting continuum, in $10^{-3}$ counts per 1.6 arcsec pixel.
    (Bottom-right panel) Fe-K exposure map in log s, including the effect of vignetting and the Cu hole.
  }
  \label{fig:coma_regions}
\end{figure*}

\begin{figure*}
  \centering
  \includegraphics[width=0.96\textwidth]{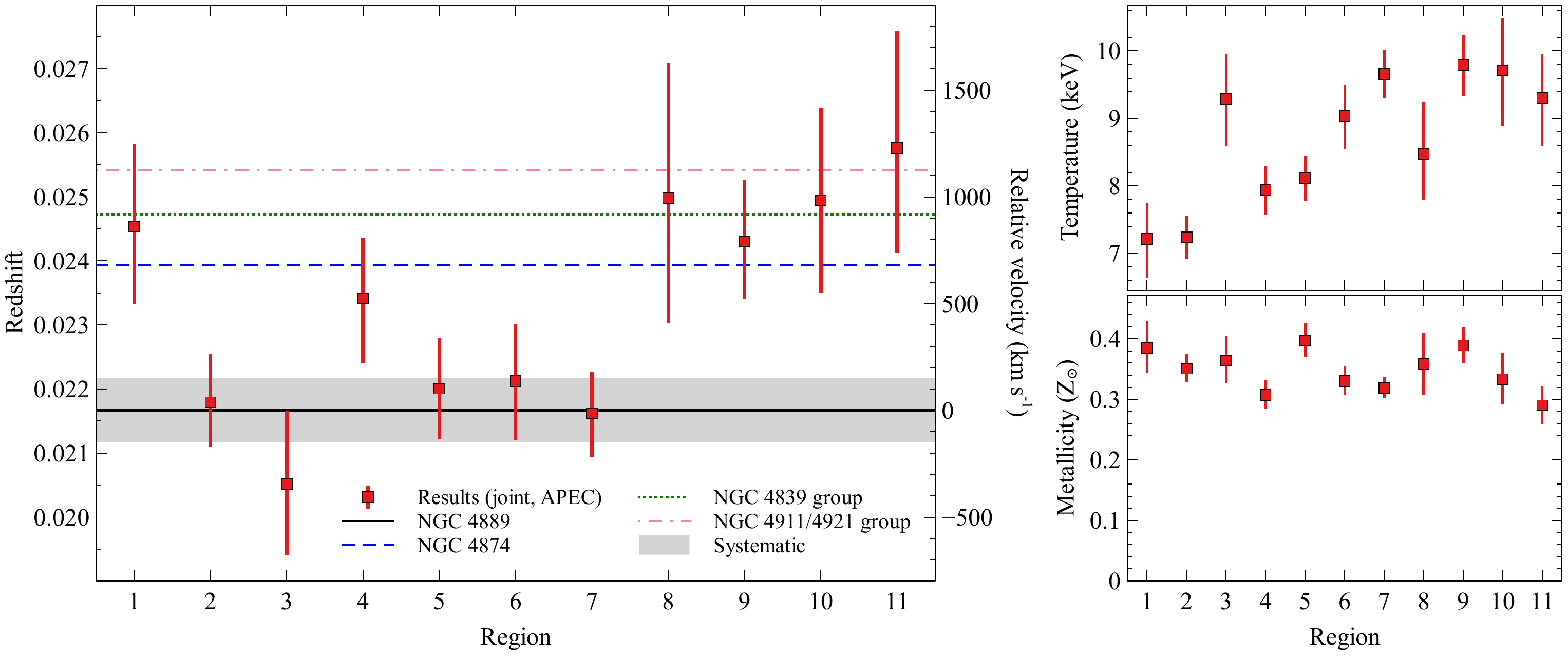}
\caption{
  Results for the Coma regions.
  Shown are the best-fitting values for the joint fits using the APEC model.
  (Left panel)
  Redshifts and velocities.
  The right axis shows the redshifts as velocities relative to the redshift of \object{NGC 4889}.
  (Right panels)
  Temperatures and metallicities.
}
\label{fig:coma_vel}
\end{figure*}

\begin{figure*}
  \centering
  \begin{tabular}{cc}
    \includegraphics[width=0.4\textwidth]{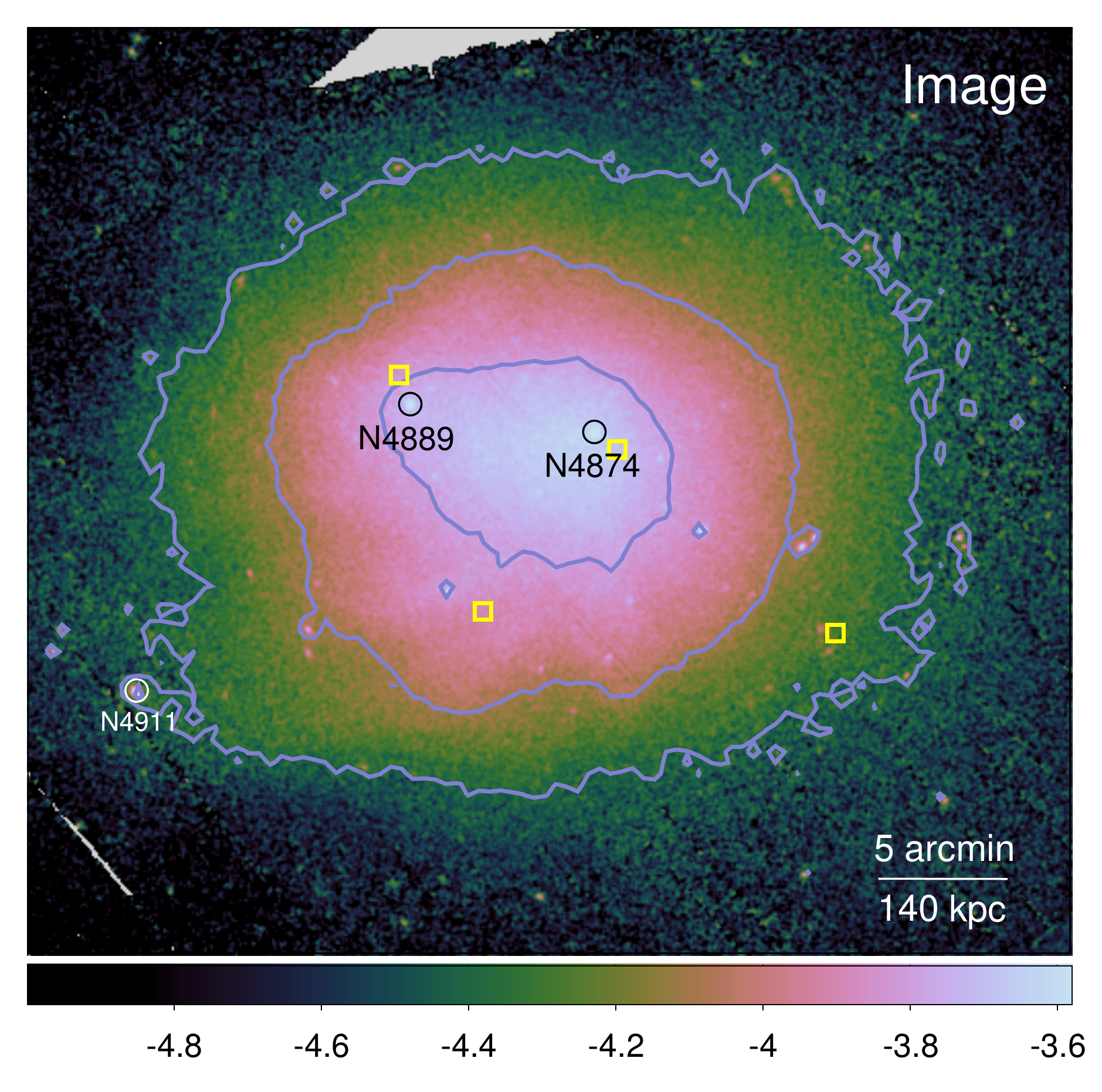} &
    \includegraphics[width=0.4\textwidth]{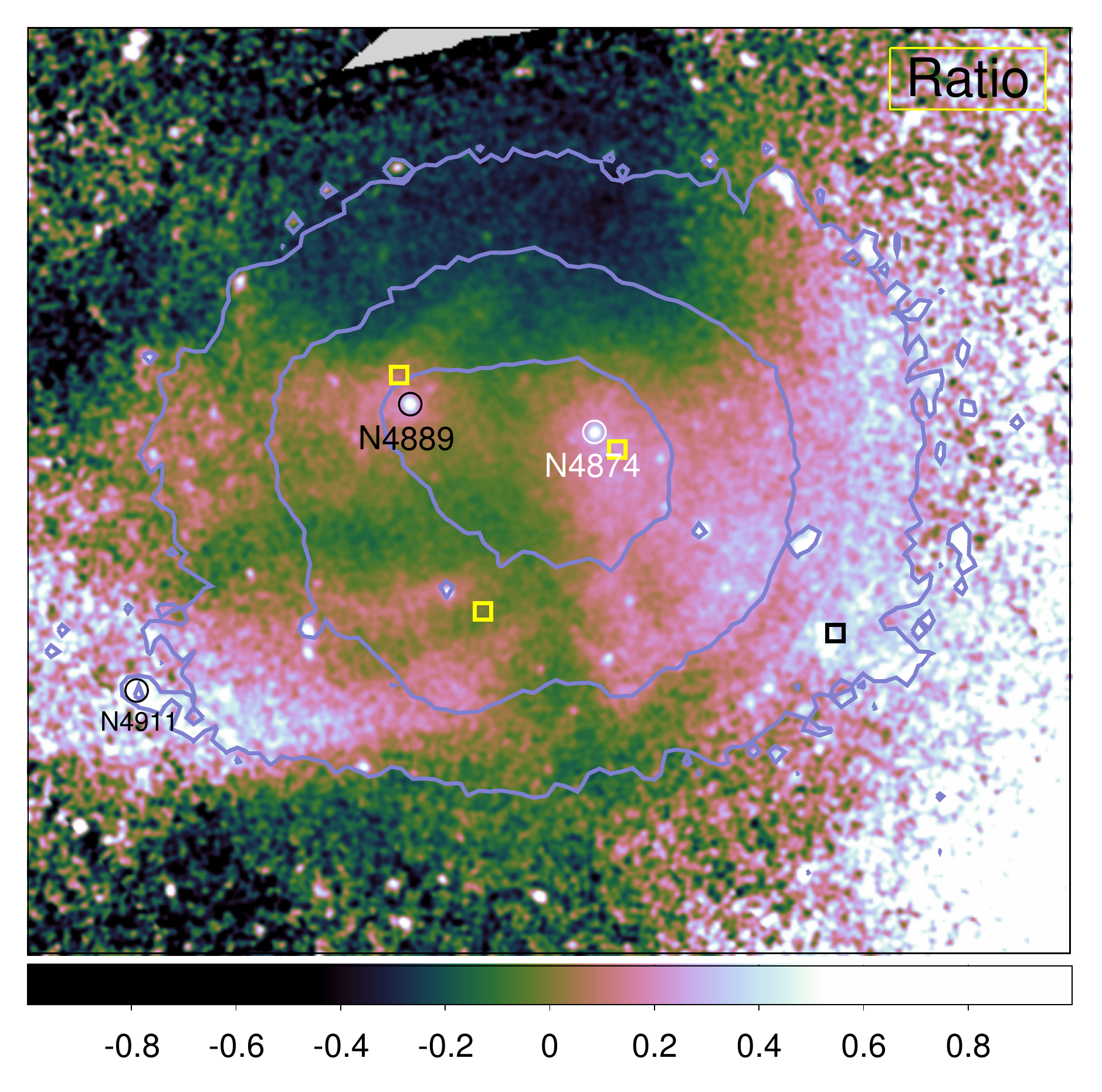} \\
    \includegraphics[width=0.4\textwidth]{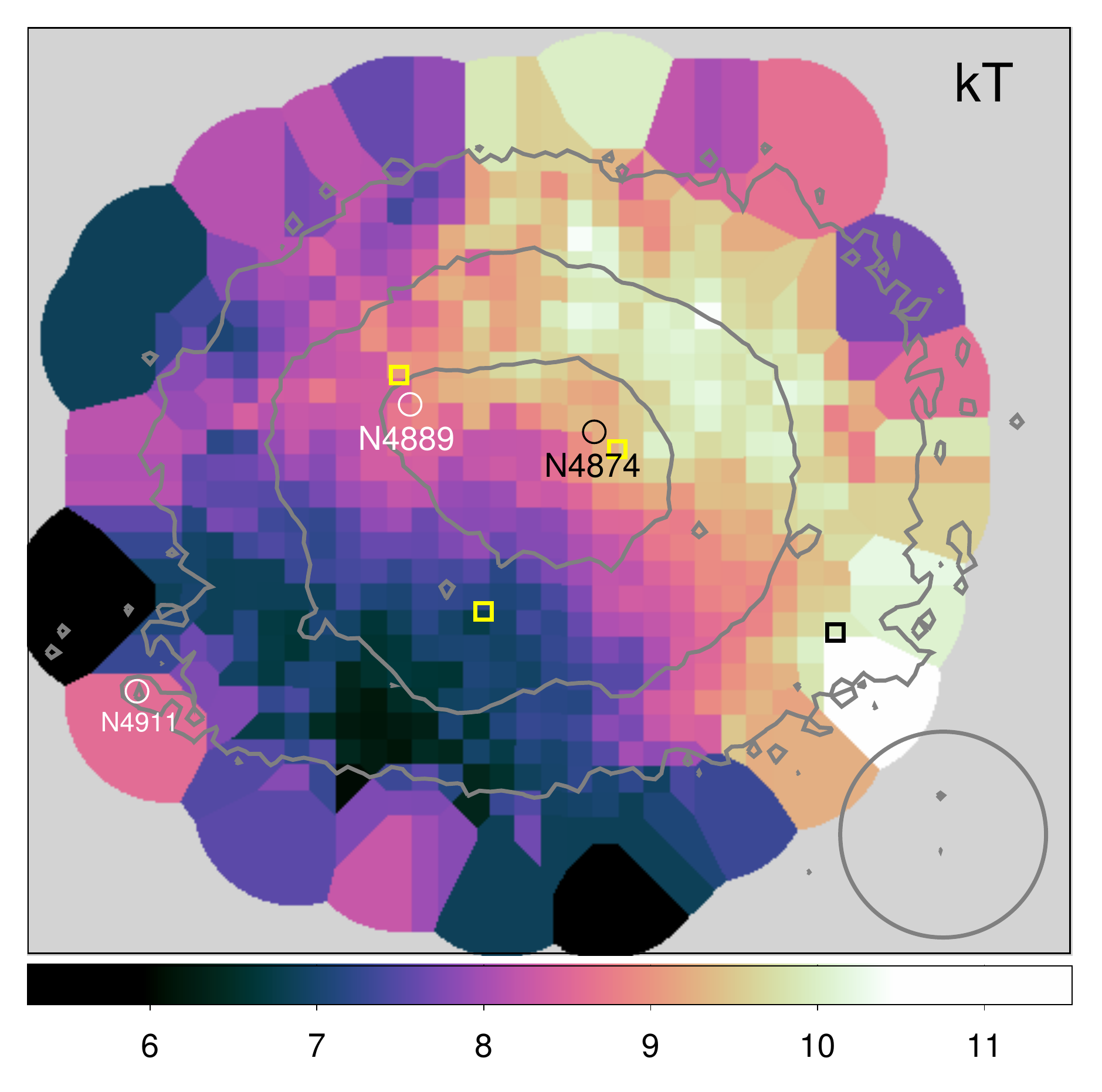} &
    \includegraphics[width=0.4\textwidth]{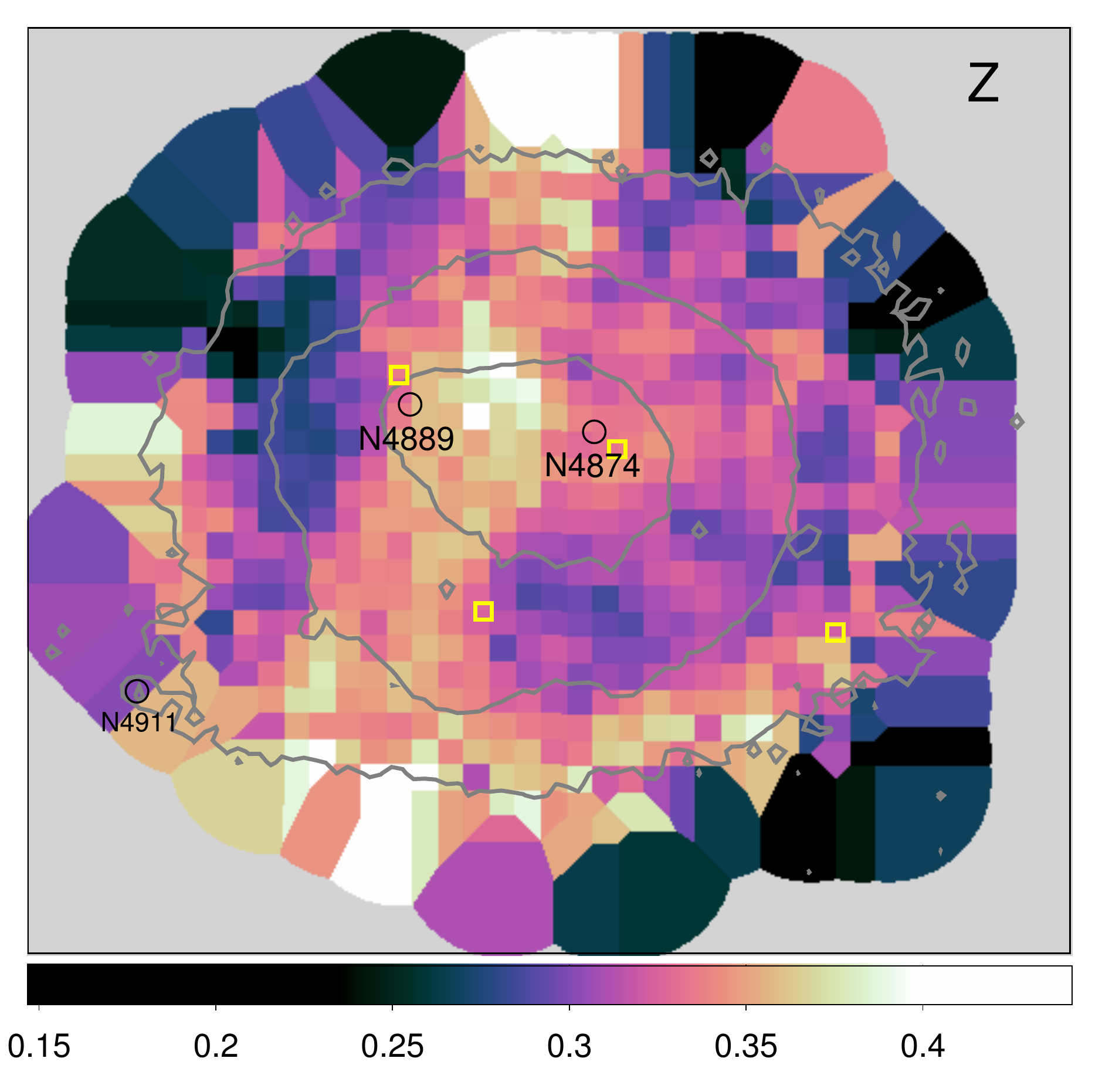} \\
    \includegraphics[width=0.4\textwidth]{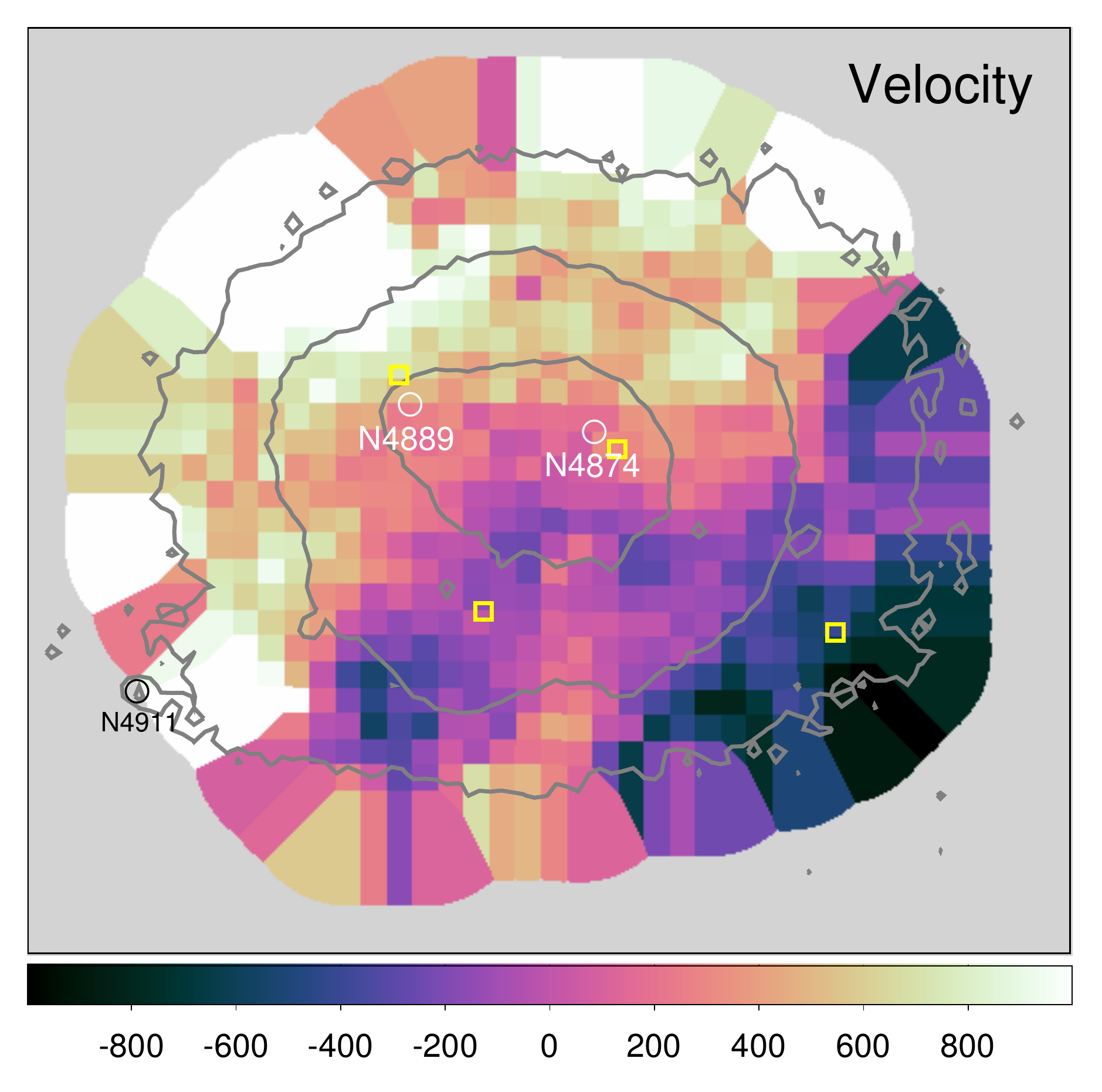} &
    \includegraphics[width=0.4\textwidth]{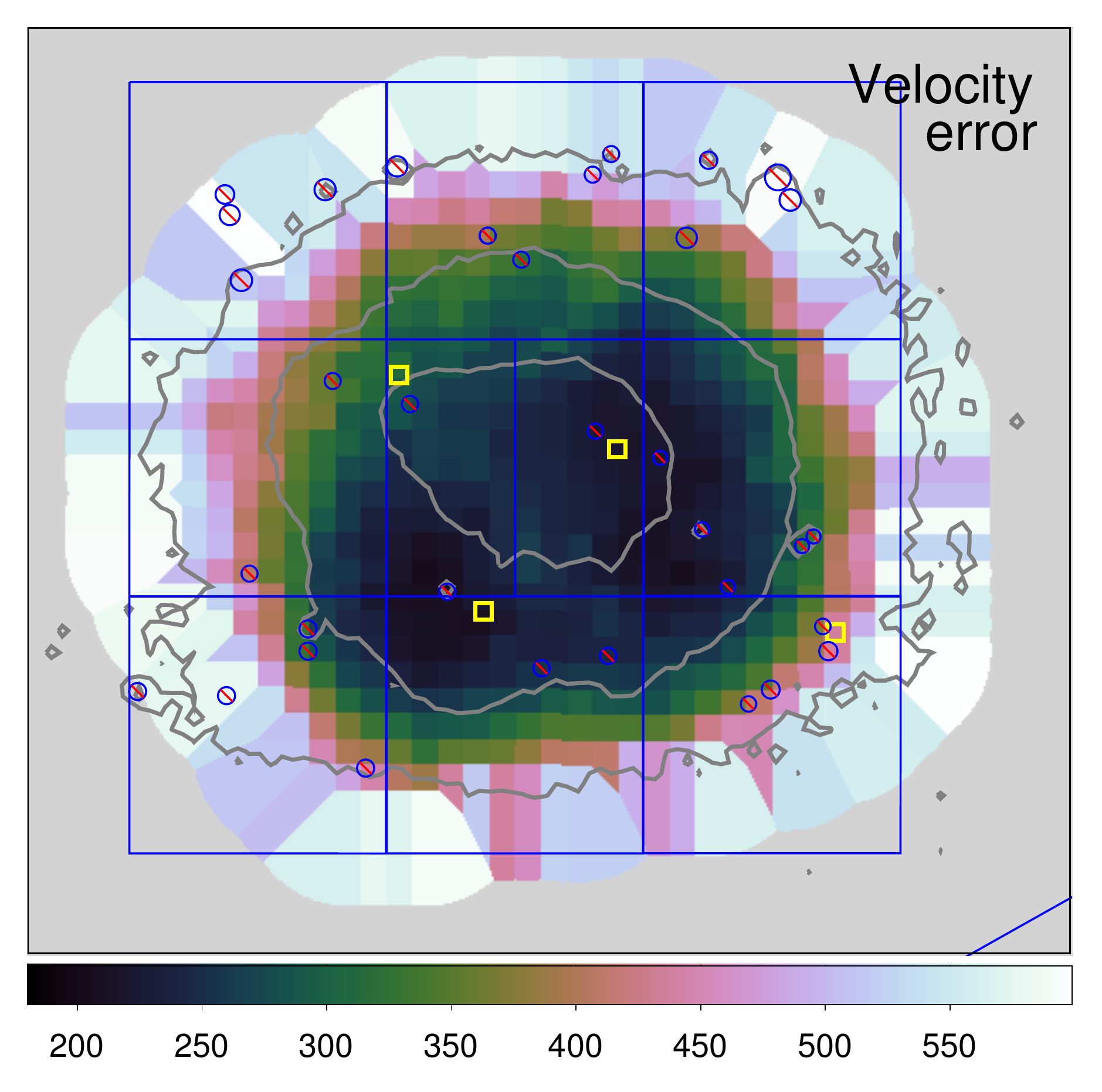} \\
  \end{tabular}
\caption{
  Coma cluster maps. Shown is the 0.5 to 2.0 keV X-ray surface brightness with contours (log cts~s$^{-1}$ per 1.6 arcsec pixel; top left), fractional difference of surface brightness to radial average (top right), temperature (keV; centre left), metallicity ($\textrm{Z}_{\odot}$; centre right), velocity relative to \object{NGC 4889} (km~s$^{-1}$; bottom left) and $1\sigma$ statistical uncertainty on velocity (km~s$^{-1}$; bottom right).
  Spectral results were obtained from spectra extracted from a sliding 4 arcmin radius extraction region (circle) over a grid.
  Regions where the uncertainty on the temperature is $>1\:\textrm{keV}$, metallicity $>0.07\:\textrm{Z}_{\odot}$ or velocity $>600 \:\textrm{km s}^{-1}$ are not shown.
  The position of the four central lensing-map peaks of \cite{Okabe10} are indicated by squares.
}
\label{fig:coma_map}
\end{figure*}

Owing to the lower total number of counts in the Fe-K complex in Coma compared to Perseus, we split the cluster into 11 regions, with ten in the centre and one in the merging \object{NGC 4839} group (Fig.~\ref{fig:coma_regions} left panel).
The group region also contains gas ahead of the group in the cluster, as there were not enough counts to make a pure extraction region, and it is therefore not a clean measurement of its properties.
In this system we are able to get complete coverage of the central region (Fig.~\ref{fig:coma_regions} right panels), because of the large number of separate pointings with different positions.
We arrange our central regions in a $3\times3$ grid, with the central box split into two to separate the two central galaxies.

Figure \ref{fig:coma_vel} (left panel) shows our obtained velocities for the different regions, compared to the velocities of the central galaxies and merging group.
Plotted are the velocities of the two central galaxies, \object{NGC 4889} and \object{NGC 4874}, and modelled values for the \object{NGC 4839} and \object{NGC 4911} groups using a substructure analysis \citep{Adami05}.
The temperatures and metallicities we obtain for the regions are shown in Fig.~\ref{fig:coma_vel} (right panels).
We list the results of the spectral fits in Table \ref{tab:comafit}.

The results show that we obtain gas velocities which are very similar to the optical velocities of the two central galaxies, appearing to cluster around these values.
The regions in the centre, to the west, to the south-west and to the south have velocities consistent with \object{NGC 4889}, while those to the north-west, north, north-east, east and south-east appear to better match \object{NGC 4874}.
The reduced-$\chi^2$ for a single velocity value for regions 1 to 10 is poor ($1.7=15/9$).
If we model the intrinsic velocity distribution of the jointly-fitted results for these regions by a Gaussian, we obtain a central redshift of $0.02250 \pm 0.00045$ and a width of $0.00083^{+0.00062}_{-0.00050}$ ($100-440\:\textrm{km s}^{-1}$), although a Gaussian distribution appears to be a poor model given the data.

We can see the structure in more detail by mapping the parameters on smaller spatial scales.
We moved a 4 arcmin radius circular region across the cluster in increments of 1 arcmin.
The spectra were added in each region to make total spectra and fitted using the same model as previously.
Similarly to Perseus, we used an average response matrix for all regions but created separate weighted ancillary responses for each region.
Figure \ref{fig:coma_map} shows the best-fitting temperature, metallicity and velocity for the nearest overlapping position, when the data are sufficient to constrain the parameters to the level given.
For the velocity parameter we also show the statistical uncertainty for each position.
Statistical uncertainties on the other quantities, such as temperature, increase in the outer parts, because of the declining surface brightness.
We also show images of the cluster to the same scale, and a map generated from the fractional difference of the average surface brightness at each radius.
Values in the maps are not statistically independent, unless they are separated by more than 8 arcmin.

There is interesting structure in the maps, with the temperature, metallicity and velocity providing complementary information.
The gradient in temperature across the core of the cluster has been seen using \emph{ASCA} \citep{Donnelly99}, \emph{XMM} \citep{Arnaud01} and \emph{Chandra} \citep{SandersComa13}.
The boundary between the hot and cold emission has a similar shape to the surface brightness ratio image.
The hotter gas appears to lie outside of the northern surface brightness edge.
Using \emph{Chandra} we found cool arms of dense gas which are hinted at with this analysis.

There is a region extending north-south of high metallicity gas which lies between the two central galaxies.
As the individual regions showed, the gas to north and west of the cluster has a higher redshift than that to the south and east.
The north-south higher metallicity gas appears to have lower velocity than the surrounding material.
There is also a weaker extension of high metallicity gas from \object{NGC 4889} to \object{NGC 4874} and beyond, which also has lower velocity.

To the south-east and south-west of the two central galaxies are regions with around $400\:\textrm{km s}^{-1}$ lower velocities, while to the north the velocity jumps up by around $700\:\textrm{km s}^{-1}$, in line with the measurements for the individual regions.

\subsection{Power spectrum of density fluctuations}

\begin{figure}
  \includegraphics[width=\columnwidth]{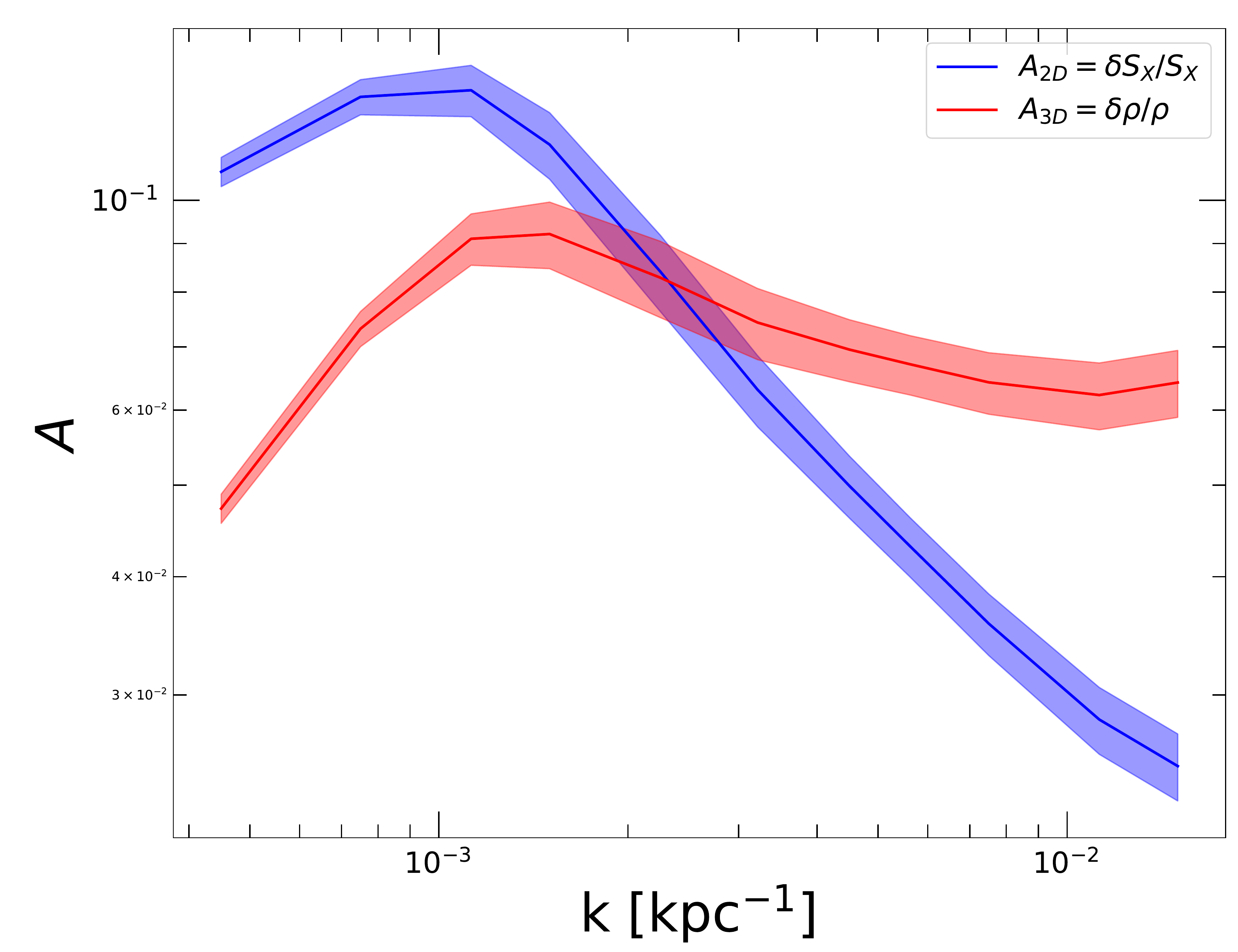}
  \caption{
Amplitude of gas density fluctuations in the Coma cluster as a function of the wave number $k=1/\ell$.
The curves show the amplitude of projected surface brightness fluctuations (red) and the amplitude of 3D density fluctuations (blue).
The uncertainties are indicated by the shaded areas.
  }
  \label{fig:a2d_coma}
\end{figure}

Recent theoretical progress has shown that the velocity field in galaxy clusters can be constrained using the power spectrum of density fluctuations \citep{Schuecker04,Churazov12,Gaspari13}.
In the presence of random, isotropic gas motions, the amplitude of density fluctuations is linearly related to the velocity dispersion of turbulent motions \citep{Gaspari14,Zhu14}, provided other contributions to surface brightness fluctuations are not important \citep{WalkerSlosh18}.
The velocity dispersion indirectly recovered through this method can be compared with the direct measurements we present.

We extracted count images in the 0.5--2.0 keV band from all available \emph{XMM} pointings.
We used the XMMSAS task EEXPMAP to compute an effective exposure map of each observation, taking vignetting and quantum efficiency into account.
To model the particle background, we extracted count images in the same energy band as for our observations from a large set of filter-wheel-closed (FWC) data.
We rescaled the FWC images to match the count rates measured in the unexposed corners of the three EPIC detectors.
We then combined the data from all observations and from the three EPIC cameras to create a mosaic image of the cluster, and we applied the same procedure to the exposure and background maps.
Point sources were detected using the XMMSAS task EWAVELET and excised from the image.
We also masked the region surrounding the infalling galaxy group around \object{NGC 4839} to avoid inducing additional fluctuations in the main cluster.

To determine the amplitude of density fluctuations in Coma, we followed the procedure outlined in \citet{Eckert17}.
Namely, we determined the ellipticity and the ellipse angle of the observed emission using a principal component analysis, and we fitted the large-scale gas distribution with an elliptical beta model, which was found to describe the gas distribution accurately.
We then constructed a model image by convolving the best-fitting model with the exposure map and adding the model particle background.
Finally, we used the modified $\Delta$-variance method of \citet{Arevalo12} to reconstruct the power spectrum from the residual image in a region of 1.5 Mpc radius.
For more details on the reconstruction of the power spectrum we refer to \citet{Arevalo12}.

In Fig. \ref{fig:a2d_coma} we show the amplitude of surface-brightness fluctuations $A_{2D}(k)=\sqrt{2\pi k^2P_{2D}(k)}$ as a function of the wave number $k=1/\ell$.
On the same figure we show the three-dimensional amplitude of density fluctuations $A_{3D}=\delta\rho/\rho$, which is recovered after correcting for projection effects using the window function $W(k)$  \citep[see Eq. 11 of][]{Churazov12}.
We calculated the window function numerically by simulating fluctuations on top of the best-fitting three-dimensional model and computing the ratio between 3D and projected power spectra.

The Coma power spectrum recovered here agrees well with the measurements of \citet{Churazov12} over the common range of scales. The fractional amplitude of density fluctuations peaks at $1/k\approx1$~Mpc.
This agrees with the results of \citet{Khatri16}, albeit with an overall lower normalisation.
The difference with the latter analysis probably lies in the treatment of the NGC 4839 group, which the authors did not mask out from their residual image.
We measure a maximum 3D amplitude $A_{\max}(\delta\rho/\rho)=0.092\pm0.008$.
In case the detected fluctuations can be entirely ascribed to random gas motions, the maximum amplitude is linearly related to the 1D Mach number $\mathcal{M}_{1D}=\sigma_v/c_s$ as $\mathcal{M}_{1D}\approx2.3A_{\max}(\delta\rho/\rho)=0.22\pm0.02$ \citep{Gaspari14}.
Given the average sound speed in the medium $c_s=(\gamma kT/\mu m_p)^{1/2}\approx1480 \:\textrm{km s}^{-1}$, surface-brightness fluctuations in the Coma cluster imply a velocity dispersion $\sigma_{1D}=310\pm25 \:\textrm{km s}^{-1}$, which matches well with the dispersion of the line-of-sight bulk velocities we present.

\subsection{Coma discussion}
The Coma cluster is a complex merger environment.
In addition to the two-subclusters which presumably hosted the two central galaxies, \object{NGC 4889} and \object{NGC 4874}, there is the well-known NGC 4839 group to the south and a group containing NGC 4911/4921 to the east.

Previous studies have suggested that the \object{NGC 4839} group is currently merging with the cluster and has not yet passed through the cluster centre \citep{Neumann01}.
Evidence of this includes the high temperature region between the group and cluster, and its morphology, with the most of the group galaxies on the cluster-side of the group, and a tail of X-ray emission on the other side.
However, it has been recently argued that a post-merger scenario better matches the observed features when compared to simulations \citep{Lyskova19}.
We find that the gas velocity of the group (including some material between the group and cluster) is consistent with the optical velocity of the group.
The gas in the group should be slowed down as it interacts with the cluster by around $540\:\textrm{km s}^{-1}$ \citep{Neumann01}, while the galaxies are not affected.
We do not find evidence of this, although our measurement uncertainties are large, and the spectrum also includes some cluster emission.

Examining the individual regions in the Coma cluster, we find the X-ray gas appears to trace the optical velocities of the two central galaxies.
The central cluster regions suggest velocities which are lower than that of the NGC 4911/4921 group.
A map of the gas velocity (Fig.~\ref{fig:coma_map} bottom-left panel), shows that there is lower velocity material extending from \object{NGC 4889} to the south-east and to the south-west of the cluster core, while the material to the north and east of the core better matches the \object{NGC 4874} velocity.
This material with velocities similar to \object{NGC 4889} also has higher metallicity and lower temperature.
An alternative origin for the high velocity gas is that it is material stripped from NGC 4911/4921 group, although the cool arms seen in the cluster core appear to be lower velocity material.

The higher metallicity north-south filament might be stripped material or a residual core from a merging subcluster, possibly the one which contained \object{NGC 4889}, considering the similarity in velocity.
In addition, there are the arms of cool material extending towards the east which are likely to be stripped from the NGC 4911 subcluster \citep{Neumann03,SandersComa13}.
These arms make a U-shaped structure, however, with \object{NGC 4889} at the tip of one of the arms and the group at the bend.

\cite{Adami05} found that the galaxies within their cluster-core subcomponent have a velocity distribution peaking between the velocities of the two central galaxies.
In contrast, \cite{Gerhard07} found that the main component of the intracluster planetary nebulae within the cluster core have a peak velocity of $\sim 6500$~km~s$^{-1}$ ($z\sim 0.022$), close to that of \object{NGC 4889}, rather than NGC~4874.
Our velocity for the gas in the central region (regions 5 and 6) is similar to \object{NGC 4889}.
However, if we take the whole examined cluster centre (regions 1--10), the modelled Gaussian distribution has a centre of 0.0225, or a weighted average of 0.0227; both of these values are close to the average velocity of the two central galaxies.

\cite{Biviano96} also examined the velocity distribution of the galaxies in the cluster, finding a velocity gradient in the faint galaxies, decreasing from the north-east to south-west.
This is similar to the morphology of the gas velocity seen in our maps, although the galaxy velocity gradient covers a wider spatial area.
This velocity distribution is reminiscent of rotation, although the connection between structures in the velocity map and the other X-ray properties argues against this simple interpretation.

\cite{Okabe10} found several weak-lensing subclumps within Coma, the inner four of which are indicated in Fig.~\ref{fig:coma_map}.
Two of these clumps are clearly associated with the central galaxies.
\cite{AndradeSantos13} modelled the X-ray emission using the inner three mass peaks under the assumptions of slow motions and no gas, and found a good agreement with the ICM distribution.
This model predicts that the south-east subclump lies around 400~kpc from the cluster mid-plane, in contrast with the two central galaxies, which would lie at $\sim 120$ and $\sim 10$~kpc, for NGC 4889 and 4874, respectively.
Examining our maps, it is possible that the subclumps may be associated with some of the features we see.
For example, the south-east subclump is coincident with a region of lower velocity, lower temperature and high metallicity.
It is, however, unclear without further work whether subclumps could be entirely responsible for the structures.

If we take the line-of-sight velocity width of the bulk motions ($V_{\:\textrm{los}}$), we can estimate the ratio of energy density in bulk motions to the thermal energy density, using $f_\textrm{bulk} = V^2_{\:\textrm{los}}\,\mu m_\textrm{p} / kT$ \citep{Werner09}, where $\mu$ is the mean particle mass, $m_\textrm{p}$ is the proton mass and $kT$ is the temperature ($\sim 8.5$ keV).
This measure excludes any turbulent component along the line of sight.
Taking our Gaussian width of $100-440\:\textrm{km s}^{-1}$, implies that $f_\textrm{bulk} < 0.15$, at the $1\sigma$ confidence level.
This estimate, however, assumes a Gaussian distribution of the gas velocity, which may not be the case.
In addition, simulations show that the assumption of a symmetric velocity distribution lead to overestimates of the pressure support from motions \citep{Vazza18}.

\section{Discussion}
\label{sect:discuss}
\subsection{Further possible systematic uncertainties}
\label{sect:systematics}

\begin{figure*}
  \centering
  \includegraphics[width=0.48\textwidth]{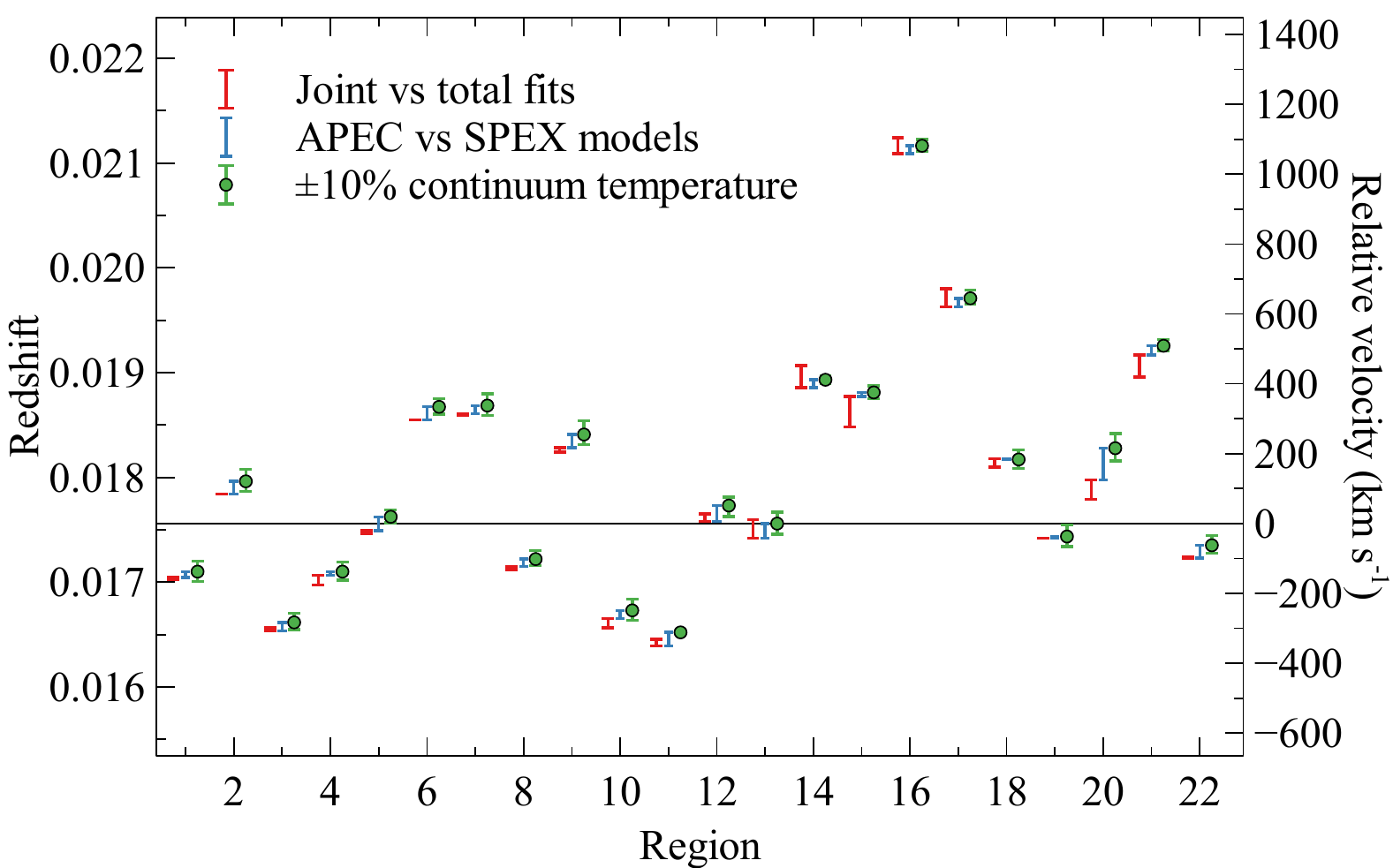}
  \hspace{3mm}
  \includegraphics[width=0.48\textwidth]{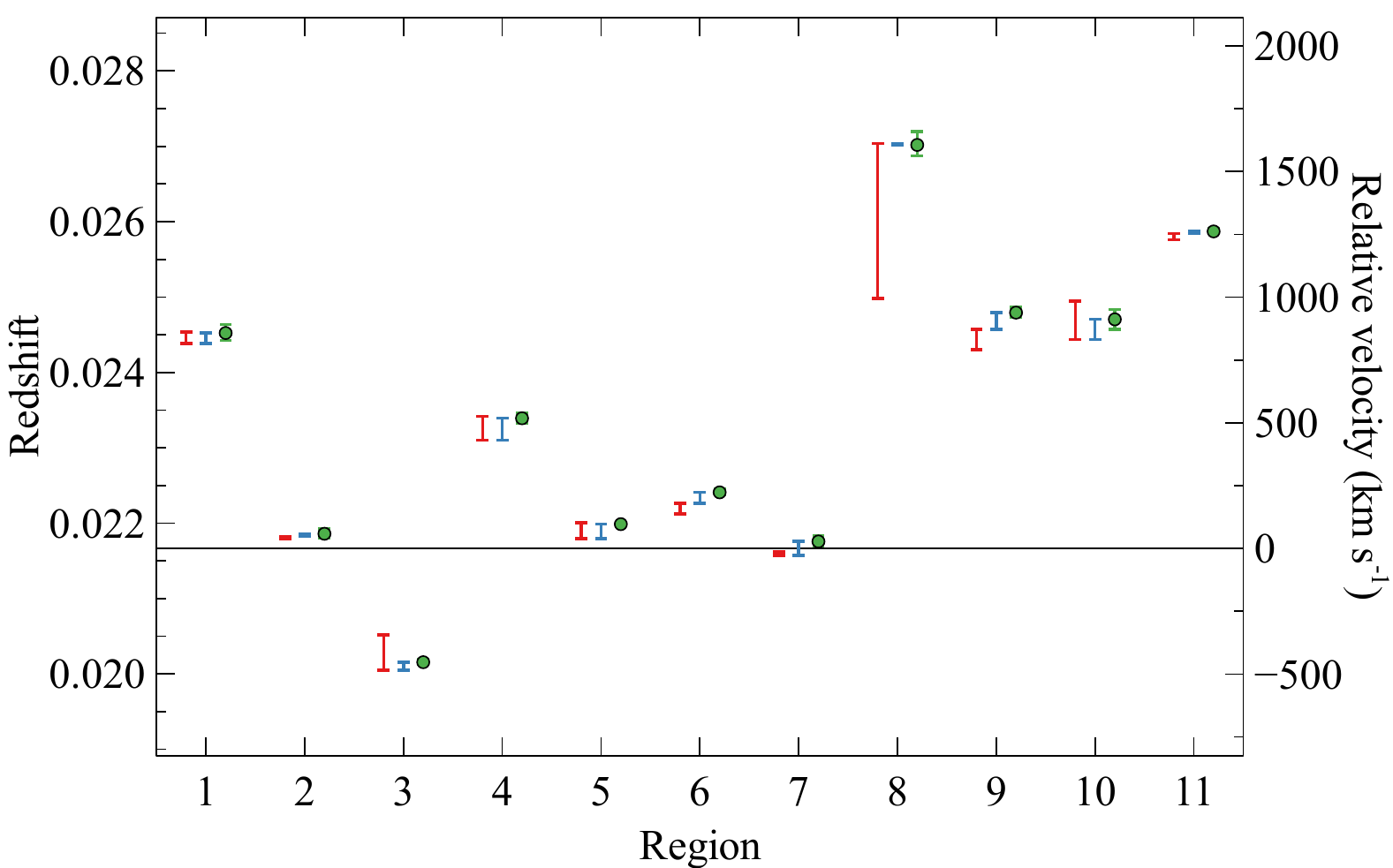}
  \caption{
        How further systematic effects change the obtained Perseus (left panel) and Coma (right panel) redshifts.
    For each region, the ends of the leftmost (red) bars are the results for the joint and for the total spectral fits, using the APEC model.
    The ends of the centre (blue) bars are the results using the APEC and using the SPEX spectral models, when fitting the total spectra.
    The right (green) points show the effect of using a continuum for the SPEX model which is for a temperature which is 10 per cent greater or less than the line temperature, fitting the total spectra.
  }
  \label{fig:systematics}
\end{figure*}

There are possible systematic uncertainties which may be present in addition to the energy calibration uncertainty.
As stated in Sect.~\ref{sect:datanalysis}, we fitted the spectra jointly (producing the results above) or fitted the total spectra.
Figure \ref{fig:systematics} shows the difference between the two by plotting bars between the two sets of results.
The variation in redshift is small (much smaller than the statistical error bars), except for region 8 in Coma.

Systematic errors may also be caused by the choice of spectral model.
Inaccuracies in modelling may lead to shifts in the convolved line shape.
We plot the results for the APEC and SPEX models when fitting the total spectra in Fig.~\ref{fig:systematics}.
There is little difference between the results using the two models, suggesting this effect is not important.
However, deviations from a thermal plasma could theoretically affect our results.

Another effect which may give systematic errors is if the temperatures obtained by spectral fitting are systematically too large or small.
This may occur, for example, if instrument calibration inaccuracies lead to a too flat or too steep spectrum, changing the best-fitting temperature.
\cite{Schellenberger15} found that the temperatures of hot galaxy clusters were different in \emph{Chandra} and \emph{XMM} data, probably caused by calibration differences.
However they found that this effect is relatively small in hard energy bands like we have used, with offsets of less than 10 per cent for high-temperature systems.

To test for errors caused by incorrect temperatures we created modified versions of our SPEX table model, where for a particular line temperature we calculated the model continuum using a temperature which was 10 per cent greater or smaller than this.
We refitted the spectra with these two models, showing the difference between them and the standard model in Fig.~\ref{fig:systematics}.
The effect of continuum inaccuracies appears small for these data.
We found that the obtained temperatures changed by around 3 per cent from the temperatures using the standard SPEX model, indicating that the obtained temperature was dominated by the spectral line shape rather than the continuum.

Another potential systematic is that we previously assumed a slope for the powerlaw background component which was fixed.
To check the effect of this assumption, we refitted the spectra using powerlaws which differed in their index by 0.2.
The changes in the obtained redshifts was very small, with a maximum redshift difference of $10^{-4}$, and an average of $2\times 10^{-5}$.

Our analysis does not apply a heliocentric correction for the motion of the \emph{XMM} spacecraft in the solar system to the energy scale.
This correction, however, is relatively small as the maximum heliocentric velocity is $\sim 30 \:\textrm{km s}^{-1}$.

The fluorescent X-ray emission line originates below the detector, unlike the astrophysical and calibration lines which come from the top.
The detector could have a different response to these two X-ray sources.
Figure \ref{fig:residual-corrfactor} suggests that this effect is not important, as we see consistent averages and widths of residual correction factor between the astrophysical observations (in quadrants) and calibration observations (for each CCD).

These results demonstrate that the above systematic effects are small compared with the measurement and energy calibration uncertainty.

\subsection{Future improvements to the method}
\label{sect:improve}
There are several possible future improvements that could be made to map velocities in extended objects with EPIC-pn.
Firstly, it should be possible to calibrate the energy of events made of doubles, where the signal is split between two pixels.
Including doubles would approximately increase the number of X-ray events by 80 per cent, although they would have poorer spectral resolution (by up to 10 per cent for Fe-K) than the single-pixel events.

We have not attempted to model the Al-K$\alpha$ lines around 1.5 keV from the calibration source.
This would probably not be directly useful for analysing Fe-K from clusters, given the large difference in energy.
However, monitoring the shift of this line may provide more information about the change in energy scale we see at larger energies, allowing us to model its effects better.
The Al-K$\alpha$ is also a detector background line.
We also could examine the fluorescent detector lines from Ti, V, Cr, Au and Mo, although these are much weaker than the other lines we used.
The Mo-K$\alpha$ line is at 17.5 keV and so could provide an extreme reference value for the energy scale.
Ti, V and Mo is also visible in the Cu hole, so would allow some sort of calibration for the central part of the detector.

Furthermore, it may be possible to calibrate the Cu hole region using calibration observations.
Calibrating the central region would allow the analysis of a many archival observations where the cluster is placed in the centre.
If we corrected the overall gain of the detector using Cu-K$\alpha$ as we currently do, we could fit for gain and the energy correction factor in the centre using Mn-K$\alpha$ and $\beta$ in calibration observations.
This technique could be tested in other regions of the detector where we have the other reference lines.
It may, however, be difficult to constrain these parameters in later times in the mission, when the calibration source has weakened.

One could redo the EPIC-pn energy calibration from scratch, working from raw PHA values to build up a complete model of the energy scale and avoiding the multiple correction factors currently employed.
This would probably be a very time-consuming project, however.

Observations where the target is offset from the centre of the detector best make use of our data analysis method.
We have obtained data for the Virgo and the Centaurus cluster using targeted observations of this nature and plan to publish them in future works.

Although \emph{XRISM} with its high-spectral resolution X-ray spectrometer should launch in a few years, \emph{XMM} EPIC-pn has the advantage of a much larger field of view,  allowing larger fractions of a cluster to be probed without many pointings.
In addition, the better angular resolution of \emph{XMM} enables point sources to be more easily excluded from the analysis and for different structures to be separated.

\section{Conclusions}
We describe and demonstrate a novel technique to use background X-ray lines seen in the spectra of the \emph{XMM} EPIC-pn detector, calibrating the absolute energy scale of the outer part of the detector to better than $150\:\textrm{km s}^{-1}$ at Fe-K.
Using this technique we map the bulk velocity distribution of the ICM over a large fraction of the central Mpc of two nearby clusters of galaxies, Perseus and Coma.

We are able to independently confirm our calibration procedure using a a 65~kpc-square region in Perseus, where we find a velocity consistent with results using an overlapping pointing by \emph{Hitomi}.
There is no evidence of significant motions over much of the cluster, and our average velocity measurement is close to the \emph{Hitomi} central pointing.
However, to the east of the cluster core we observe a shift in velocity of $480 \pm 210\:\textrm{km s}^{-1}$ (statistical).
This is spatially coincident with a cold front, seen as an edge in surface brightness and temperature and is direct evidence of gas sloshing in the potential well of the cluster.
Excluding this region, we find that the width ($\sigma$) of the bulk velocity distribution is $20-150\:\textrm{km s}^{-1}$, if modelled with a Gaussian distribution.
This width is consistent to the \emph{Hitomi} line width measurements in the core of the cluster, excluding the regions containing strong feedback.

In the Coma cluster, we find that the velocity of the gas is close to the optical velocities of the two central galaxies.
The ICM to the north and east of the core has a velocity close to that of \object{NGC 4874}, while that to the south and south-west has a lower velocity similar to \object{NGC 4889}.
The NGC 4911/4921 group, which may have passed through the centre of the cluster, has a consistent optical velocity with some of the higher velocity regions to the north and east.
Cooler and denser filaments which were thought be stripped from the subgroup, appear, however, to have a substantially lower velocity close to that of \object{NGC 4889}.
The velocity of material located near the merging subgroup \object{NGC 4839} is consistent with its optical velocity, although the measurement uncertainties are large.

Our results highlight the differences between a relatively relaxed galaxy cluster, such as Perseus, and the product of an ongoing merger, such as Coma.
In the relaxed system, where \emph{Hitomi} previously determined a low amount of turbulence, we see a narrow range in velocity, except for some evidence of the ICM sloshing in the potential well at the level of a few hundred km~s$^{-1}$.
In the merging system, where one would expect high levels of turbulence, we see a large $\sim 1000$~km~s$^{-1}$ range in velocity, with significant structure in our maps.

\begin{acknowledgements}
  JSS thanks M.~Freyberg for helpful discussions.
  HRR is supported by an STFC Rutherford Fellowship and an Anne McLaren Fellowship.
  CP is supported by European Space Agency (ESA) Research Fellowships.
  ACF acknowledges ERC Advanced Grant 340442.
  Based on observations obtained with \emph{XMM-Newton}, an ESA science mission with instruments and contributions directly funded by ESA Member States and NASA.
\end{acknowledgements}

\begin{table}
  \caption{Cluster observations analysed. Shown are the observation identifier, date, nominal pointing position (J2000) and exposure after cleaning.
We note that short observations were effectively excluded from the joint spectral analysis because of the minimum count criterion.
  }
  \centering
  \small
  \begin{tabular}{cccccc}
    OBSID & Date & RA & Dec & Exposure \\
    & & (deg) & (deg) & (ks) \\ \hline
    \multicolumn{5}{|c|}{Perseus cluster} \\
    0085110101 & 2001-01-30 & $49.9507$ & $41.5117$ & $22.4$ \\
    0085590201 & 2001-02-10 & $49.9570$ & $41.0964$ & $35.3$ \\
    0151560101 & 2003-02-26 & $49.1792$ & $41.3247$ & $19.3$ \\
    0204720101 & 2004-02-04 & $50.4108$ & $41.5286$ & $6.0$ \\
    0204720201 & 2004-02-04 & $50.8483$ & $41.5281$ & $20.7$ \\
    0305690101 & 2006-02-10 & $49.5112$ & $41.2833$ & $22.9$ \\
    0305690301 & 2006-02-11 & $49.9608$ & $41.8929$ & $15.6$ \\
    0305690401 & 2006-02-11 & $50.4738$ & $41.8250$ & $22.7$ \\
    0305780101 & 2006-01-29 & $49.9500$ & $41.5113$ & $72.2$ \\
    0405410101 & 2006-08-03 & $50.2500$ & $41.9347$ & $11.4$ \\
    0405410201 & 2006-08-03 & $49.6664$ & $41.1087$ & $8.3$ \\
    0673020201 & 2011-09-10 & $50.7818$ & $41.2099$ & $17.9$ \\
    0673020301 & 2011-08-19 & $50.5417$ & $41.3833$ & $12.0$ \\ \hline
    \multicolumn{5}{|c|}{Coma cluster} \\
    0058940701 & 2003-06-10 & $193.8542$ & $27.2294$ & $12.0$ \\
    0124710101 & 2000-06-21 & $194.1987$ & $27.4019$ & $24.2$ \\
    0124710201 & 2000-06-11 & $194.4271$ & $27.7272$ & $30.0$ \\
    0124710301 & 2000-06-27 & $194.6342$ & $27.4033$ & $14.2$ \\
    0124710401 & 2000-06-23 & $195.0192$ & $27.5233$ & $8.4$ \\
    0124710501 & 2000-05-29 & $194.8646$ & $27.7814$ & $19.9$ \\
    0124710601 & 2000-06-12 & $194.7083$ & $27.9811$ & $4.7$ \\
    0124710701 & 2000-06-24 & $194.3654$ & $28.1447$ & $0.4$ \\
    0124710801 & 2000-12-10 & $195.3567$ & $27.7314$ & $18.7$ \\
    0124710901 & 2000-06-12 & $195.1362$ & $27.9497$ & $16.5$ \\
    0124711101 & 2000-06-24 & $194.6521$ & $28.3989$ & $16.7$ \\
    0124711401 & 2000-05-29 & $194.9446$ & $27.9500$ & $14.2$ \\
    0124712001 & 2000-12-10 & $194.7083$ & $27.9811$ & $10.8$ \\
    0124712101 & 2000-12-10 & $194.3654$ & $28.1447$ & $20.2$ \\
    0124712401 & 2002-06-05 & $195.4592$ & $28.1578$ & $12.1$ \\
    0124712501 & 2002-06-07 & $195.1521$ & $28.4208$ & $21.1$ \\
    0153750101 & 2001-12-04 & $194.9446$ & $27.9500$ & $17.3$ \\
    0300530101 & 2005-06-19 & $194.9300$ & $27.9833$ & $17.0$ \\
    0300530201 & 2005-06-17 & $194.9561$ & $27.9708$ & $1.5$ \\
    0300530301 & 2005-06-11 & $194.9625$ & $27.9426$ & $22.9$ \\
    0300530401 & 2005-06-09 & $194.9445$ & $27.9200$ & $12.8$ \\
    0300530501 & 2005-06-09 & $194.9155$ & $27.9200$ & $18.1$ \\
    0300530601 & 2005-06-07 & $194.8975$ & $27.9426$ & $15.5$ \\
    0300530701 & 2005-06-07 & $194.9039$ & $27.9708$ & $17.3$ \\
    0403150101 & 2006-06-14 & $194.4272$ & $27.3194$ & $27.9$ \\
    0403150201 & 2006-06-11 & $194.4272$ & $27.3194$ & $36.2$ \\
    0403150301 & 2006-06-17 & $194.4198$ & $26.9371$ & $37.6$ \\
    0403150401 & 2006-06-21 & $194.4198$ & $26.9371$ & $45.4$ \\
    0652310201 & 2010-06-18 & $194.3512$ & $27.4978$ & $0.1$ \\
    0652310401 & 2010-06-24 & $194.3512$ & $27.4978$ & $8.7$ \\
    0652310701 & 2010-06-16 & $194.3512$ & $27.4978$ & $3.2$ \\
    0652310801 & 2010-12-03 & $194.3512$ & $27.4978$ & $4.4$ \\
    0652310901 & 2010-12-05 & $194.3512$ & $27.4978$ & $3.4$ \\
    0652311001 & 2010-12-11 & $194.3512$ & $27.4978$ & $0.5$ \\
    0691610201 & 2012-06-02 & $194.3527$ & $27.4952$ & $26.4$ \\
    0691610301 & 2012-06-04 & $194.3527$ & $27.4952$ & $18.3$ \\ \hline
    \label{tab:obs}
  \end{tabular}
\end{table}

\begin{figure*}
  \includegraphics[width=\textwidth]{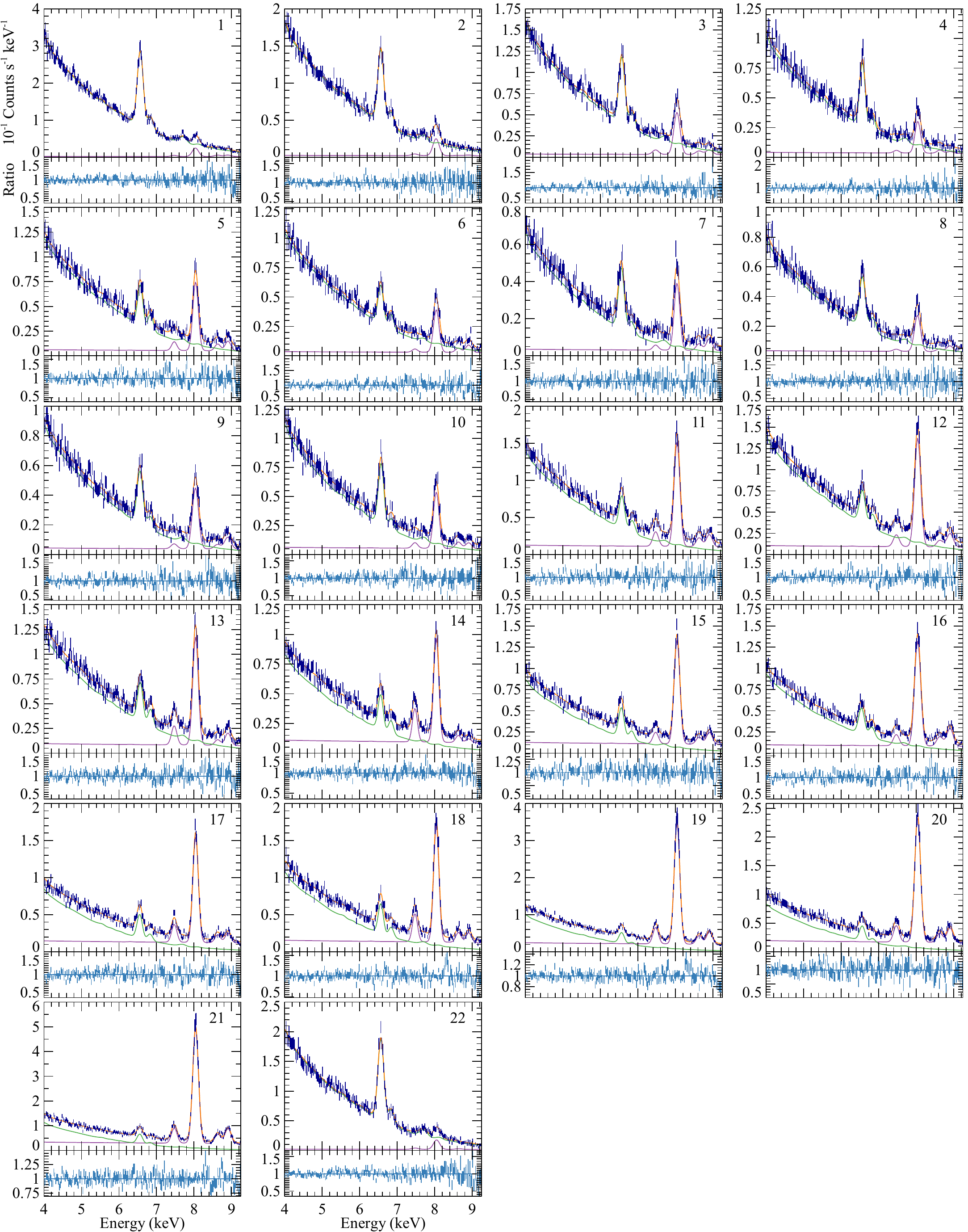}
  \caption{
    Total spectra and APEC fits for the \object{Perseus cluster} regions.
    The top panels shows the spectra (blue), total best-fitting model (orange), foreground components (green) and background components (purple) in units of $10^{-1}\:\textrm{cts}^{-1}\:\textrm{keV}^{-1}$.
    The bottom panels show the ratio between the spectrum and model.
}
\label{fig:perfit}
\end{figure*}

\begin{figure*}
  \includegraphics[width=\textwidth]{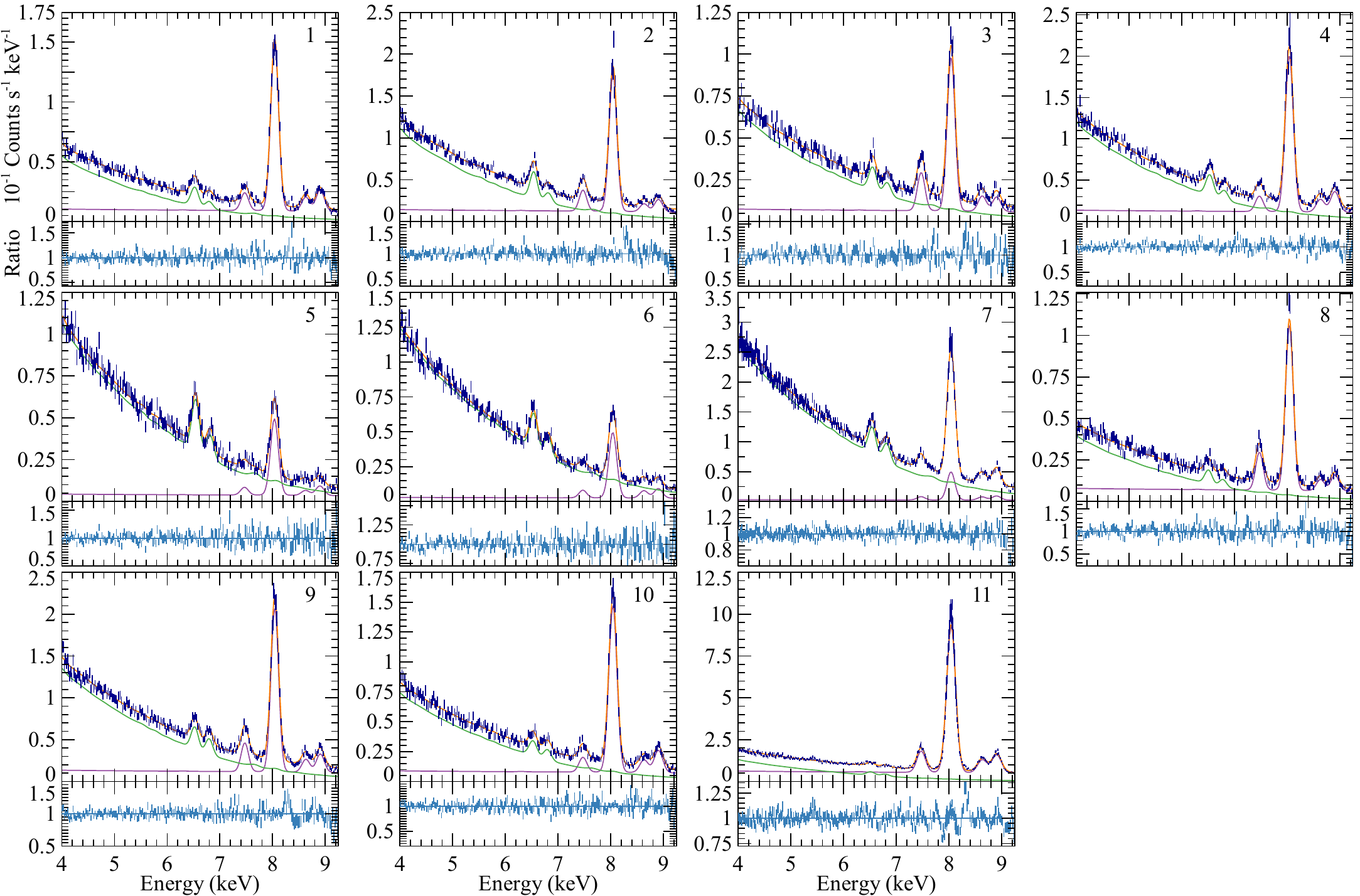}
  \caption{
    Total spectra and APEC fits for the \object{Coma cluster} regions.
    The top panels shows the spectra, total fit, foreground fit and background fit in units of $10^{-1}\:\textrm{cts}^{-1}\:\textrm{keV}^{-1}$.
    The bottom panels show the ratio between the spectrum and model.
}
\label{fig:comafit}
\end{figure*}

\begin{table*}
  \caption{
    Results from fits for the \object{Perseus cluster} regions.
    The columns show temperatures ($kT$; keV), metallicities ($Z$; $Z_\odot$) and redshifts ($z$).
    $^1$ Jointly-fitted results using APEC model.
    $^2$ Total-fitted results using APEC model.
    $^3$ Total-fitted results using SPEX model.
    The spectral fits to the total spectra with the APEC model are shown in Fig.~\ref{fig:perfit}.
  }
  \centering
  \tiny
\begin{tabular}{cccccccccc}
    Region & $kT^1$ & $kT^2$ & $kT^3$ & $Z^1$ & $Z^2$ & $Z^3$ & $z^1$ & $z^2$ & $z^3$ \\ \hline
    $  1$ & $5.4 \pm 0.1$ & $5.4 \pm 0.1$ & $5.6 \pm 0.1$ & $0.55 \pm 0.01$ & $0.55 \pm 0.01$ & $0.58 \pm 0.02$ & $0.0170 \pm 0.0004$ & $0.0170 \pm 0.0004$ & $0.0171 \pm 0.0004$ \\
$  2$ & $5.7 \pm 0.2$ & $5.7 \pm 0.2$ & $5.9 \pm 0.2$ & $0.52 \pm 0.03$ & $0.52 \pm 0.02$ & $0.54 \pm 0.02$ & $0.0178 \pm 0.0005$ & $0.0178 \pm 0.0005$ & $0.0180 \pm 0.0005$ \\
$  3$ & $5.9 \pm 0.2$ & $5.9 \pm 0.2$ & $6.1 \pm 0.2$ & $0.49 \pm 0.02$ & $0.49 \pm 0.02$ & $0.51 \pm 0.03$ & $0.0166 \pm 0.0006$ & $0.0165 \pm 0.0006$ & $0.0166 \pm 0.0006$ \\
$  4$ & $5.8 \pm 0.3$ & $5.8 \pm 0.3$ & $6.0 \pm 0.3$ & $0.48 \pm 0.03$ & $0.47 \pm 0.03$ & $0.49 \pm 0.03$ & $0.0170 \pm 0.0007$ & $0.0171 \pm 0.0007$ & $0.0171 \pm 0.0007$ \\
$  5$ & $7.4 \pm 0.4$ & $7.5 \pm 0.4$ & $7.5 \pm 0.4$ & $0.44 \pm 0.03$ & $0.42 \pm 0.03$ & $0.44 \pm 0.03$ & $0.0175 \pm 0.0008$ & $0.0175 \pm 0.0008$ & $0.0176 \pm 0.0008$ \\
$  6$ & $7.4 \pm 0.4$ & $7.4 \pm 0.4$ & $7.4 \pm 0.4$ & $0.39 \pm 0.03$ & $0.38 \pm 0.03$ & $0.39 \pm 0.03$ & $0.0185 \pm 0.0009$ & $0.0185 \pm 0.0009$ & $0.0187 \pm 0.0009$ \\
$  7$ & $5.6 \pm 0.3$ & $5.6 \pm 0.3$ & $5.8 \pm 0.3$ & $0.54 \pm 0.04$ & $0.54 \pm 0.04$ & $0.56 \pm 0.04$ & $0.0186 \pm 0.0008$ & $0.0186 \pm 0.0008$ & $0.0187 \pm 0.0008$ \\
\hline
  \end{tabular}
  \label{tab:perfit}
\end{table*}

\begin{table*}
  \caption{
    Results from fits for the \object{Coma cluster} regions.
    The columns show temperatures ($kT$; keV), metallicities ($Z$; $Z_\odot$) and redshifts ($z$).
    $^1$ Jointly-fitted results using APEC model.
    $^2$ Total-fitted results using APEC model.
    $^3$ Total-fitted results using SPEX model.
    The spectral fits to the total spectra with the APEC model are shown in Fig.~\ref{fig:comafit}.
  }
  \centering
  \tiny
  \begin{tabular}{cccccccccc}
    Region & $kT^1$ & $kT^2$ & $kT^3$ & $Z^1$ & $Z^2$ & $Z^3$ & $z^1$ & $z^2$ & $z^3$ \\ \hline
    $  1$ & $7.2 \pm 0.6$ & $7.4 \pm 0.5$ & $7.5 \pm 0.5$ & $0.38 \pm 0.04$ & $0.38 \pm 0.04$ & $0.40 \pm 0.04$ & $0.0245 \pm 0.0012$ & $0.0244 \pm 0.0012$ & $0.0245 \pm 0.0012$ \\
$  2$ & $7.2 \pm 0.3$ & $7.2 \pm 0.3$ & $7.3 \pm 0.3$ & $0.35 \pm 0.02$ & $0.35 \pm 0.02$ & $0.37 \pm 0.02$ & $0.0218 \pm 0.0007$ & $0.0218 \pm 0.0007$ & $0.0219 \pm 0.0007$ \\
$  3$ & $9.3 \pm 0.7$ & $8.5 \pm 0.6$ & $8.8 \pm 0.5$ & $0.36 \pm 0.04$ & $0.32 \pm 0.03$ & $0.33 \pm 0.03$ & $0.0205 \pm 0.0011$ & $0.0200 \pm 0.0010$ & $0.0202 \pm 0.0010$ \\
$  4$ & $7.9 \pm 0.4$ & $8.0 \pm 0.3$ & $8.2 \pm 0.4$ & $0.31 \pm 0.02$ & $0.31 \pm 0.02$ & $0.32 \pm 0.03$ & $0.0234 \pm 0.0010$ & $0.0231 \pm 0.0009$ & $0.0234 \pm 0.0009$ \\
$  5$ & $8.1 \pm 0.3$ & $8.0 \pm 0.3$ & $8.1 \pm 0.4$ & $0.40 \pm 0.03$ & $0.38 \pm 0.03$ & $0.40 \pm 0.03$ & $0.0220 \pm 0.0008$ & $0.0218 \pm 0.0008$ & $0.0220 \pm 0.0008$ \\
$  6$ & $9.0 \pm 0.5$ & $8.8 \pm 0.5$ & $8.9 \pm 0.4$ & $0.33 \pm 0.02$ & $0.31 \pm 0.03$ & $0.32 \pm 0.02$ & $0.0221 \pm 0.0009$ & $0.0223 \pm 0.0009$ & $0.0224 \pm 0.0009$ \\
$  7$ & $9.7 \pm 0.3$ & $9.6 \pm 0.3$ & $9.7 \pm 0.3$ & $0.32 \pm 0.02$ & $0.32 \pm 0.02$ & $0.33 \pm 0.02$ & $0.0216 \pm 0.0007$ & $0.0216 \pm 0.0007$ & $0.0218 \pm 0.0007$ \\
\hline
  \end{tabular}
  \label{tab:comafit}
\end{table*}

\bibliographystyle{aa}
\bibliography{refs}

\clearpage

\tiny
\longtab{
\begin{longtable}{llcc|llcc|llcc|llcc}
\caption{\label{tab:datasets} List of astrophysical datasets used for the calibration, showing revolution, observation identifier, mode (M; E is EFF and F is FF) and clean exposure (ks).}\\
Rev. & OBSID & M & Exp. & Rev. & OBSID & M & Exp. & Rev. & OBSID & M & Exp. & Rev. & OBSID & M & Exp. \\
\hline
\endfirsthead
\caption{continued.}\\
Rev. & OBSID & M & Exp. & Rev. & OBSID & M & Exp. & Rev. & OBSID & M & Exp. & Rev. & OBSID & M & Exp. \\
\hline
\endhead
\hline
\endfoot
0029 & 0116320801 &   F &  18.8 & 0033 & 0117710701 &   F &  12.3 & 0033 & 0117710601 &   F &  13.1 & 0048 & 0095011101 &   F &  28.0 \\
0057 & 0122520201 &   E &  24.8 & 0070 & 0123700101 &   F &  30.5 & 0071 & 0123700201 &   F &  29.7 & 0073 & 0123700401 &   F &  12.8 \\
0077 & 0123900101 &   E &  28.4 & 0086 & 0124711401 &   E &  14.3 & 0086 & 0124710501 &   E &  20.5 & 0089 & 0125960101 &   F &  31.4 \\
0093 & 0124710901 &   E &  16.8 & 0093 & 0124710201 &   E &  30.0 & 0098 & 0124710101 &   F &  24.6 & 0099 & 0124711101 &   E &  17.8 \\
0100 & 0112570401 &   F &  23.6 & 0100 & 0097820101 &   F &  34.5 & 0101 & 0124710301 &   F &  14.6 & 0106 & 0110980401 &   E &  27.7 \\
0113 & 0127921001 &   E &  44.6 & 0118 & 0112370101 &   F &  36.9 & 0119 & 0112371001 &   F &  38.2 & 0120 & 0112370301 &   F &  36.1 \\
0122 & 0112371701 &   F &  20.2 & 0128 & 0109920101 &   E &  22.9 & 0129 & 0109930101 &   F &  31.6 & 0131 & 0110660101 &   E &  10.2 \\
0132 & 0112230701 &   E &  11.8 & 0133 & 0112230801 &   E &  15.2 & 0137 & 0112231001 &   E &  20.2 & 0139 & 0112970701 &   E &  16.9 \\
\end{longtable}
}

\longtab{
\begin{longtable}{ccccc|ccccc}
\caption{\label{tab:calib_datasets} List of calibration datasets used, showing revolution, observation identifier, exposure identifier, mode and clean exposure (ks).}\\
    Rev. & OBSID & Exp. ID & Mode & Exp. & Rev. & OBSID & Exp. ID & Mode & Exp. \\
\hline
\endfirsthead
\caption{continued.}\\
    Rev. & OBSID & Exp. ID & Mode & Exp. & Rev. & OBSID & Exp. ID & Mode & Exp. \\
\hline
\endhead
\hline
\endfoot
0042 & 0119100201 & 030 &  FF &  17.6 & 0055 & 0122310201 & 003 &  FF &  31.2 \\
0055 & 0122310301 & 003 &  FF &  49.4 & 0080 & 0124700101 & 003 &  FF &  61.3 \\
0124 & 0128720901 & 001 &  FF &  30.0 & 0125 & 0110010101 & 002 &  FF &  15.3 \\
0125 & 0110010101 & 015 &  FF &  55.0 & 0131 & 0129121101 & 011 &  FF &  16.6 \\
0150 & 0105262301 & 003 &  FF &  23.8 & 0249 & 0108860301 & 010 &  FF &  19.4 \\
0309 & 0060740801 & 002 &  FF &  21.3 & 0332 & 0068940101 & 003 &  FF &  34.2 \\
0402 & 0153950201 & 009 &  FF &  19.9 & 0407 & 0154150201 & 006 &  FF &  19.4 \\
0429 & 0134521501 & 002 &  FF &  24.7 & 0440 & 0153950801 & 002 &  FF &  16.6 \\
\end{longtable}
}

\appendix

\section{Additional consistency checks}
\label{sect:checks}
As there are multiple observations for each object, we can check for consistency between them in different regions.
For these comparisons we employed the same spectral modelling as used in the astrophysical cluster analysis.

There are two observations of the Perseus cluster, 0085110101 and 0305780101, which are aimed close to the central nucleus (with an offset of 10s of arcsec) and have similar roll angles.
We extracted spectra from five different regions.
These include a total comparison region made of an annulus centred on the coordinate 03:19:56.9, +41:31:35 between radii of 6.70 and 12.75 arcmin, designed to exclude the Cu-hole and include most of the detector.
We also made four further regions consisting of this annulus split into the different detector quadrants.
We fitted these spectra in each region with the APEC code plus background model, allowing independent redshifts and normalisations for the observations, but jointly fitting the temperature and metallicity.

\begin{figure}
  \includegraphics[width=\columnwidth]{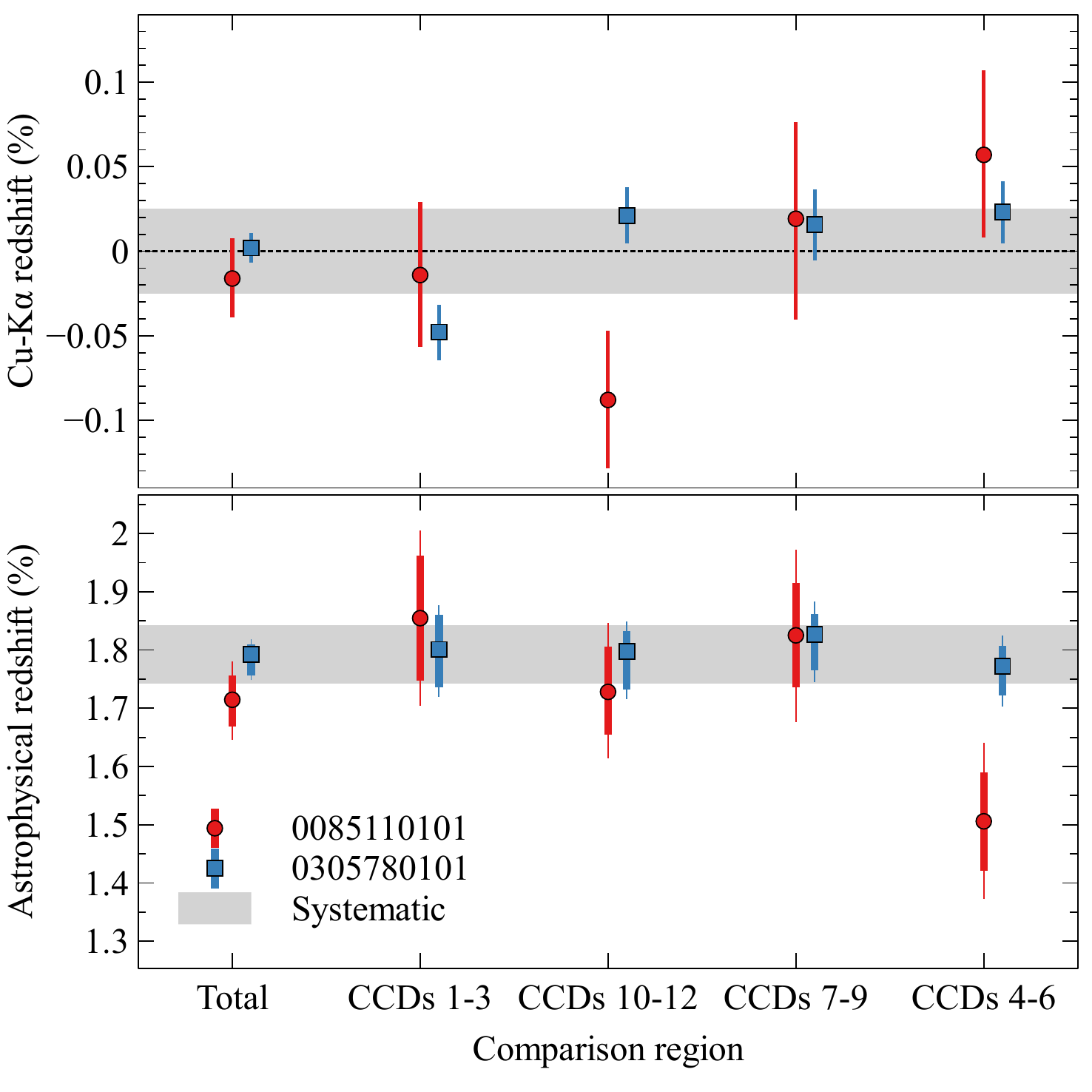}
  \caption{
Comparison of redshift between the two central observations of Perseus, 0085110101 and 0305780101, for different detector regions.
(Top panel) Comparison of Cu-K$\alpha$ redshifts after correction, with an estimated systematic of $75\:\textrm{km s}^{-1}$.
(Bottom panel) Comparison of astrophysical redshifts, with two sets of error bars showing the statistical uncertainty on the redshift after correcting for the Cu-K$\alpha$ best-fitting redshift, and the error bars extended by the uncertainty on the Cu-K$\alpha$ redshift.
Shown is the systematic of $150\:\textrm{km s}^{-1}$.
}
\label{fig:per_consist_check}
\end{figure}

Figure \ref{fig:per_consist_check} shows the Cu-K$\alpha$ redshifts (top panel) and astrophysical redshifts (bottom panel) for each of these different regions.
The corrected Cu-K$\alpha$ redshifts for each of the observations are consistent with zero.
The astrophysical redshifts for each of the observations should also be consistent in each of the regions.

The results agree well between the observations, except for a $\sim 2\sigma$ difference for CCDs 4-6, ignoring the systematic uncertainty.
The original disagreement may be caused by the short 16~ks exposure time for CCDs 4-6 (the exposure for CCDs 1-3 is 31~ks).
Including the systematic uncertainties on the two data points makes the difference insigificant.

\begin{figure*}
  \includegraphics[width=0.48\textwidth]{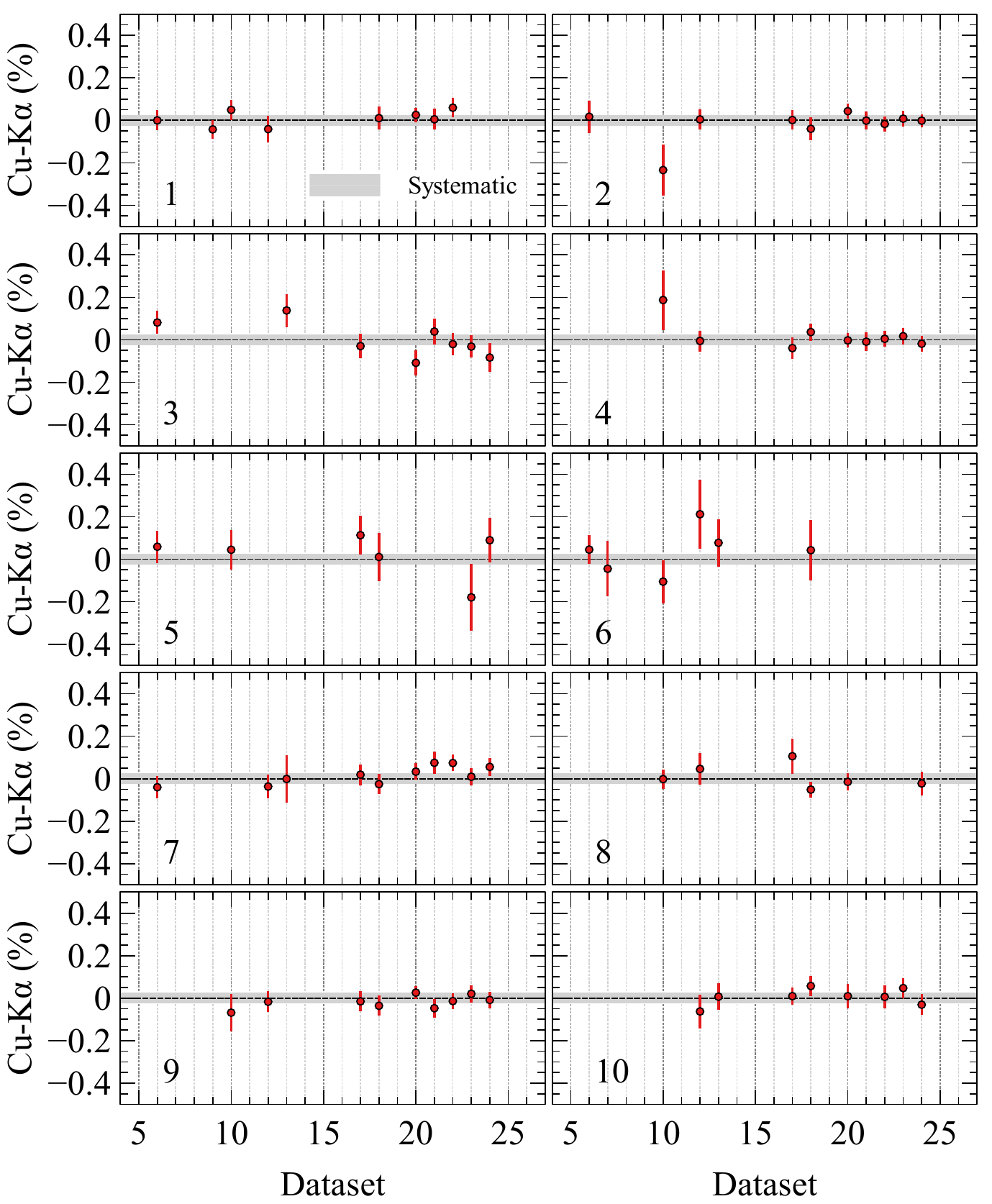}
  \includegraphics[width=0.48\textwidth]{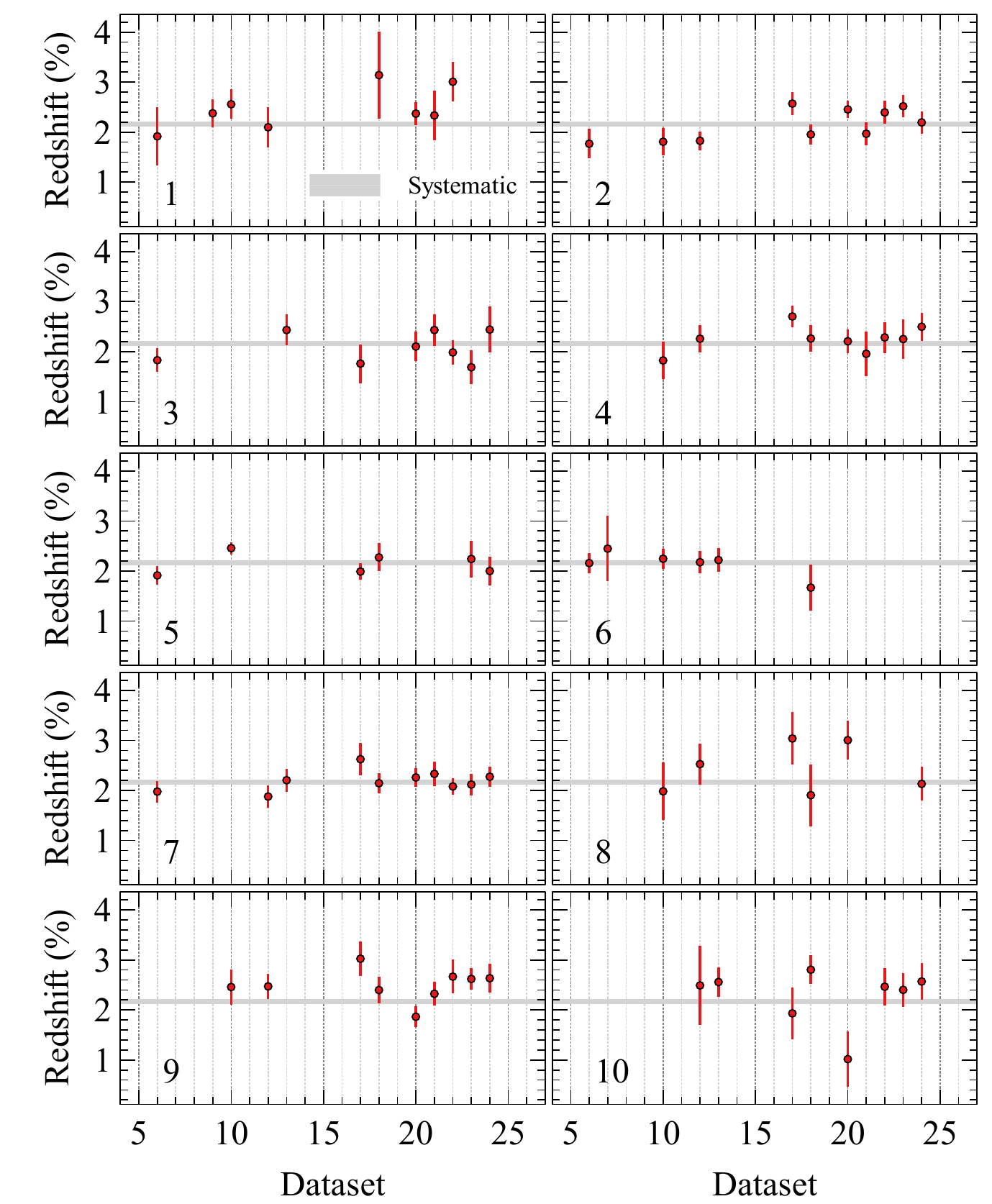}
  \caption{
Cu-K$\alpha$ (left side) and astrophysical (right side) redshifts for regions 1--10 in the Coma cluster, as a function of observation.
The dataset number is the number of the observations in Table \ref{tab:obs}.
Systematic uncertainties of $75$ and $150\:\textrm{km s}^{-1}$, respectively, are shown as a grey shaded region.
\label{fig:coma_z_check}
}
\end{figure*}

\begin{figure}
  \includegraphics[width=\columnwidth]{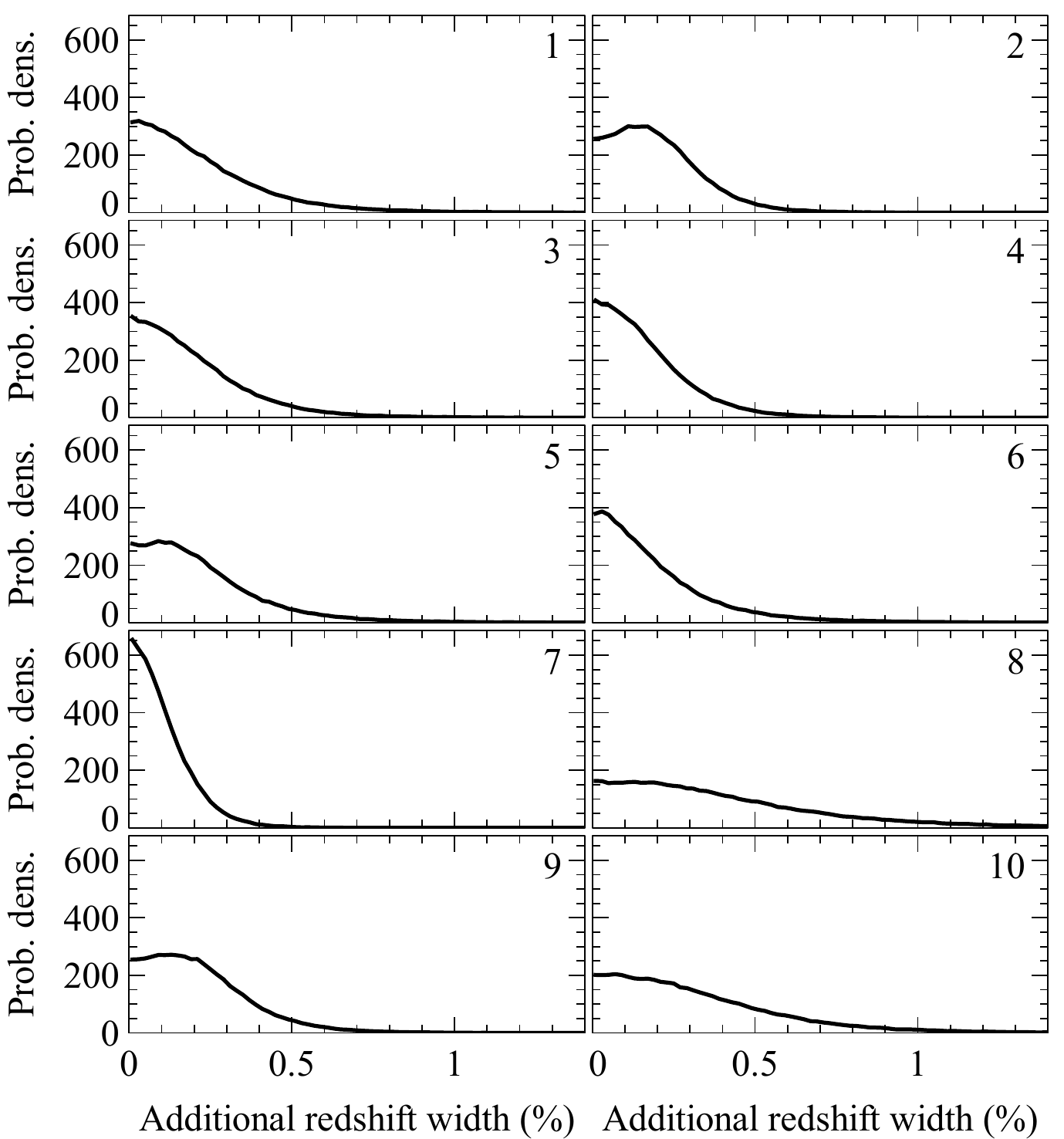}
  \caption{
Posterior probability density distributions for the additional velocitity width ($\sigma$) between observations, for regions 1--10 in the Coma cluster.
}
\label{fig:coma_width_check}
\end{figure}

For the Coma cluster we cannot make exact consistency tests as the observations have a different pointing positions, so any differences may be caused by astrophysical signal.
However, despite this we compared the redshifts for each observation in the extraction regions 1--10 (Fig.~\ref{fig:coma_regions}) and looked for any additional evidence of variation within the region.
In this analysis, we jointly fitted the temperature and abundance for each spatial region, but allowed the redshifts and normalisations to be free for each observation.
Figure \ref{fig:coma_z_check} (left side) shows the obtained Cu-K$\alpha$ redshifts after correction.
The horizontal axis shows the observation, indexed by its order in the Coma observations listed in Table \ref{tab:obs}.
As during the joint fits of astrophysical spectra, we ignored those observations for which there were less than 2000 counts between $4.00$ and $9.25$~keV in the extraction region.
Similarly, Fig.~\ref{fig:coma_z_check} (right side) shows the astrophysical redshifts for the regions and observations.

Both the Cu-K$\alpha$ and astrophysical redshifts appear consistent for each region.
We note again that for the astrophysical results, there may be real astrophysical differences in these datapoints.
There are no obvious trends observed with observation number, however.
To quantify the scatter for the astrophysical redshifts, we modelled the additional width of points beyond the uncertainties for each dataset in a region by a Gaussian model.
We increased the statistical uncertainties on the astrophysical redshift input datapoints by the uncertainties on the Cu-K$\alpha$ redshift and by a systematic of $150 \:\textrm{km s}^{-1}$, adding in quadrature.
Figure \ref{fig:coma_width_check} shows the posterior probability distribution on the Gaussian $\sigma$, obtained from a Markov Chain Monte Carlo analysis assuming a Gaussian distribution.
There appears to be no significant evidence of a width beyond the statistical uncertainties in any region.
In regions 2, 5 and 9 there is very marginal evidence of additional spread, which may be because of the velocity structure seen in the maps (Fig.~\ref{fig:coma_map}).

\end{document}